\documentclass[preprint,secnumarabic,amssymb,superscriptaddress, nobibnotes, aps, pre]{revtex4-2}

\usepackage{bm}
\usepackage{setspace}
\usepackage{tikz}
\usetikzlibrary{automata,patterns,positioning}
\usepackage{graphicx}
\usepackage{epstopdf, epsfig}
\usepackage{makecell}
\usepackage{bm}
\usepackage{mathrsfs}
\usepackage{amsmath}
\usepackage{array}
\usepackage{enumitem}

\graphicspath{ {Figures/} }

\newcommand\CouPois{MHD-Cou\-ette--Shercliff}

\newcommand\TolSchl{Toll\-mien--Schl\-icht\-ing}

\newcommand\AdamBash{Adams--Bash\-forth}
\newcommand\threed{three-di\-men\-sion\-al}

\newcommand\twod{two-di\-men\-sion\-al}

\newcommand\qtwod{quasi-two-di\-men\-sion\-al}

\newcommand\TS{TS}

\newcommand\Fig{Figure} 
\newcommand\Figs{Figures} 
\newcommand\fig{Fig.}  
\newcommand\figs{Figs.}  
\newcommand\tbl{Table}  

\newcommand{\bnabla}{\boldsymbol{\nabla}}
\newcommand{\bcdot}{\boldsymbol{\cdot}}
\newcommand{\p}{\partial}

\newcommand{\matbm}[1]{\bm{#1}} 

\newcommand\Rez{\Re}
\newcommand\Imz{\Im}

\newcommand\alphaCrit{\alpha_\mathrm{c}}
\newcommand\alphaE{\alpha_\mathrm{E}}
\newcommand\alphaMax{\alpha_\mathrm{max}}
\newcommand\alphaOpt{\alpha_\mathrm{opt}}
\newcommand\deltaS{\delta_\mathrm{S}}

\newcommand\tauOpt{\tau_\mathrm{opt}}
\newcommand\Gmax{G_\mathrm{max}}
\newcommand\Nc{N_\mathrm{c}}
\newcommand\Np{N_\mathrm{p}}
\newcommand\Nf{N_\mathrm{f}}
\newcommand\Rey{\textit{Re}}  
\newcommand\Har{\textit{Ha}}  
\newcommand\ReyD{\Rey_\Delta}
\newcommand\ReyCrit{\Rey_\mathrm{c}}
\newcommand\ReyDCrit{\Rey_\mathrm{\Delta c}}
\newcommand\ReyE{\Rey_\mathrm{E}}
\newcommand\ReyDE{\Rey_\mathrm{\Delta E}}

\newcommand\ReyMarg{\Rey_\mathrm{marg}}
\newcommand\ReyS{\Rey_\mathrm{S}}

\newcommand\UsubR{U_\mathrm{R}}

\newcommand\UD{U_\Delta}
\newcommand\timeD{t_\Delta}

\newcommand\tauDOpt{\tau_{\Delta\mathrm{opt}}}
\newcommand\meanfoco{\bar{c}_\kappa}
\newcommand\meanfocoti{\langle \bar{c}_\kappa \rangle_t}

\newcommand{\vect}[1]{\bm{#1}}
\newcommand\ii{\mathrm{i}}
\newcommand\dUP{\mathrm{d}}
\newcommand{\de}[2]{\frac{\dUP #1}{\dUP #2}}
\newcommand{\pde}[2]{\frac{\partial #1}{\partial #2}}
\newcommand{\pdesqr}[2]{\frac{\partial^2 #1}{\partial #2^2}}
\newcommand\bigO[1]{\textit{O}\!\left(#1\right)}

\begin{document}

\title{Transition to turbulence in \qtwod\ MHD flow driven by lateral walls}%

\author{Christopher J. Camobreco}%
\email{christopher.camobreco@monash.edu}
\affiliation{Department of Mechanical and Aerospace Engineering, Monash University, VIC 3800, Australia}
\author{Alban Poth{\'e}rat}%
\email{alban.potherat@coventry.ac.uk}
\affiliation{Fluid and Complex Systems Research Centre, Coventry University, Coventry CV15FB, United Kingdom}
\author{Gregory J. Sheard}%
\email{greg.sheard@monash.edu}
\affiliation{Department of Mechanical and Aerospace Engineering, Monash University, VIC 3800, Australia}
\date{\today}%

\begin{abstract}
This manuscript has been accepted for publication in Physical Review Fluids, see https://journals.aps.org/prfluids/accepted/d5074S28J6b11905012b7cb06505e8f2149dd5f20 for the online abstract.

This work investigates the mechanisms that underlie transitions to turbulence in a three-dimensional domain in which the variation of flow quantities in the out-of-plane direction is much weaker than any in-plane variation. This is achieved using a model for the \qtwod\ magnetohydrodynamic flow in a duct with moving lateral walls and an orthogonal magnetic field, where three-dimensionality persists only in regions of asymptotically small thickness. In this environment, conventional subcritical routes to turbulence, which are highly three-dimensional (with large variations from non-zero out-of-plane wavenumbers), are prohibited. To elucidate the remaining mechanisms involved in quasi-two-dimensional turbulent transitions, the magnetic field strength and degree of antisymmetry in the base flow are varied, the latter via the relative motion of the lateral duct walls. Introduction of any amount of antisymmetry to the base flow drives the critical Reynolds number infinite, as the \TolSchl\ instabilities take on opposite signs of rotation, and destructively interfere. However, an increasing magnetic field strength isolates the instabilities, which, without interaction, permits finite critical Reynolds numbers. The transient growth obtained by similar \TolSchl\ wave perturbations only mildly depends on the base flow, with negligible differences in growth rate for friction parameters $H\gtrsim30$. Weakly nonlinear analysis determines the local bifurcation type, which is always subcritical at the critical point, and along the entire neutral curve just before the magnetic field strength becomes too low to maintain finite critical Reynolds numbers. Direct numerical simulations, initiated with random noise, indicate that a subcritical bifurcation is difficult to achieve in practice, with only supercritical behavior observed. For $H \leq 1$, supercritical exponential growth leads to saturation, but not turbulence. For higher $3 \leq H \leq 10$, a turbulent transition occurs, and is maintained at $H=10$. For $H \geq 30$, the turbulent transition still occurs, but is short lived, as the turbulent state quickly collapses. In addition, for $H \geq 3$, an inertial subrange is identified, with the perturbation energy exhibiting a $-5/3$ power law dependence on wave number.
\end{abstract}


\maketitle

\section{Introduction}\label{sec:introduction}
This work is concerned with the mechanisms that underpin transitions to turbulence in \qtwod\ (Q2D) shear flows; specifically, flow in a rectangular duct pervaded by a transverse magnetic field. A number of natural and industrial flows exhibit \qtwod\ dynamics, where departures from two-dimensionality are either asymptotically small in amplitude or only occur in regions of asymptotically small thickness (for example boundary layers). This invariably raises the challenge of understanding the appearance of turbulence.
In the context of magnetohydrodynamics (MHD), motivation arises from the search for an efficient design of liquid metal cooling blankets, which extract heat from the adjacent plasma in proposed nuclear fusion reactors \citep{Smolentsev2008characterization}. The strength of the plasma-confining magnetic field, which extends into the adjacent  blanket ducts, makes the flow there mostly \qtwod.  Furthermore, turbulence is rapidly damped via the Lorentz force \citep{Davidson2001introduction}. Though less pertinent to this problem, a second motivation to study Q2D MHD flows has been their remarkable ability to reproduce at laboratory scale the main features of two-dimensional turbulence observed in shallow channel and atmospheric flows
\citep{Sommeria1986experimental,sommeria1988_jfm,lindborg1999_jfm}.

Two- or \qtwod\ MHD turbulence was first encountered as a limit state of three-dimensional MHD turbulence at low magnetic Reynolds number \citep{moffatt1967_jfm,Alemany1979influence,Kolesnikov1974experimental} in domains where out-of-plane boundaries were respectively periodic and no-slip. In this limit, the induced magnetic field can be neglected \citep{Roberts1967introduction}, and predominantly the Lorentz force diffuses momentum along the magnetic field lines \citep{Sommeria1982why}. When the Lorentz force  dominates both diffusive and inertial forces (in the ratios $\mathit{Ha}^{-2}$ and $N^{-1}$, respectively, where $\mathit{Ha}$ and $N$ are the Hartmann number and interaction parameter), the flow becomes two- or \qtwod\, depending on the boundary conditions \citep{schumann1976_jfm,Zikanov1998direct,thess2007_jfm,potherat2010_jfm}. 
Along walls perpendicular to magnetic field lines, viscous forces oppose momentum diffusion by the Lorentz force, forming Hartmann boundary layers of thickness $\sim \mathit{Ha}^{-1}$ \citep{Sommeria1982why,Potherat2015decay}. A cut-off length scale $l_\perp^c\sim N^{2/3}$ separates the larger Q2D scales from the smaller 3D ones \citep{Sommeria1982why,Baker2018inverse}.
However, this cut-off scale cannot drop below that of horizontal viscous friction, so boundary layers parallel to the magnetic field, of thickness $\sim \mathit{Ha}^{-1/2}$, remain intrinsically three-dimensional \citep{Potherat2007quasi}.


The conditions at which 3D MHD turbulence becomes \qtwod\, and the formation of three-dimensionality in Q2D turbulence have been clarified \citep{Sommeria1982why,thess2007_jfm,klein2010_prl,Potherat2014why}.
However, a clear path to Q2D turbulence from a \qtwod\ laminar state is yet to be established.
This question is specifically important in the context of duct flows, and particularly in fusion blanket design.
Indeed, if \qtwod\ turbulence is to arise in blankets, it is unlikely to do so out of  three-dimensional turbulence \citep{Smolentsev2008characterization}.

Research on transition to turbulence in MHD conduits has been mostly experimental \citep{Moresco2004experimental} or based on fully three-dimensional simulations at moderate values of $\mathit{Ha}$ ($<100$) and $N$, when the turbulent 
state can be expected to remain three-dimensional \citep{Krasnov2010optimal,Krasnov2012numerical}.
However, these regimes stand very far from fusion relevant regimes ($\mathit{Ha} \simeq 10^4$). The only study to date approaching these regimes indicated that the 
growth of three-dimensional perturbations in electrically insulating ducts was impeded at Hartmann numbers as low as $\mathit{Ha} \simeq 300$, where the less efficient, \qtwod\ Orr-mechanism remains the only source of transient growth \citep{Cassels2019from3D}.
The corresponding optimal growth stood at least one order of magnitude below its 3D counterpart, raising the question as to whether the sort of subcritical transition normally associated with shear flows may indeed take place in the \qtwod\ limit.


With these limitations in mind, a number of shallow water models can be derived to represent MHD flows in a \qtwod\ state \citep{Sommeria1982why, Buhler1996instabilities,Potherat2000effective,Potherat2011shallow} very much in the spirit of shallow water models in rotating flows \citep{pedlosky87}. 
Such models have proved to be accurate, sometimes surprisingly so, for a number of complex flows ranging from simple straight ducts \citep{Potherat2007quasi,young2014_fed,Cassels2019from3D}, vortex lattices \citep{thess1992_pf1,thess1992_pf,thess1993_pf}, sheared turbulence \citep{Buhler1996instabilities,Potherat2005numerical}, flows around obstacles \citep{Dousset2008numerical,Hussam2012enhancing,Hussam2012optimal,hamid2015_pf,Cassels2016heat} and convective flows \citep{Vo2017linear}, linearly and nonlinearly. The clear advantage of these models is their low computational cost, as full three-dimensional numerics are prohibitively expensive for large $\Rey$, $\Har$ and $N$.
As such, they offer a unique chance to identify and obtain insight into laminar to turbulent transitions in duct flows in these regimes. 

In these regimes, traditional subcritical routes to turbulence may be obstructed, which would be detrimental to the efficient extraction of heat in the blanket coolant ducts \cite{Smolentsev2008characterization}. Hence, beyond the classical Shercliff profile of insulating ducts \citep{hunt1971_arfm}, it is legitimate to consider whether alternative profiles may more efficiently generate turbulence, or be less prone to suppressing it. 
As modifications to the base flow appear to be a more promising 
direction for turbulence suppression than influencing turbulent fields directly \citep{Marensi2019stabilisation}, it is instead worth exploring whether it is more efficient to select an optimal base flow, rather than an optimal perturbation, to generate and sustain turbulence.
Although the flow was not natively \qtwod, Ref.~\cite{Marensi2019stabilisation} and Ref.~\cite{Kuhnen2018destabilizing} applied  forces designed to flatten the base flow away from the walls in an attempt to suppress turbulence. In both cases, the preferred force accelerates flow near the walls, and decelerates flow in the bulk.
Flatter base flows noticeably reduce turbulence production \citep{Budanur2020upper}, and if sufficiently flattened, can relaminarize the flow.
This may take place in plug-like Shercliff flows. Linear transient growth was also found to be a good proxy for turbulent production far from the wall \citep{Budanur2020upper}. 
A different strategy was taken by Ref.~\cite{Hof2010eliminating}, where base flow inflexion points were smoothed to eliminate turbulence. Conversely, Ref.~\citep{young2014_fed} applied the inverse strategy of introducing inflexion points for the promotion of turbulence in MHD duct flows. As such, understanding the role of the base flow in the transition process appears to be crucial both in the fusion context and more generally.
%
In particular, the questions we set out to answer are the following:
\begin{enumerate}[label=(\arabic*)]
\item What are the \qtwod\ linear mechanisms promoting the growth of perturbations in \qtwod\ duct 
flows?
\item What is the nature of the bifurcation to any turbulent states that ensue?
\item Can a subcritical transition take place at fusion-relevant parameters?
\item Do the answers to these questions change, as the base flow profile is varied?
\end{enumerate}

We address these questions by studying a \qtwod\ wall-driven duct flow using the shallow water (SM82) model proposed in Ref.~\cite{Sommeria1982why}, where electromagnetic forces reduce to a linear friction exerted by the Hartmann layers on the bulk flow.
The relative velocity of the walls can be continuously varied to achieve a range of base flows from symmetric to antisymmetric with an inflexion point.
These flows are introduced in Sec.~\ref{sec:problem_formulation}.
We then perform  linear modal and non-modal analyses to identify the linear growth mechanisms (Sec.~\ref{sec:lin} and Sec.~\ref{sec:tgp}).
A lower bound for their activation is obtained via the energy stability method (Sec.~\ref{sec:eng}).
The nature of the bifurcation is then sought through weakly nonlinear stability analysis (Sec.~\ref{sec:wnl}) before addressing the question of the fully nonlinear transition by means of two-dimensional DNS (Sec.~\ref{sec:dns}) over a limited range of parameters.

\section{Problem formulation}\label{sec:problem_formulation}
\subsection{Problem setup}\label{sec:prob_setup}
%
\begin{figure}
    \centering
\begin{tikzpicture}
\draw[dashed,line width = 0.2mm] (-8,-2) -- (-8,0);  
\draw[line width = 0.5mm] (-8,0)  -- (0,0); 
\draw[dashed,line width = 0.2mm] (0,0)   -- (0,-2);  
\draw[line width = 0.5mm] (-8,-2)  -- (0,-2);     
\draw[line width = 0.2mm,<->] (-8,-2.7) -- (0,-2.7) node[anchor=north] {$W=2\pi/\alpha$};
\draw[line width = 0.2mm,<->] (-8.4,-2) node[anchor=east] {$-L$} -- (-8.4,0) node[anchor=east] {$L$};
\draw[line width = 0.3mm,->] (-8,-1) -- (-7.5,-1) node[anchor=north west] {};
\draw[line width = 0.3mm,->] (-8,-1) -- (-8,-0.5) node[anchor= west] {$y$};
\node at (-8,-1) [circle,draw=black,inner sep=0.6mm,line width = 0.3mm] {};
\node at (-7.4,-1.2) {$x$};
\node at (-8.2,-1.2) {$z$};
\draw[line width = 0.1mm] (-8,-1.8) -- (0,-1.8);
\draw[line width = 0.1mm] (-8,-0.2) -- (0,-0.2);
\draw[line width = 0.1mm,->] (-0.8,-2.2) -- (-0.8,-2);
\draw[line width = 0.1mm,<-] (-0.8,-1.8) -- (-0.8,-1.6) node[anchor=south west] {$\deltaS$};
\draw[line width = 0.1mm,->] (-0.8,-0.4) -- (-0.8,-0.2);
\draw[line width = 0.1mm,<-] (-0.8,0) -- (-0.8,0.2) node[anchor=south west] {$\deltaS$};
\node at (-1.8,-0.7) [circle,draw=black,fill=black,inner sep=0.4mm] {};
\node at (-1.8,-0.7) [circle,draw=black,inner sep=2mm] {};
\node at (-2.4,-0.7) {$\vect{B}$};
\node at (-4,0.4) {$u=U_0/U_0=1$, $v=0$, $\hat{\vect{u}}=0$};
\node at (-4,-2.4) {$u=U_1/U_0=\UsubR$, $v=0$, $\hat{\vect{u}}=0$};
\node at (1.4,-1) {\makecell{$\vect{u}(0)=\vect{u}(W)$\\
 $p(0) = p(W)$}};
\fill[pattern=north east lines, pattern color=black] (-8,0) rectangle (0,0.1);
\fill[pattern=north east lines, pattern color=black] (-8,-2) rectangle (0,-2.1);
\end{tikzpicture}
    \caption{Schematic diagram of the problem setup, with characteristic length of the duct half height $L$. The vertical dashed lines denote a periodicity constraint. The thick horizontal lines represent impermeable, no-slip boundaries, where the velocity is fixed (non-dimensional boundary conditions provided), with an extent based on the streamwise wavelength, or corresponding wave number, being considered. Fully developed Shercliff boundary layers form on these walls, of  thickness $\deltaS$, which is a function of the friction parameter $H$. A uniform magnetic field is imposed normal to the page. The fixed out-of-plane Hartmann walls are the sources of the linear friction (not drawn).}
    \label{fig:prob_setup}
\end{figure}
%
An incompressible Newtonian fluid, with density $\rho$, kinematic viscosity $\nu$ and electrical conductivity $\sigma$, flows through a duct of height $2L$ ($y-$direction) and width $a$ ($z-$direction). The flow over a streamwise length $W$ is periodic in the $x-$direction. The duct walls are impermeable, no-slip and electrically insulating. Fluid motion is generated by the streamwise motion of the walls at $y=\pm L$, at dimensional velocities $U_0$ (top) and $U_1$ (bottom). 
A homogeneous magnetic flux density (hereafter magnetic field for brevity) $B\mathbf e_z$ pervades the entire domain. In the limit where the Lorentz force outweighs viscous and inertial forces, the flow is \qtwod, with $z-$variation of pressure and 
velocity exclusively localized in boundary layers on the out-of-plane walls. The bulk velocity outside these layers is $O(\mathit{Ha})$ close to the local $z-$averaged velocity along the duct and accurately represented by the SM82 model \citep{Sommeria1982why},
\begin{equation}\label{eq:SM82_continuity}
\bnabla_\perp \bcdot \vect{u} = 0,
\end{equation}
\begin{equation}
\label{eq:SM82_momentum}
\pde{\vect{u}}{t} + (\vect{u} \bcdot \bnabla_\perp) \vect{u} = - \bnabla_\perp p + \frac{1}{\Rey}\bnabla_\perp^2 \vect{u} - \frac{H}{\Rey} \vect{u},
\end{equation}
where the last term on the RHS of equation (\ref{eq:SM82_momentum}) represents the source of friction. Here, the non-dimensional variables $t$, $p$ and $\vect{u}=(u,v)$ represent time, pressure and the 2D $z-$averaged velocity vector, respectively, while $\bnabla_\perp = (\p_x,\p_y)$ and $\nabla_\perp^2=\p_x^2+\p_y^2$ are the 2D gradient and Laplacian operators, respectively. These were scaled by $L/U_0$, $\rho U_0^2$, $U_0$, $1/L$ and $1/L^2$, respectively. The relevant non-dimensional groupings are the Reynolds number (representing the ratio of inertial to viscous forces at the duct scale)
\begin{equation}
\Rey = \frac{U_0L}{\nu},
\end{equation}
and the friction parameter (representing the ratio of friction in the Hartmann layers to viscous forces at the duct scale)
\begin{equation}
H = 2\frac{L^2}{a^2}\Har=2\frac{L^2}{a^2}aB\left(\frac{\sigma}{\rho\nu}\right)^{1/2}.
\end{equation}
%
The SM82 approximation assumes $\Har \gg 1$ and $\Har^2/\Rey \gg 1$, 
which are obtainable for any $H$ with appropriate choice of $a$, as discussed in Ref.~\cite{Vo2017linear}. 
The last governing  non-dimensional grouping is the dimensionless bottom wall velocity
%
\begin{equation}
\UsubR = \frac{U_1}{U_0}.
\end{equation}
$\UsubR$ varies in the range $[-1,1]$, where the \qtwod\ counterpart of MHD-Couette flow is represented by $\UsubR=-1$ and Shercliff flow by $\UsubR=1$.

\subsection{Base flows}\label{sec:base_flows}

\begin{figure}
\begin{center}
\addtolength{\extrarowheight}{-10pt}
\addtolength{\tabcolsep}{-2pt}
\begin{tabular}{ llll } 
\footnotesize{(a)} & \footnotesize{\hspace{6mm} $H=0.1$}  & \footnotesize{(b)}  & \footnotesize{\hspace{6mm} $H=1$} \\
\makecell{ \vspace{25mm}  \\ \rotatebox{90}{\footnotesize{$y$}} \vspace{33mm}} & \makecell{\includegraphics[width=0.458\textwidth]{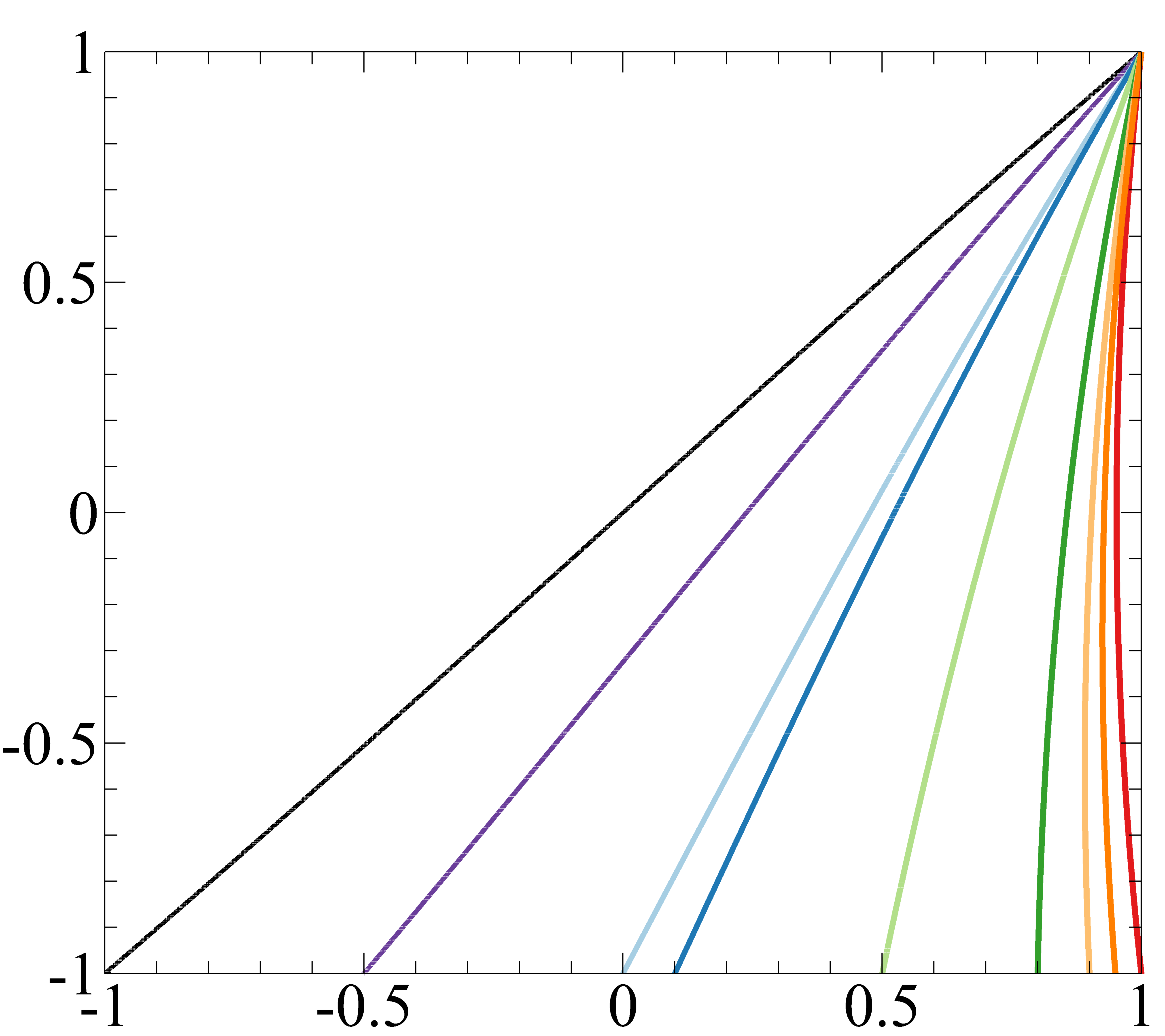}} &
\makecell{\vspace{25mm}  \\ \rotatebox{90}{\footnotesize{$y$}} \vspace{33mm} } & \makecell{\includegraphics[width=0.458\textwidth]{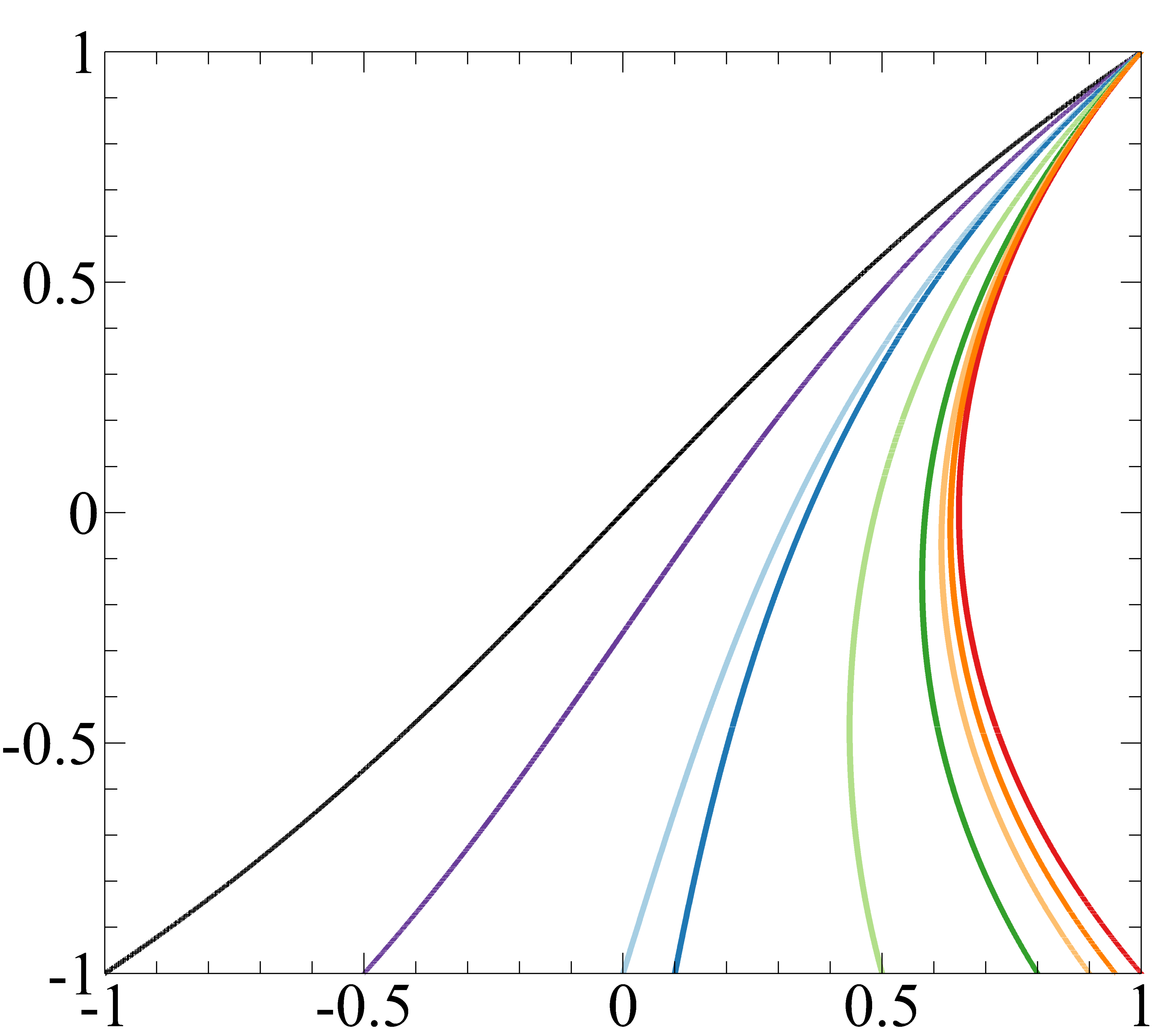}} \\
 & \hspace{36mm} \footnotesize{$U(y)$} &  & \hspace{36mm} \footnotesize{$U(y)$} \\
\footnotesize{(c)} & \footnotesize{\hspace{6mm} $H=10$} & \footnotesize{(d)}  & \footnotesize{\hspace{6mm} $H=100$} \\
\makecell{ \vspace{25mm}  \\ \rotatebox{90}{\footnotesize{$y$}} \vspace{33mm}} & \makecell{\includegraphics[width=0.458\textwidth]{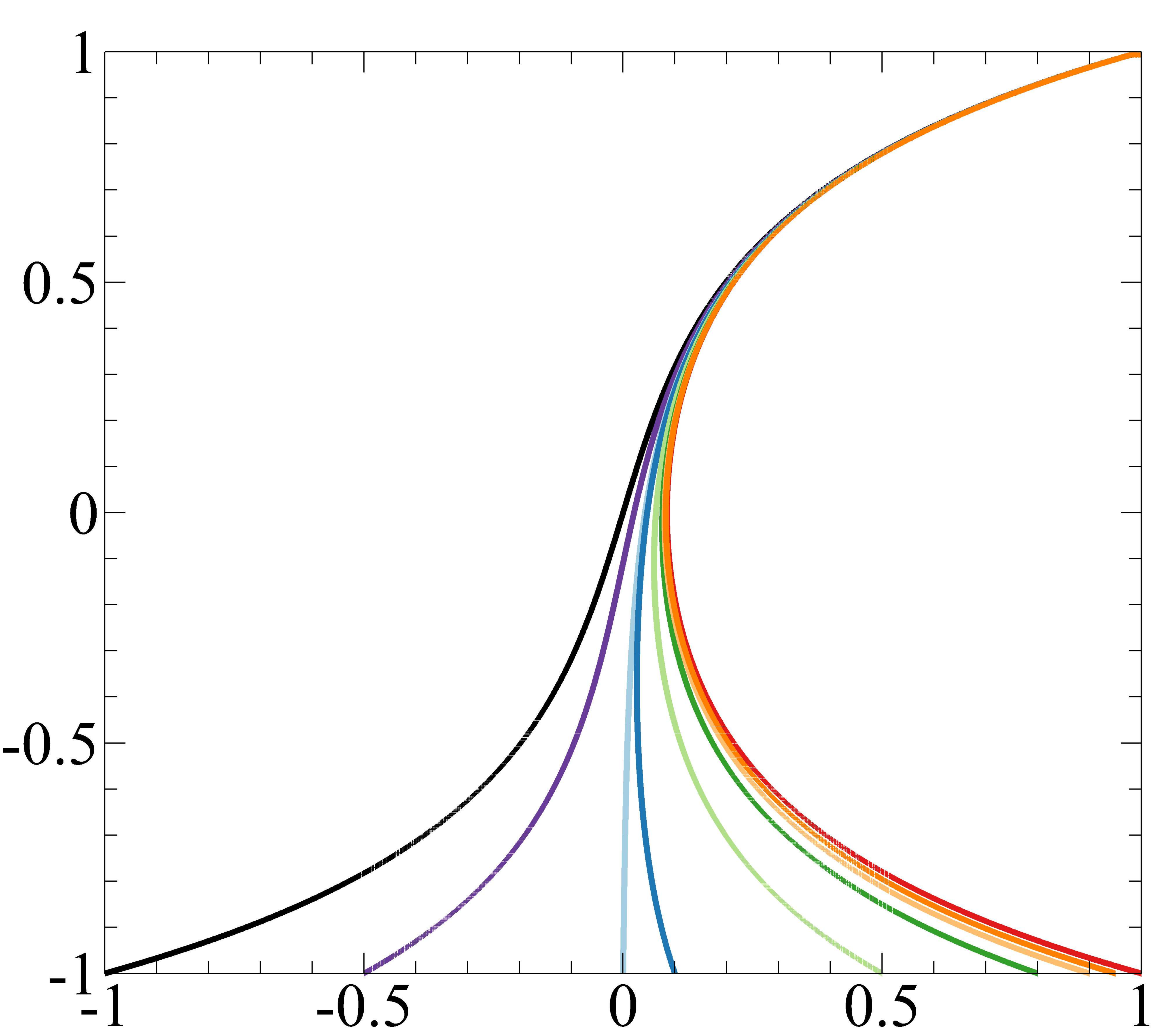}} &
\makecell{\vspace{25mm}  \\ \rotatebox{90}{\footnotesize{$y$}} \vspace{33mm} } & \makecell{\includegraphics[width=0.458\textwidth]{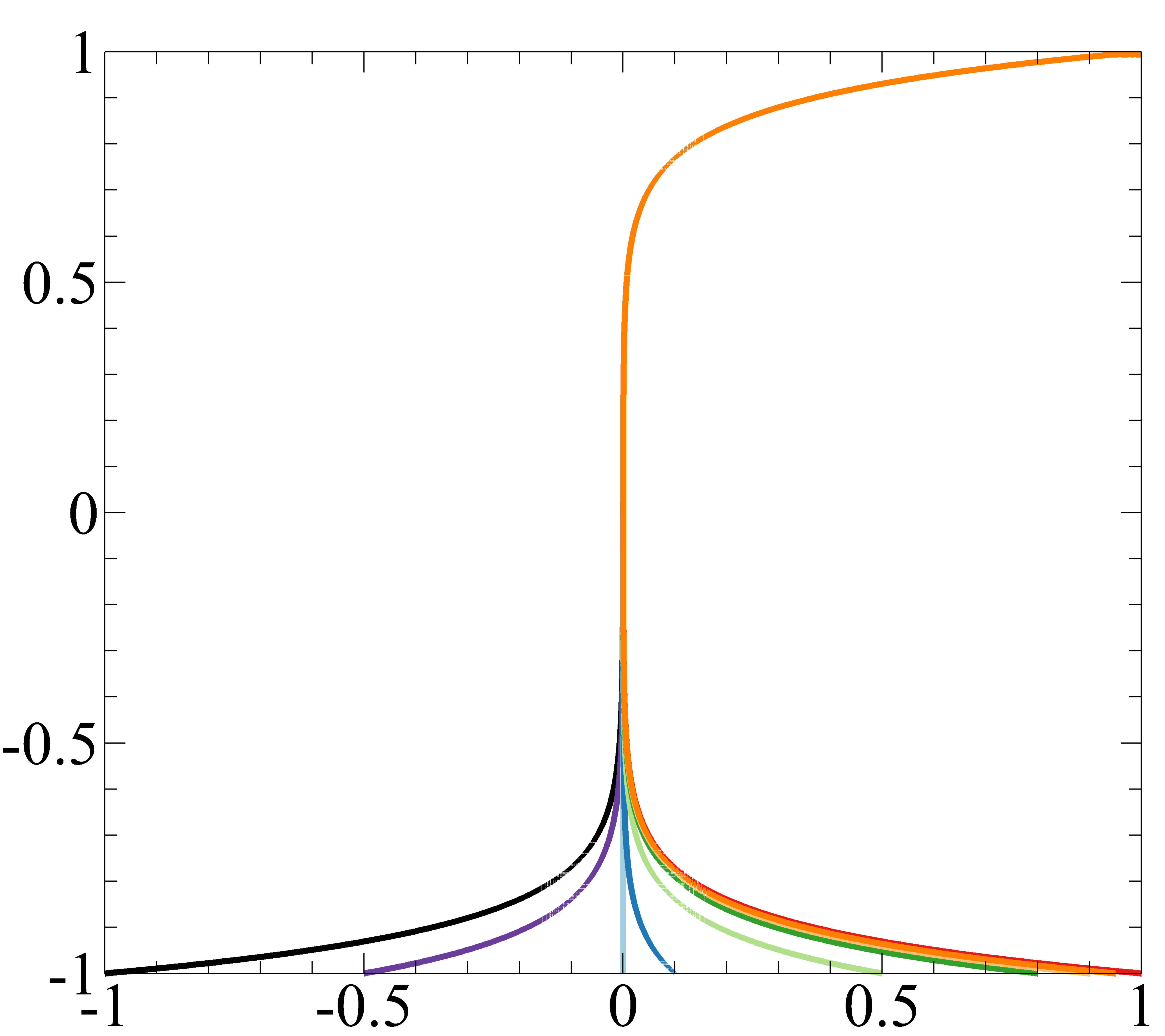}} \\
 & \hspace{36mm} \footnotesize{$U(y)$}  &  & \hspace{36mm} \footnotesize{$U(y)$}  \\
\end{tabular}
\addtolength{\tabcolsep}{+2pt}
\addtolength{\extrarowheight}{+10pt}
\end{center}
    \caption{Base flow profiles $U(y)$, at various $H$, for $\UsubR$ = $-1$, $-0.5$, $0$, $0.1$, $0.5$, $0.8$, $0.9$, $0.95$, $1$.}
    \label{fig:vel_profiles}
\end{figure}
The steady, fully developed solution for the parallel base flow, $\vect{U}=U(y)\vect{e_x}$, without a driving pressure gradient, is
\begin{equation}
U(y) = C_1\exp(-H^{1/2}y)+C_2\exp(H^{1/2}y),
\end{equation}
where
\begin{equation}
C_1 = \frac{\UsubR\exp(H^{1/2}) - \exp(-H^{1/2})}{\exp(2H^{1/2})-\exp(-2H^{1/2})}, \, C_2 = \frac{\exp(H^{1/2}) - \UsubR\exp(-H^{1/2})}{\exp(2H^{1/2})-\exp(-2H^{1/2})}.
\end{equation}
Example base flows for various values of $\UsubR$ are provided in \fig\ \ref{fig:vel_profiles}.
$\UsubR=-1$ constitutes the MHD-Couette limit, in which $U(y) = \sinh(H^{1/2}y)/\sinh(H^{1/2})$. This simplifies to pure Couette flow in the hydrodynamic case: as  $H\rightarrow 0$, $U(y) = y$.  $\UsubR=1$ constitutes the Shercliff limit, in which $U(y) = \cosh(H^{1/2}y)/\cosh(H^{1/2})$. This expression differs from the Shercliff profile derived by Ref.~\cite{Potherat2007quasi} for pressure driven flows, by the finite wall velocity (an unavoidable translation), and a negative multiplicative factor reflecting different ratios of centreline to bottom wall velocity in pressure-driven and wall-driven flows (the coefficient of Ref.~\cite{Potherat2007quasi} can be matched with appropriate choice of $\UsubR$, or by redefining $\Rey$). The Shercliff profile, with $\UsubR=1$, does not simplify to the Poiseuille flow solution in the limit $H\rightarrow0$ because of the absence of a pressure gradient, unlike the profile derived in Ref.~\cite{Potherat2007quasi}. In the hydrodynamic wall-driven flow, viscous diffusion is unopposed and the momentum imparted by the walls is fully diffused across the channel, unlike in finite pressure gradient Poiseuille flow. Interestingly, when $H > 0$ Hartmann friction balances diffusion in both wall- or pressure-driven flows, in an identical fashion, which explains the similarity between the profiles in this work, and those in Ref.~\cite{Potherat2007quasi}.

Varying $\UsubR$ therefore varies the base flow through the family of \CouPois\ profiles. Unlike in the classical MHD-Couette or Shercliff flows, the non-dimensional velocity $1-U_{\rm min}$, where $U_{\rm min}=\min\{U(y)\}$, depends on the friction parameter $H$ (recalling that velocities are non-dimensionalized by $U_0$). Therefore, it is useful to express our results using an alternative definition of the Reynolds number
\begin{equation}
Re_\Delta=\frac{\UD L}{\nu} = Re\,(1-U_{\rm min}),
\end{equation}
based on a velocity scale $\UD = U_0\,(1-U_{\rm min})$. Similarly, a non-dimensional timescale  $\timeD= t\,U_0/\UD  = t/(1-U_{\rm min})$ is also defined.

\subsection{Perturbation equations}\label{sec:pert_eq}
Much of this work is dedicated to analysing infinitesimal perturbations $(\hat{\vect{u}},\,\hat{p})$ about the base flow,
\begin{equation}\label{eq:pert_dec}
\vect{u}=U(y)\mathbf e_x+\hat{\vect{u}}\, , \, p = \hat{p}.
\end{equation}
The equations governing $\hat{\vect{u}}$ are obtained by substituting equation~(\ref{eq:pert_dec}) into equations~(\ref{eq:SM82_continuity}) and (\ref{eq:SM82_momentum}) and neglecting terms of $\bigO{|\hat{\vect{u}}|^2}$, yielding
\begin{equation}\label{eq:pert_c}
\bnabla_\perp \bcdot \hat{\vect{u}} = 0,
\end{equation}
\begin{equation}\label{eq:pert_m}
\pde{\hat{\vect{u}}}{t} + (\hat{\vect{u}} \bcdot \bnabla_\perp ) \vect{U} + (\vect{U} \bcdot \bnabla_\perp ) \hat{\vect{u}}= - \bnabla_\perp \hat{p} + \frac{1}{\Rey}\bnabla_\perp^2 \hat{\vect{u}} - \frac{H}{\Rey} \hat{\vect{u}}.
\end{equation}
On the lateral walls, $\hat{\vect{u}}=\partial_y\hat{\vect{u}}=0$ boundary conditions are applied.

\section{Linear stability}\label{sec:lin}
\subsection{Formulation}\label{sec:lin_form}
A sufficient condition for the base flow to be unstable is determined by seeking the least stable infinitesimal perturbation. Taking twice the curl of equation (\ref{eq:pert_m}), substituting equation (\ref{eq:pert_c}), and projecting along $\vect{e_y}$, provides an equation for the wall-normal component of the velocity perturbation
\begin{equation} \label{eq:linearised_v}
\pde{}{t} \nabla_\perp^2 \hat{v} = \pdesqr{U}{y}\pde{}{x} \hat{v} - U\pde{}{x} \nabla_\perp^2 \hat{v} + \frac{1}{\Rey}\nabla_\perp^4 \hat{v} - \frac{H}{\Rey}\nabla_\perp^2 \hat{v}.
\end{equation}
As linearity is assumed, each mode evolves independently, with perturbations decomposed into plane waves (by virtue of the problem's invariance in the streamwise direction)
\begin{equation}\label{eq:plane_wave}
\hat{v}(y) = \Rez\{\tilde{v}(y)\mathrm{e}^{\ii\alpha x}\mathrm{e}^{-\ii\lambda t}\},
\end{equation}
%
with eigenvalue $\lambda$, eigenvector $\tilde{v}(y)$, streamwise wave number $\alpha$, exponential growth rate $\Imz(\lambda)$ and wave speed $\Rez(\lambda)/\alpha$. 
Substituting equation (\ref{eq:plane_wave}) into equation (\ref{eq:linearised_v}) provides an SM82 modification to the Orr--Sommerfeld equation \citep{Schmid2001stability},
\begin{equation} \label{eq:critical}
\left[-\ii\lambda(D^2-\alpha^2)\right]\tilde{v} = \left[\ii\alpha  U'' -\ii\alpha U (D^2-\alpha^2) + \frac{1}{\Rey}(D^2-\alpha^2)^2 - \frac{H}{\Rey}(D^2-\alpha^2)\right]\tilde{v},
\end{equation}
where, respectively, primes and $D^n$ represent derivatives and the $n^{\rm th}$ order derivative operator, with respect to $y$. Boundary conditions are now $\tilde{v} = D\tilde{v} = 0$.

\begin{table}
\begin{center}
\begin{tabular}{ ccc|ccc } 
\hline
$\Nc$ & $\alphaMax$ & $\max (\Imz(\lambda_1)) \times 10^{1}$  & $\Nc$ & $\alphaMax$ & $\max (\Imz(\lambda_1)) \times 10^{2}$ \\
\hline
20 & 6.38246470 & $-1.53187927830825 $ & 200 & 3.48248937 & $1.78999418074040$ \\
40 & 6.42263964 & $-1.53055895392212 $  & 300 & 3.47528224 & $1.79276681627594$ \\
\textbf{60} & 6.42263962 & $-1.53055895418749 $  & 400 & 3.47527862 & $1.79275949928794$ \\
80 & 6.42263963 & $-1.53055895418486 $  & \textbf{500} & 3.47527864 & $1.79275949851556$ \\
100 & 6.42263954 & $-1.53055895418970 $  & 600 & 3.47527873 & $1.79275949846157$ \\
\hline
\end{tabular}
\caption{Resolution testing for eigenvalue problems. Left: energetic stability at $H=100$, $\Rey=500$, MHD-Couette flow ($\UsubR=-1$). Right: linear stability at $H=1000$, $\Rey=10^7$, Shercliff flow ($\UsubR=1$). The bold resolutions were chosen, as discussed in Sec.~\ref{sec:lin_form} and Sec.~\ref{sec:eng_form}. $\alphaMax$ is the wave number with $\max(\Imz(\lambda_1))$ for a given $\Rey$.}
\label{tab:lin_examples}
\end{center}
\end{table}

Equation \ref{eq:critical} is discretized with $\Nc$ Chebyshev collocation points \citep{Weideman2001differentiation}. Differentiation matrices $\matbm{D^n}$ and boundary conditions are implemented following Ref.~\cite{Trefethen2000spectral}. The eigenvalue problem is solved in MATLAB in the standard form at default tolerance of $10^{-14}$. $\lambda_j$ is defined as the $j$'th eigenvalue of the discretized operator, sorted by ascending growth rate, with corresponding eigenvector $\tilde{v}_j$. The critical Reynolds number is attained when $\Imz(\lambda_1)$ is zero for a single wave number $\alphaCrit$. For the linear stability analysis, for all base flows, operators are discretized with $\Nc = 200$, $350$, $500$ and $800$ for $H \leq 10^2$, $5\times10^2$, $10^3$ and $10^4$, respectively, which ensures at least 30 Chebyshev points reside within a single Shercliff boundary layer. This enables the dominant wave number and growth rate to be determined to respective precisions of $7$ and $9$ significant figures (\tbl\ \ref{tab:lin_examples}). Spurious eigenvalues \citep{Hagan2013capacitance} are not an issue for the linear analysis, as they are situated sufficiently far below the real axis. 

\subsection{Results}\label{sec:lin_results}

\begin{figure}
\begin{center}
\addtolength{\extrarowheight}{-10pt}
\addtolength{\tabcolsep}{-2pt}
\begin{tabular}{ llll }
\makecell{\footnotesize{(a)} \vspace{24mm}  \\ \rotatebox{90}{\footnotesize{$\ReyDCrit$}} \vspace{32mm}} & \makecell{\includegraphics[width=0.458\textwidth]{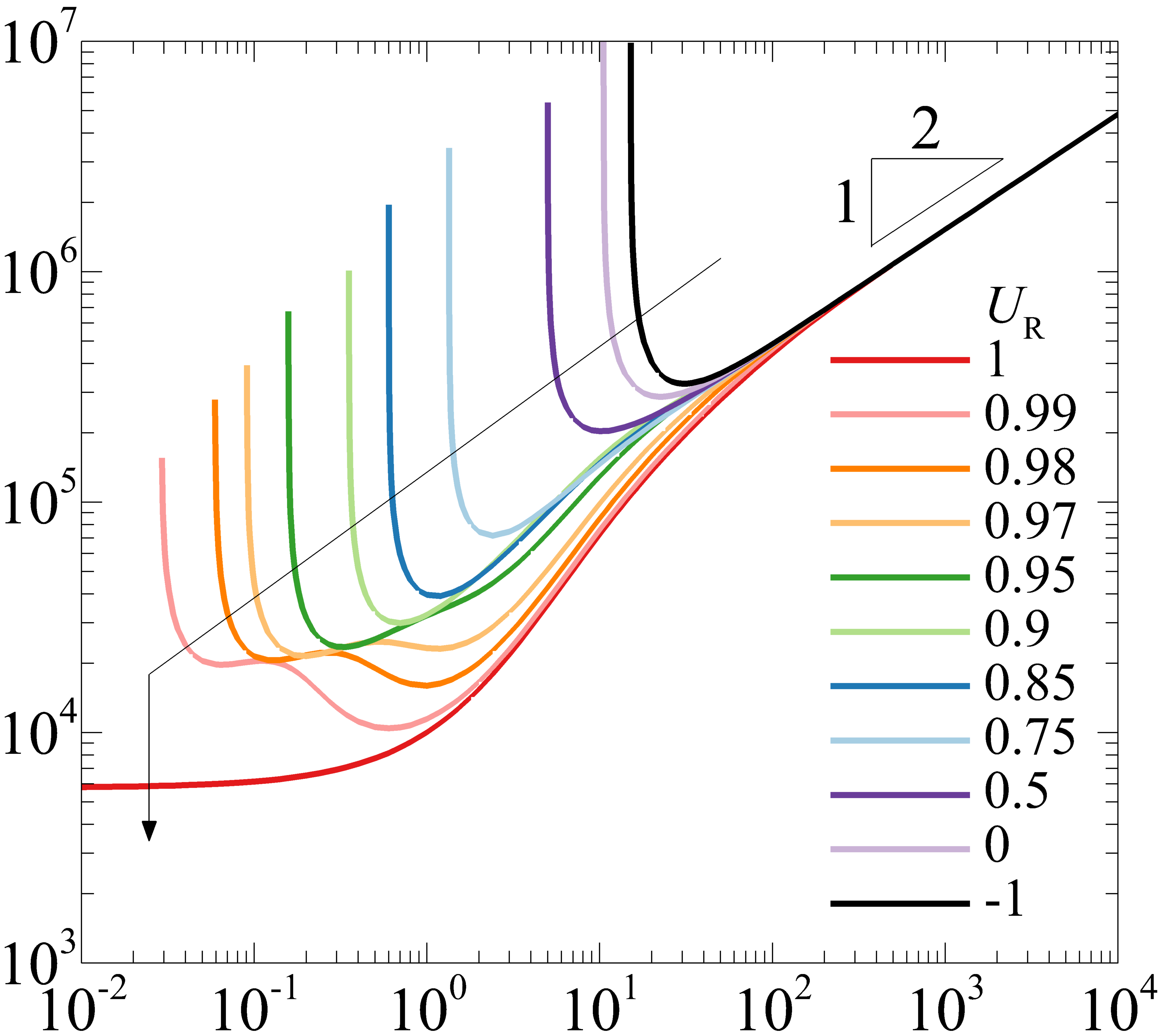}} &
\makecell{\footnotesize{(b)} \vspace{26mm}  \\ \rotatebox{90}{\footnotesize{$\alphaCrit$}} \vspace{34mm} } & \makecell{\includegraphics[width=0.458\textwidth]{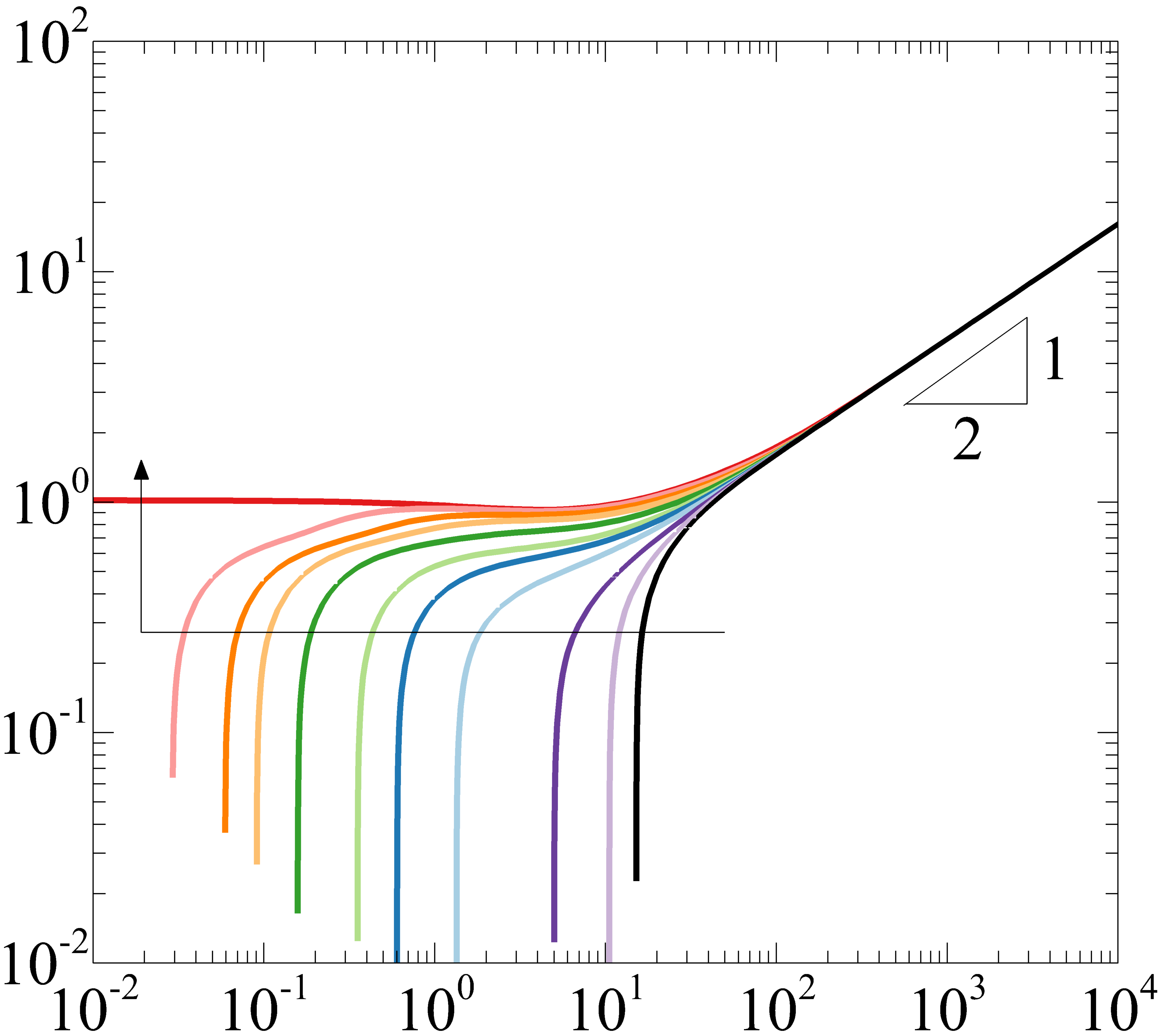}} \\
 & \hspace{36mm} \footnotesize{$H$} &  & \hspace{36mm} \footnotesize{$H$} \\
\end{tabular}
\addtolength{\tabcolsep}{+2pt}
\addtolength{\extrarowheight}{-10pt}
\end{center}
    \caption{Linear stability results, with arrows indicating increasing $\UsubR$. (a) Critical Reynolds number. (b) Critical wave number. $\ReyCrit \rightarrow \infty$ asymptotes are computed to $\Rey=10^7$. As $H \rightarrow \infty$, $\ReyDCrit=4.83468\times10^{4}\,H^{1/2}$ and $\alphaCrit = 0.161513H^{1/2}$, which agree well with Ref.~\cite{Potherat2007quasi}. As $H\rightarrow 0$, $\ReyDCrit \rightarrow 5772.22$ for $\UsubR=1$. MHD-Couette ($\UsubR=-1$) results are modified by a factor of $1/2$. The isolated boundary layer on the top wall sees an effective local minimum velocity of $U_\mathrm{min,eff}=0$, just at the edge of the boundary layer. However, the velocity profile across the entire duct still has $U_\mathrm{min}=-1$, at the bottom wall, resulting in $(1-U_\mathrm{min,eff})/(1-U_\mathrm{min})=1/2$.}
    \label{fig:lin_and_eng}
\end{figure}

The linear stability results for the family of Q2D mixed \CouPois\ flows are shown in \fig\ \ref{fig:lin_and_eng}. \Fig\ \ref{fig:lin_and_eng}(a) depicts the critical Reynolds number $\ReyDCrit$ as a function of the friction parameter $H$. The symmetric Shercliff flow \citep{Potherat2007quasi, Vo2017linear} 
has finite $\ReyDCrit$ for all non-zero $H$. Once the symmetry of the base flow is broken, a value of $H$, $H^\infty(\UsubR)$ exists, below which the critical Reynolds number is infinite. Hence, except for the symmetric Shercliff flow, $\ReyDCrit$ can initially be reduced with increasing $H$. $\ReyDCrit$ decreases to a minimum for $H>H^\infty$, so that past this minimum, increasing the friction parameter  stabilizes all flows to infinitesimal perturbations ($\ReyDCrit$ increases monotonically with increasing $H$).
A greater degree of antisymmetry ($\UsubR$ closer to $-1$) requires a larger value of $H$ before the critical Reynolds number becomes finite ($H^\infty$ monotonically increases with decreasing $\UsubR$), and provides increasing stability to infinitesimal perturbations. 
As such, the antisymmetric MHD-Couette flow is the most stable base flow for a given $H$, and has finite $\ReyDCrit$ for $H\gtrsim15.102$. The asymptotic behavior is also reflected in the critical wave numbers, \fig\ \ref{fig:lin_and_eng}(b), where $\alphaCrit \rightarrow 0$ for sufficiently small $H$. As discussed in Ref.~\cite{Kakutani1964hydromagnetic}, disturbances with finite wavelength are stable in the inviscid limit, $\ReyD \rightarrow \infty$. Hence, a finite wave number cannot be maintained as $\ReyDCrit \rightarrow \infty$. 

\begin{figure}
\begin{center}
\addtolength{\extrarowheight}{-10pt}
\addtolength{\tabcolsep}{-2pt}
\begin{tabular}{ llll }
\makecell{\footnotesize{(a)} \vspace{24mm}  \\ \rotatebox{90}{\footnotesize{$y$}} \vspace{32mm}} & \makecell{\includegraphics[width=0.458\textwidth]{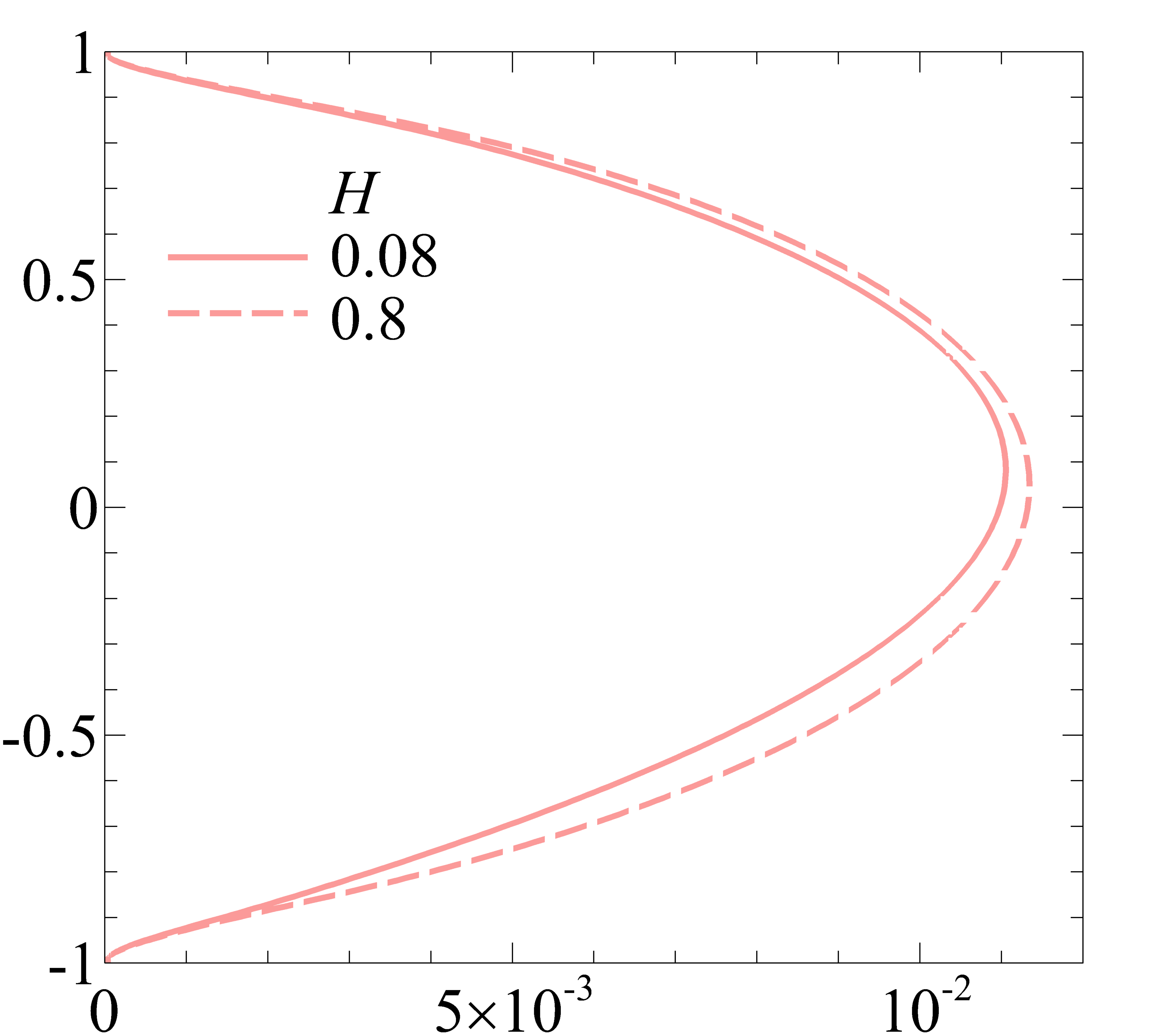}} &
\makecell{\footnotesize{(b)} \vspace{24mm}  \\ \rotatebox{90}{\footnotesize{$y$}} \vspace{32mm} } & \makecell{\includegraphics[width=0.458\textwidth]{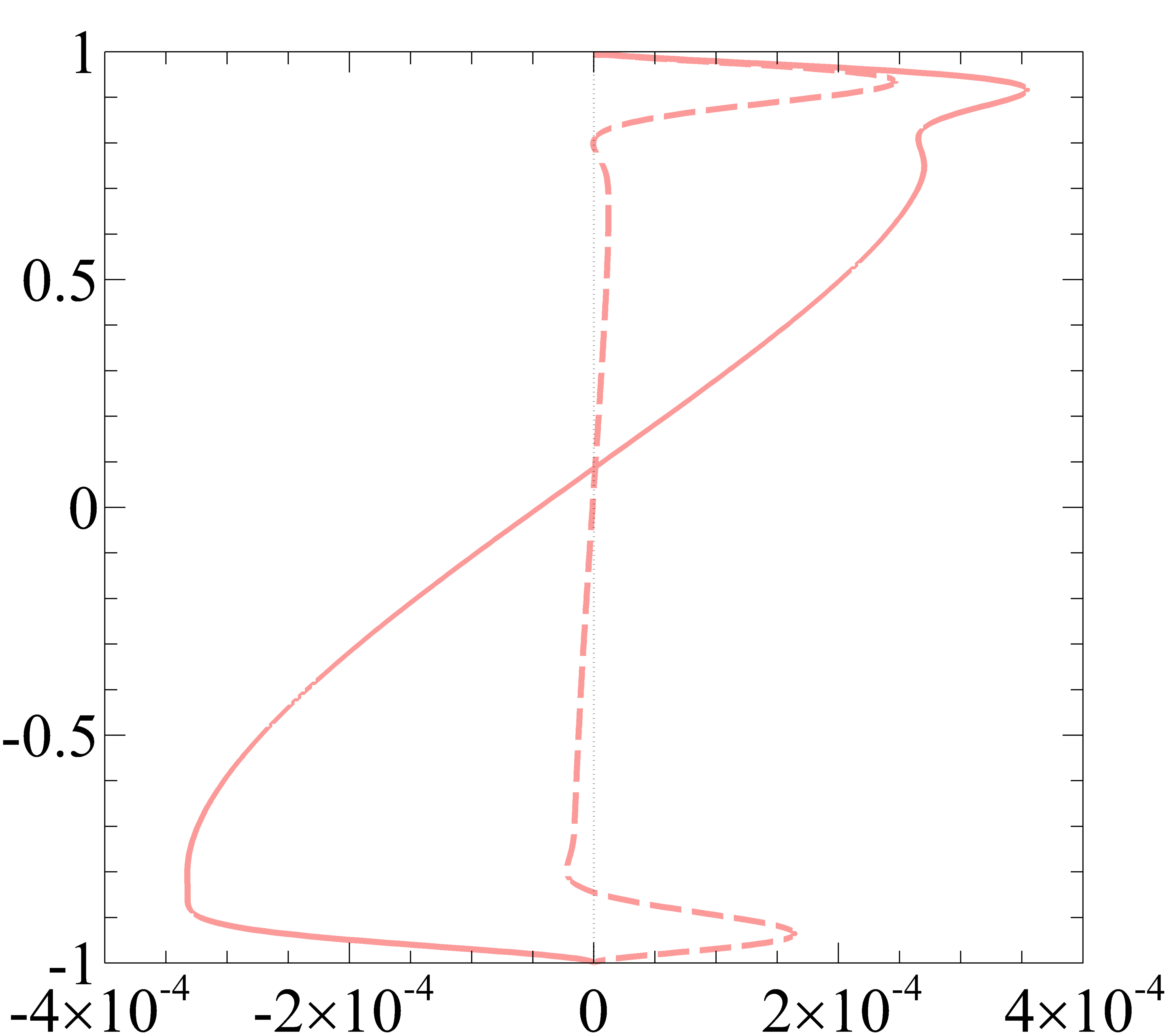}} \\
 & \hspace{34mm} \footnotesize{$\Re(\tilde{v})$} &  & \hspace{34mm} \footnotesize{$\Im(\tilde{v})$} \\
\end{tabular}
\addtolength{\tabcolsep}{+2pt}
\addtolength{\extrarowheight}{+10pt}
\end{center}
\caption{Eigenvectors $\tilde{v}_1$ from the linear stability analysis, $\UsubR = 0.99$ ($H^\infty = 0.02898$). Comparison between  $H=0.8$, $\ReyDCrit = 1.07724\times 10^4$ ($\ReyDCrit$ is smaller, due to constructive interference; indicated by approximate symmetry in imaginary component) and $H=0.08$, $\ReyDCrit = 1.99932\times 10^4$ ($\ReyDCrit$ is larger, due to destructive interference; indicated by approximate antisymmetry in imaginary component). (a) Real components. (b) Imaginary components.}
    \label{fig:lin_evecs_swap}
\end{figure}

As observed in Ref.~\cite{Hagan2013weakly} for the even and odd modes of Hartmann flow, the asymptotic behavior ($\ReyDCrit \rightarrow \infty$, $\alphaCrit \rightarrow 0$) is explained by the interaction between the \TS\ wave structures running along the top and bottom walls. Note that as the base flow is not symmetric  (resp.\ antisymmetric) unless $\UsubR=1$ (resp.\ $\UsubR=-1$), the entire domain $y \in [-1,1]$ is always simulated. This allows natural, sometimes approximate, symmetries in the dominant eigenmode to be observed. For symmetric modes, which can only be supported by symmetric base flows, the instabilities at the top and bottom walls rotate in the same direction, and constructively interfere along the centreline, causing additional destabilization (compared to an isolated \TS\ wave).
For antisymmetric modes, the instabilities rotate in the opposite direction along the top and bottom wall, and hence destructively interfere. 
The destructive interference is maximum at $\UsubR=-1$ and $H=0$, to the point of preventing the growth of any perturbation, such that $\ReyDCrit$ diverges in this limit. 
Increasing $H$ from 0, for a given value of $\UsubR$, reduces the length scale of the \TS\ waves attached to the top and bottom wall, causing them to separate from each other, which reduces interference. 
For $H>H^{\infty}$, the destructive interference between \TS\ waves is insufficient to prevent the growth of all perturbations and the flow becomes linearly unstable. Subsequent increases in $H$ further reduce the level of destructive interference, leading to a drop in $\ReyDCrit(H)$. 
Once all destructive interference has been eradicated, a subsequent increase in $H$ only results in higher friction that impedes modal growth. 
As such, $\ReyDCrit(H)$ increases. 
This explains the presence of a minimum in $\ReyDCrit(H)$.
Similarly, increasing $\UsubR$ progressively from $-1$ introduces increasingly more symmetry in the most unstable mode, which forms an alternate means of decreasing the amount of destructive interference. As such, lower values of $H$ become sufficient to suppress complete destructive interference, and $H^{\infty}(\UsubR)$ decreases monotonically with increasing $\UsubR$. For $\UsubR$ sufficiently close to 1, and for $H$ sufficiently above $H^{\infty}$, the mode can even experience noticeable constructive interference (resulting in a second set of local minima, recalling \fig\ \ref{fig:lin_and_eng}, which appear slightly above the curve for the purely symmetric $\UsubR=1$ case). A comparison of the two local minima is considered in \fig\ \ref{fig:lin_evecs_swap}, for $\UsubR=0.99$ (almost symmetric base flow). The degree of symmetry in the imaginary component of the eigenvector provides a clear indication of the type of interference. There is a much greater degree of antisymmetry in the imaginary component at $H=0.08$, near the first local minimum, indicating some destructive interference, than at $H=0.8$, near the second local minimum, which experiences significant constructive interference (the imaginary component is almost symmetric). However, as the real component has a much larger magnitude than the imaginary component, the overall mode structures look very similar. 

\begin{figure}
\begin{center}
\addtolength{\extrarowheight}{-10pt}
\addtolength{\tabcolsep}{-2pt}
\begin{tabular}{llllll}
\footnotesize{(a)} & \footnotesize{\hspace{4mm} $H=200$}  & \footnotesize{(b)}  & \footnotesize{\hspace{4mm} $H=400$}  & \footnotesize{(c)}  & \footnotesize{\hspace{4mm} $H=1000$} \\
\makecell{ \\ \rotatebox{90}{\footnotesize{$y$}} \vspace{10mm}} & \makecell{\includegraphics[width=0.31\textwidth]{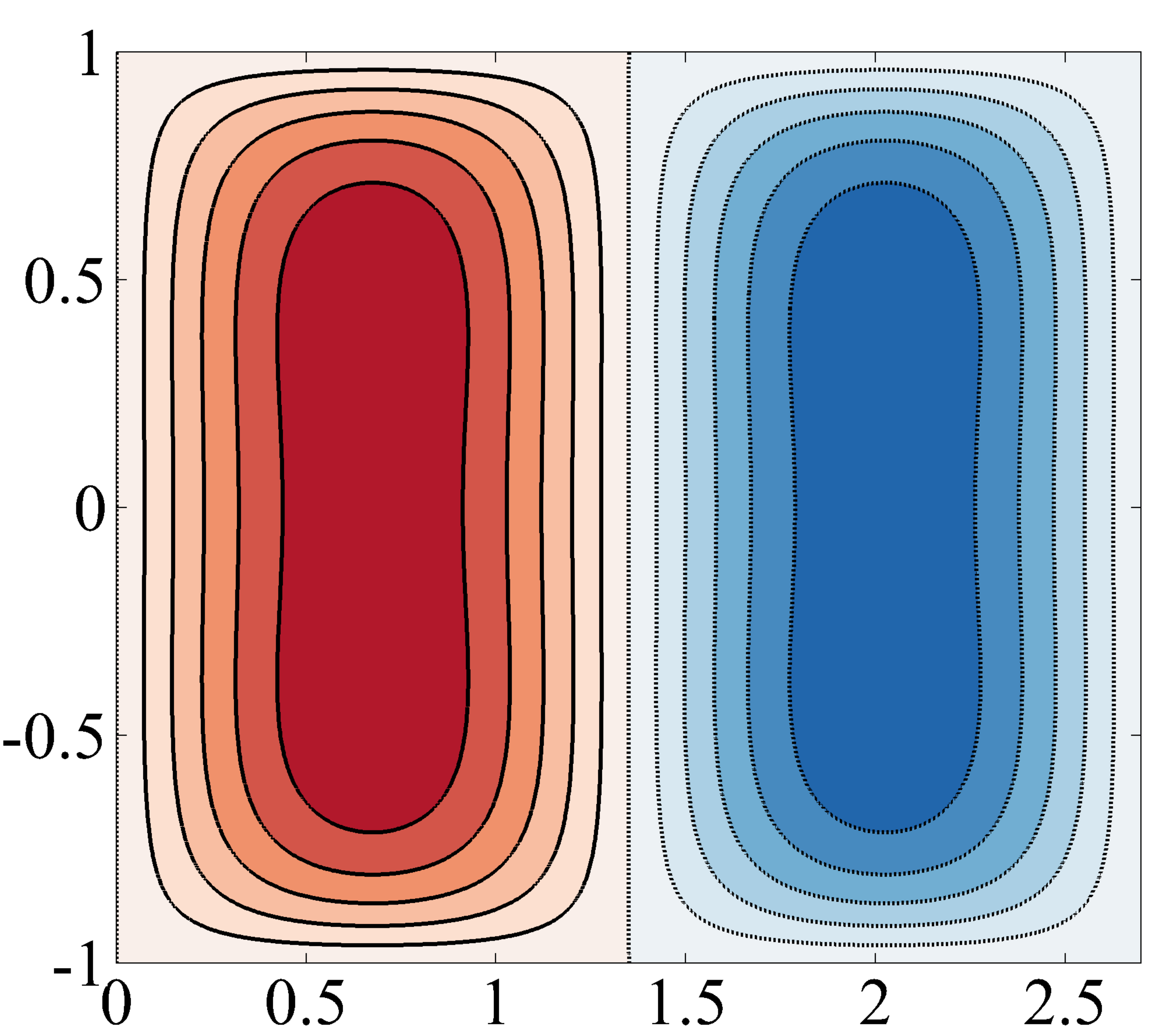}} &
\makecell{ \\ \rotatebox{90}{\footnotesize{$y$}} \vspace{10mm} } & \makecell{\includegraphics[width=0.31\textwidth]{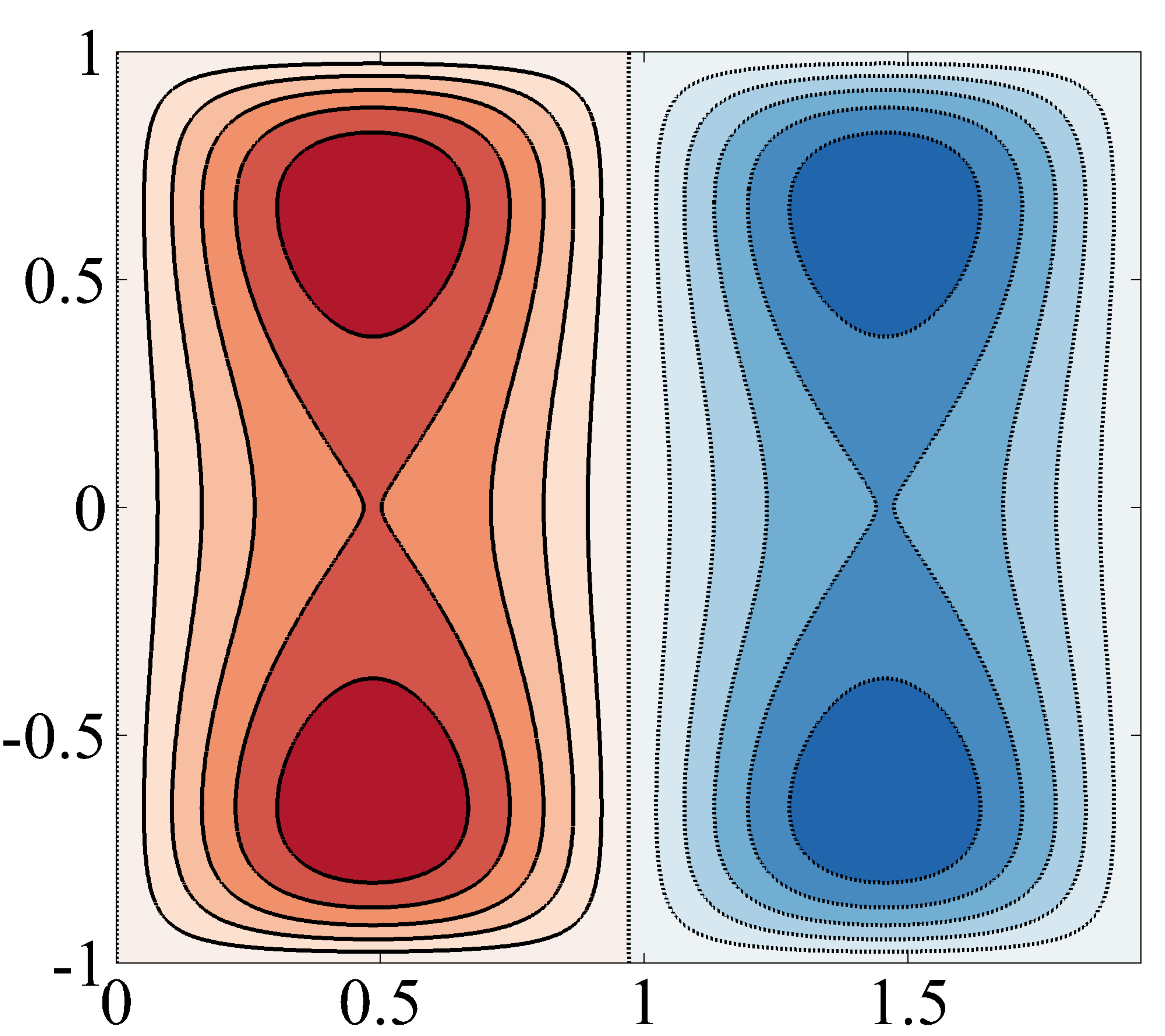}} & 
\makecell{ \\ \rotatebox{90}{\footnotesize{$y$}} \vspace{10mm}} & \makecell{\includegraphics[width=0.31\textwidth]{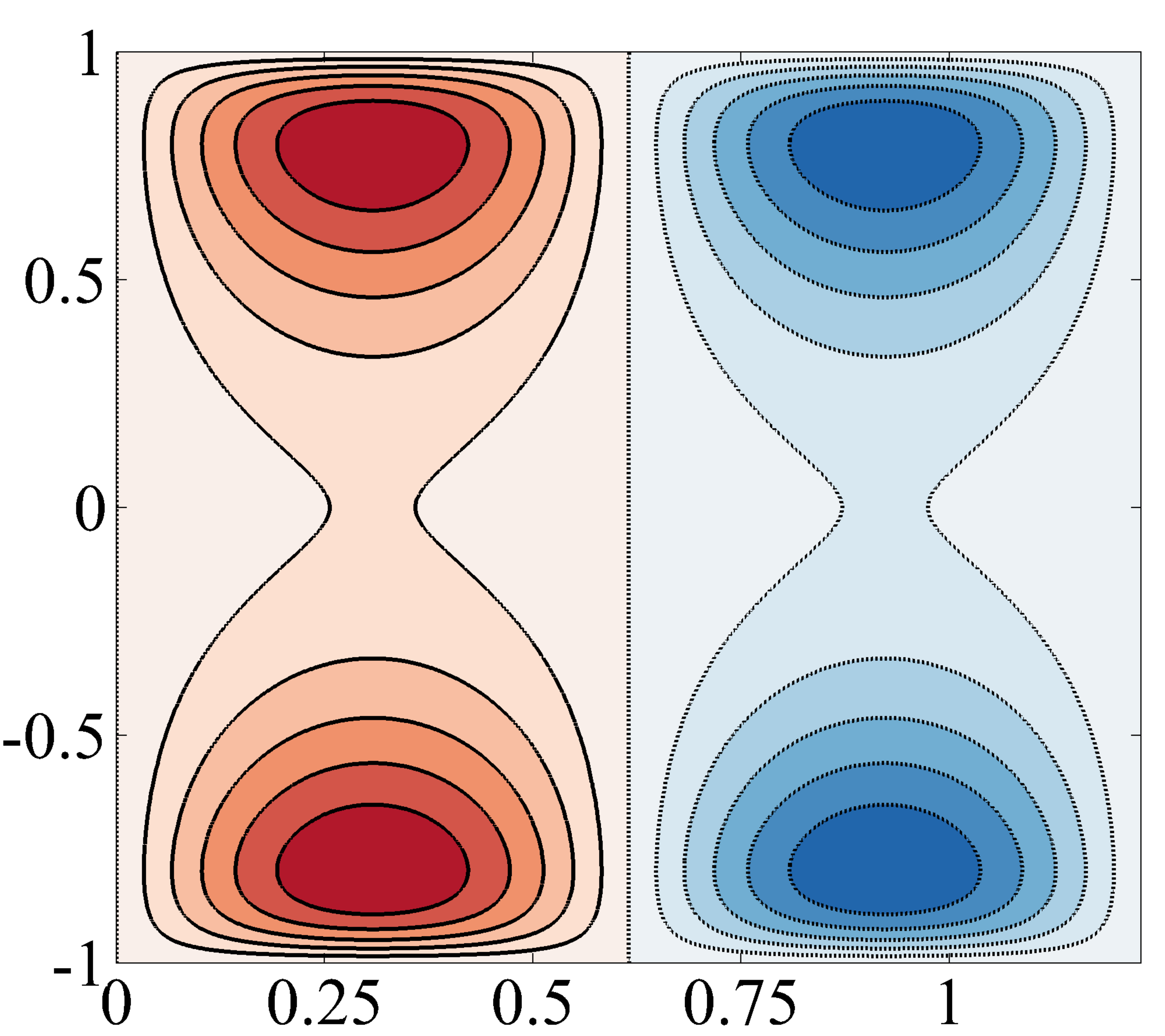}} \\
 & \hspace{25mm} \footnotesize{$x$} &  & \hspace{25mm} \footnotesize{$x$} &  & \hspace{25mm} \footnotesize{$x$} \\
 \footnotesize{(d)} & \footnotesize{\hspace{4mm} $H=50$}  & \footnotesize{(e)}  & \footnotesize{\hspace{4mm} $H=200$}  & \footnotesize{(f)}  & \footnotesize{\hspace{4mm} $H=1000$} \\
\makecell{ \\ \rotatebox{90}{\footnotesize{$y$}} \vspace{10mm}} & \makecell{\includegraphics[width=0.31\textwidth]{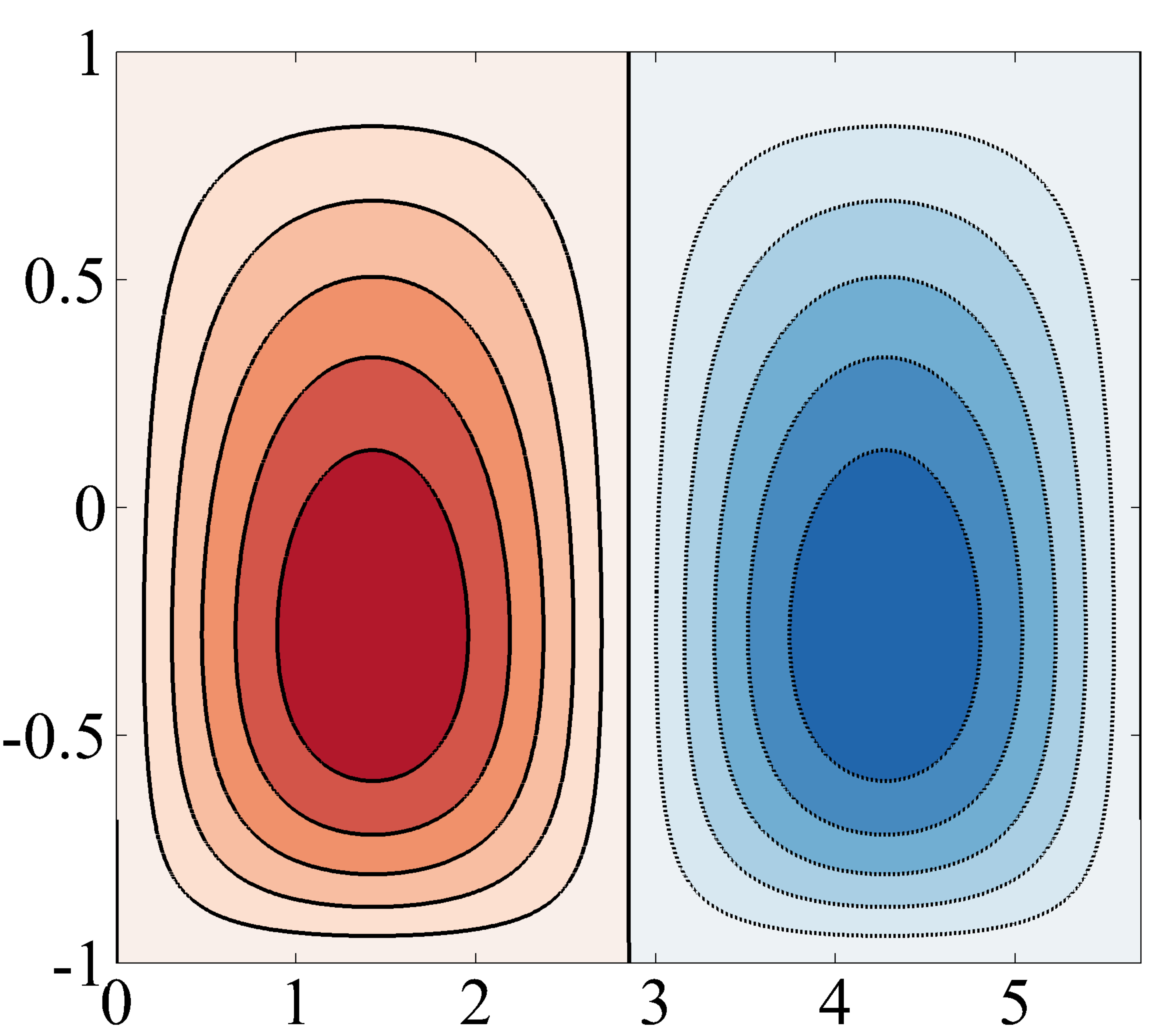}} &
\makecell{ \\ \rotatebox{90}{\footnotesize{$y$}} \vspace{10mm} } & \makecell{\includegraphics[width=0.31\textwidth]{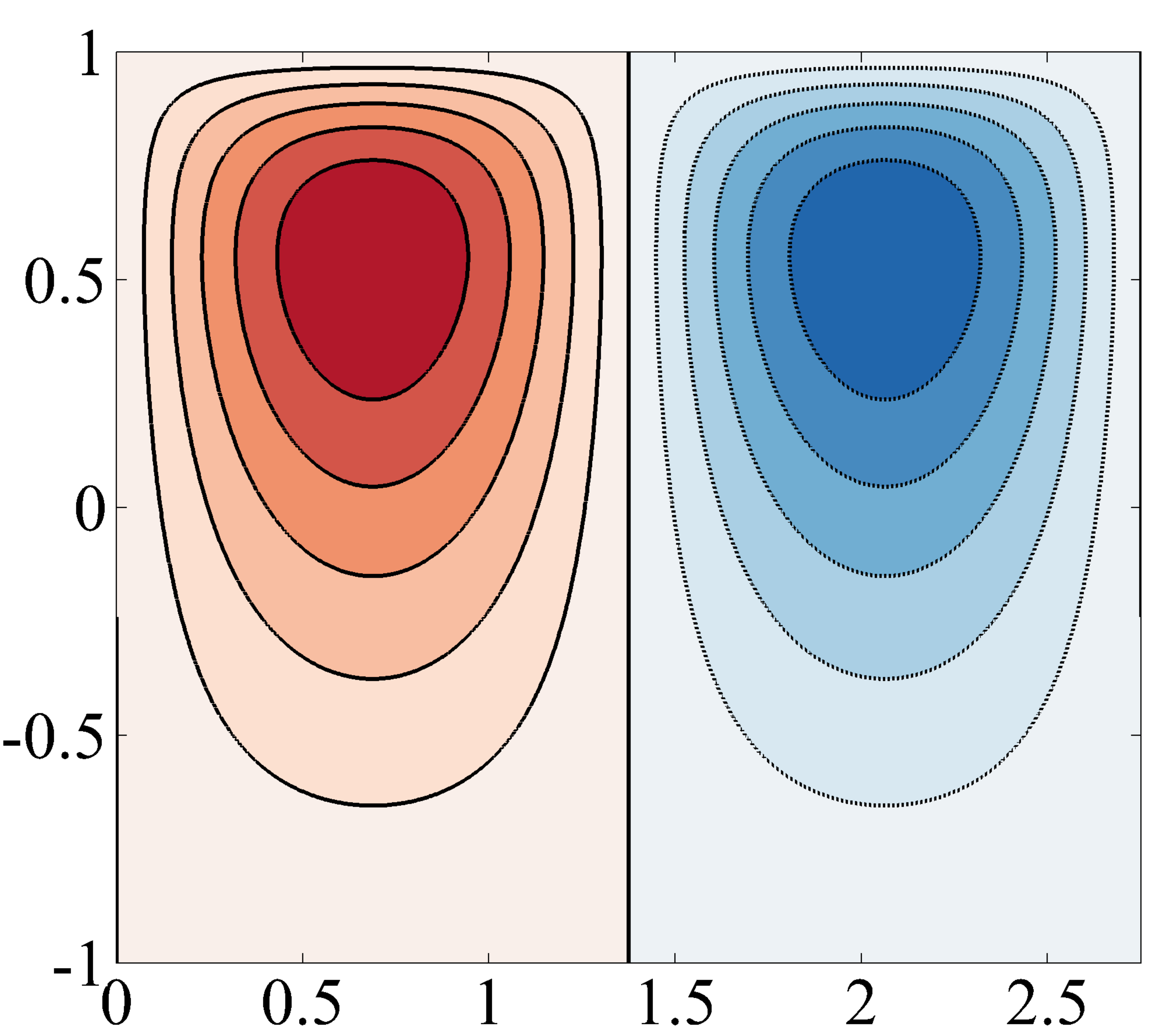}} & 
\makecell{ \\ \rotatebox{90}{\footnotesize{$y$}} \vspace{10mm}} & \makecell{\includegraphics[width=0.31\textwidth]{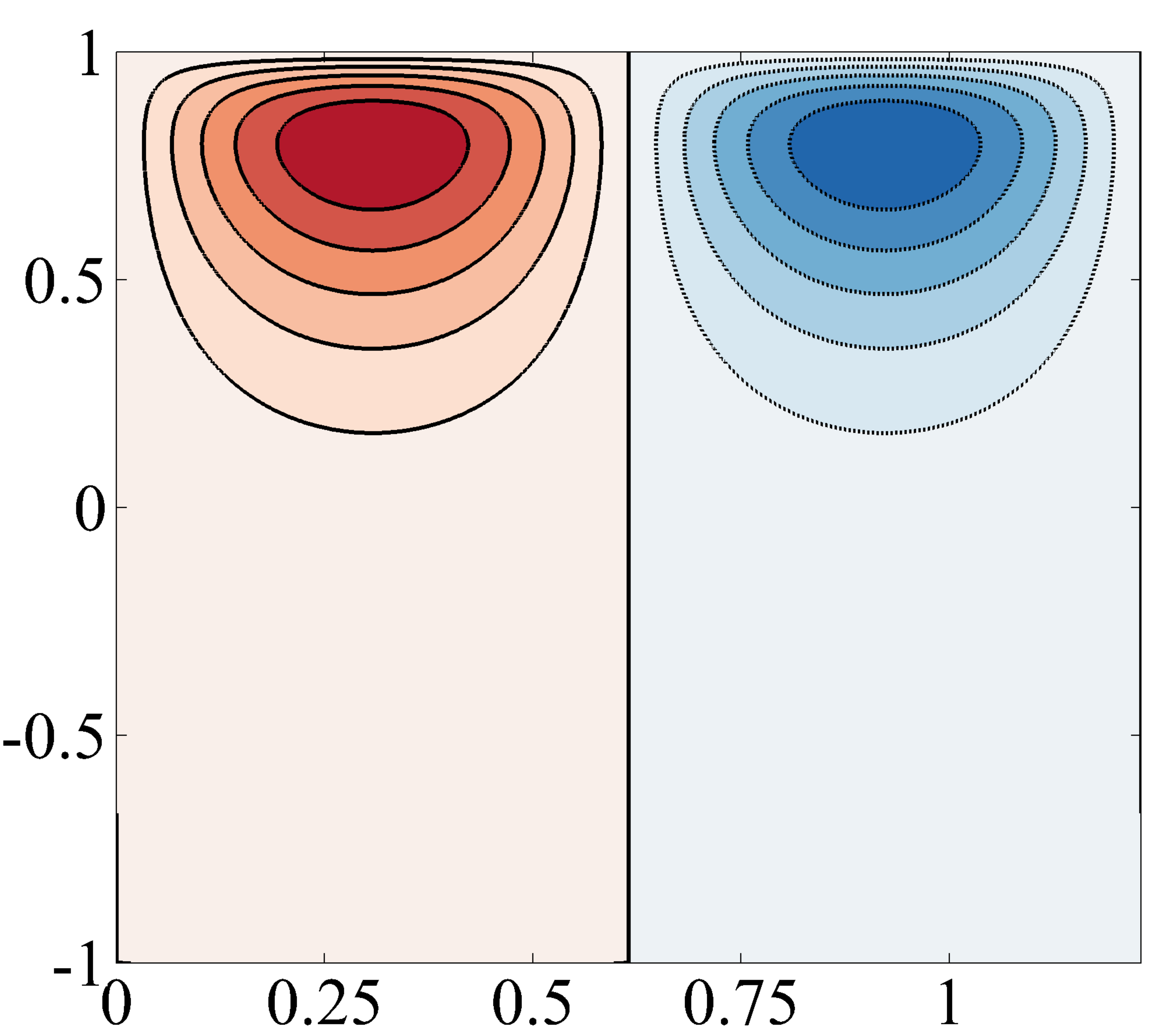}} \\
 & \hspace{25mm} \footnotesize{$x$} &  & \hspace{25mm} \footnotesize{$x$} &  & \hspace{25mm} \footnotesize{$x$} \\
\end{tabular}
\addtolength{\tabcolsep}{+2pt}
\addtolength{\extrarowheight}{+10pt}
\end{center}
\caption{Dominant eigenvectors of the linear stability analysis, $\hat{v}-$velocity contours; solid lines (red flooding) positive; dotted lines (blue flooding) negative. (a--c) Shercliff flow ($\UsubR = 1$). (d--f) MHD-Couette flow ($\UsubR = -1$).    
}
\label{fig:lin_evecs_vary_H_sym}
\end{figure}
 
The collapse of the critical Reynolds numbers and wave numbers in the limit $H\rightarrow\infty$ is due to the isolation of the boundary layers, already noted for Shercliff \citep{Potherat2007quasi} and Hartmann layers \citep{Takashima1998stability,lingwood1999_pf}.  For these large $H$, the critical Reynolds numbers and wave numbers scale with $H^{1/2}$, consistent with the thickness of a Shercliff boundary layer.
The separation mechanism for \TS\ waves at high $H$ is illustrated in \fig\ \ref{fig:lin_evecs_vary_H_sym}, for Shercliff (a--c) and  MHD-Couette (d--f) flows. 
The \TS\ wave pattern in the Shercliff flow displays the progressive separation of one central wave structure into two distinct \TS\ wave structures as $H$ increases, as found in the Hartmann flow \citep{Airiau2004amplification}.
Conversely, for flows with any degree of antisymmetry (excepting MHD-Couette flow) the velocity gradient at one wall will always be larger than at the other, drawing and confining the central mode toward the more highly sheared wall region as $H$ increases, which isolates the modes to a greater degree when $\UsubR$ is smaller, for a given $H$.



\section{Energetic analysis}\label{sec:eng}
\subsection{Formulation}\label{sec:eng_form}
%
%
The largest Reynolds number at which any perturbation would decay monotonically, $\ReyE$, is determined from the equation governing the evolution of the perturbation energy. Following Ref.~\cite{Schmid2001stability}, taking the dot product of the perturbation $u_i$ with equation (\ref{eq:pert_m}) and integrating over a volume $V$, such that all divergence terms vanish, yields
\begin{equation} \label{eq:ener_int}
\de{E}{t} = -\frac{1}{2}\int_V u_i u_j\left(\pde{U_i}{x_j}+\pde{U_j}{x_i}\right)\,\dUP V - \frac{1}{\Rey}\int_V \pde{u_i}{x_j} \pde{u_i}{x_j}\dUP V - 2E \frac{H}{\Rey}.
\end{equation}
The terms on the RHS respectively describe energy transfer from mean shear, viscous dissipation and Hartmann friction \citep{Potherat2007quasi}. The perturbation that maximizes $1/\Rey$ is found by using variational calculus and introducing a Lagrange multiplier to enforce the constraint of mass conservation \citep{Drazin2004hydrodyamic, Joseph1976stability, Schmid2001stability}, which once eliminated, and seeking plane-wave solutions, leads to the following eigenvalue problem
%
\begin{equation} \label{eq:energy}
\left[-\ii\lambda_\mathrm{E}(D^2-\alpha^2)\right]\tilde{v}_\mathrm{E} = \left[\frac{1}{2}\ii\alpha  U'' + \ii\alpha U'D + \frac{1}{\Rey}(D^2-\alpha^2)^2 - \frac{H}{\Rey}(D^2-\alpha^2)\right]\tilde{v}_\mathrm{E}.
\end{equation}
Equation (\ref{eq:energy}) is discretized and solved in an identical manner to the linear stability problem in Sec.~\ref{sec:lin_form}. $\ReyE$ is obtained when the largest imaginary component over all eigenvalues $\lambda_{\mathrm{E},j}$ is zero for a single wave number $\alphaE$. $\Nc=60$, $80$ and $140$ for $H \leq 10^2$, $10^3$ and $10^4$ again allow the dominant wave number and growth rate to be determined to respective precisions of $7$ and $9$ significant figures (\tbl\ \ref{tab:lin_examples}). 

\subsection{Results}\label{sec:eng_res}
\begin{figure}
\begin{center}
\addtolength{\extrarowheight}{-10pt}
\addtolength{\tabcolsep}{-1pt}
\begin{tabular}{ llll }
\makecell{\vspace{24mm} \footnotesize{(a)} \\  \vspace{28mm} \rotatebox{90}{\footnotesize{$\ReyDE$}}} & \makecell{\includegraphics[width=0.458\textwidth]{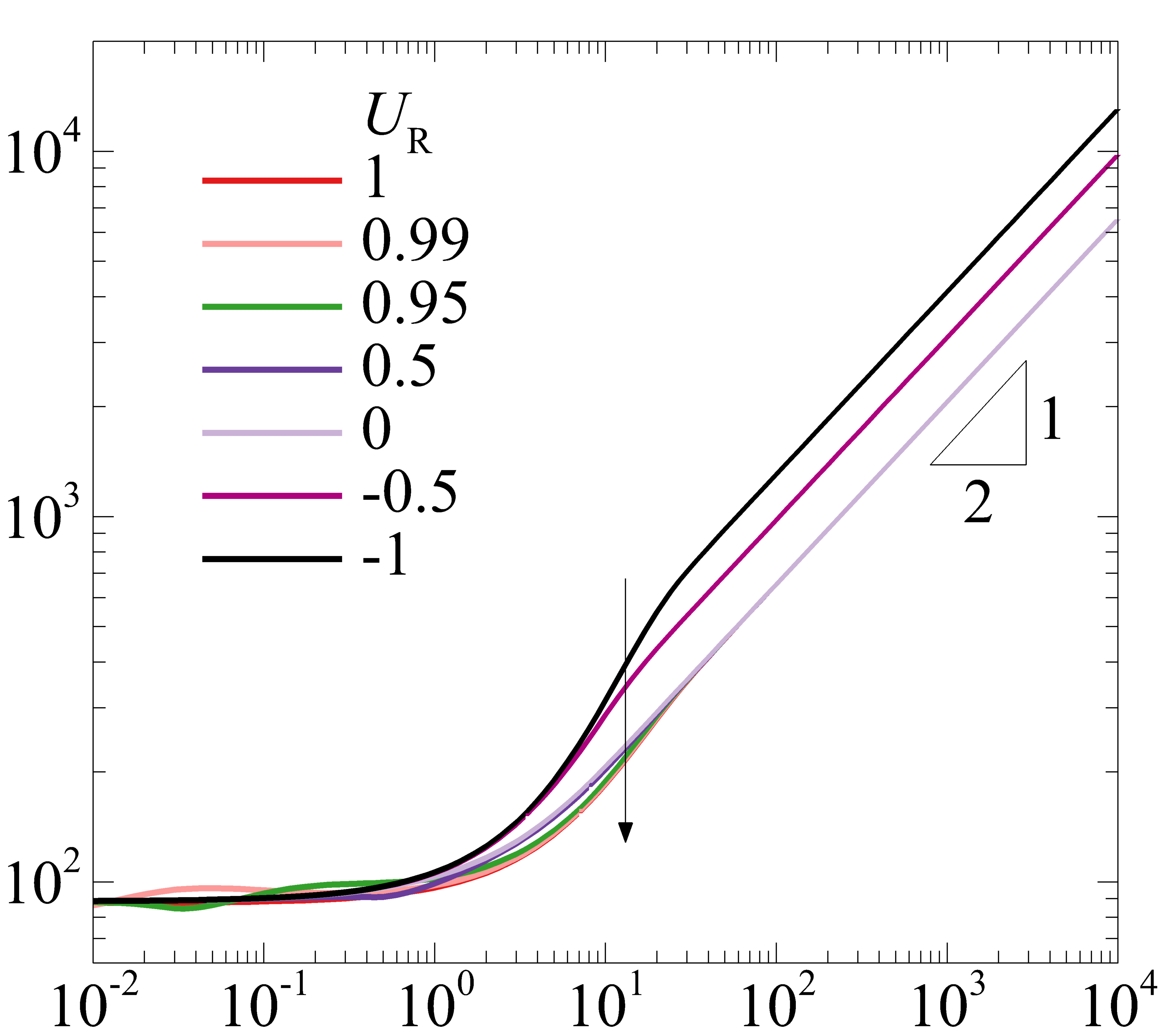}} &
\makecell{\vspace{24mm} \footnotesize{(b)} \\  \vspace{32mm} \rotatebox{90}{\footnotesize{$\alphaE$}}} & \makecell{\includegraphics[width=0.458\textwidth]{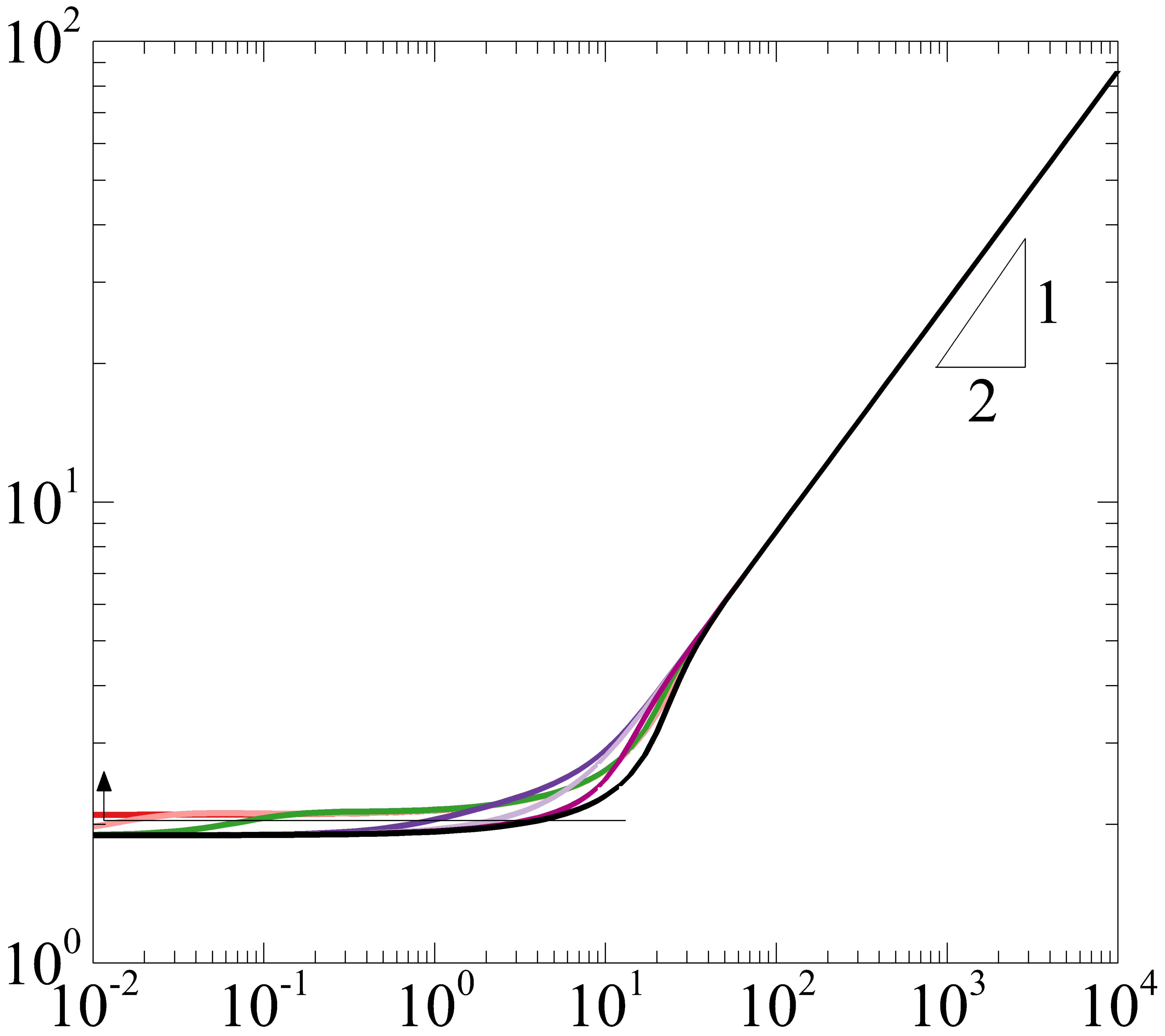}} \\
 & \hspace{38mm} \footnotesize{$H$} & & \hspace{38mm} \footnotesize{$H$} \\
\end{tabular}
\addtolength{\tabcolsep}{+1pt}
\addtolength{\extrarowheight}{+10pt}
\end{center}
\caption{Energetic analysis results, with arrows indicating increasing $\UsubR$. a) Energy Reynolds number ($\UsubR=-1$ and $-0.5$ cases collapse at high $H$ if $\ReyE$ is plotted, but appear translated with $\ReyDE$ plotted, as discussed for $\UsubR=-1$ in the caption of \fig\ \ref{fig:lin_and_eng}). b) Energy wave number. As $H \rightarrow \infty$, $\ReyDE =  65.3288H^{1/2}$ and $\alphaE=0.863470H^{1/2}$, which agree well with Ref.~\cite{Potherat2007quasi}. As $H\rightarrow 0$, $\ReyDE \rightarrow 87.5933$ for $\UsubR=1$.}
    \label{fig:eng_only}
\end{figure}

The energetic Reynolds numbers are shown in \fig\ \ref{fig:eng_only}(a). Unlike the linear stability analysis, \fig\ \ref{fig:lin_and_eng}(a), none of the curves asymptote to infinite Reynolds number, for profiles with any degree of antisymmetry, at low $H$. Overall, the energetic analysis indicates a limited influence of the base flow profile, as using the appropriate velocity scale in the Reynolds number, the results are virtually coincident for all \CouPois\ profiles, for all $H$.
Note that in the high $H$ region, the curves collapse in $\ReyE$ rather than $\ReyDE$, as only the local difference in the maximum and minimum velocity over an isolated boundary layer is important. The collapse to dynamics dominated by an isolated boundary layer occurs for all base flows simultaneously, and is initiated at much lower $H$ ($H \gtrsim 30$) than the linear analysis (which collapses between $H \gtrsim 300$ for $\UsubR=-1$ to $H \gtrsim 1000$ for $\UsubR=1$). The wave numbers from the energetic analysis, \fig\ \ref{fig:eng_only}(b), are also notably larger than those from the linear stability analysis, \fig\ \ref{fig:lin_and_eng}(b).

\begin{figure}
\begin{center}
\addtolength{\extrarowheight}{-10pt}
\addtolength{\tabcolsep}{-2pt}
\begin{tabular}{ ll ll ll ll }
\footnotesize{(a)} & \footnotesize{\hspace{2.5mm} $H=0.01$, $\UsubR=1$}  &
\footnotesize{(b)} & \footnotesize{\hspace{2.5mm} $H=0.01$, $\UsubR=-1$} & 
\footnotesize{(c)} & \footnotesize{\hspace{2.5mm} $H=100$, $\UsubR=1$} &
\footnotesize{(d)} & \footnotesize{\hspace{2.5mm} $H=100$, $\UsubR=-1$} \\
\makecell{ \\  \vspace{10mm} \rotatebox{90}{\footnotesize{$y$}}} & \makecell{\includegraphics[width=0.21\textwidth]{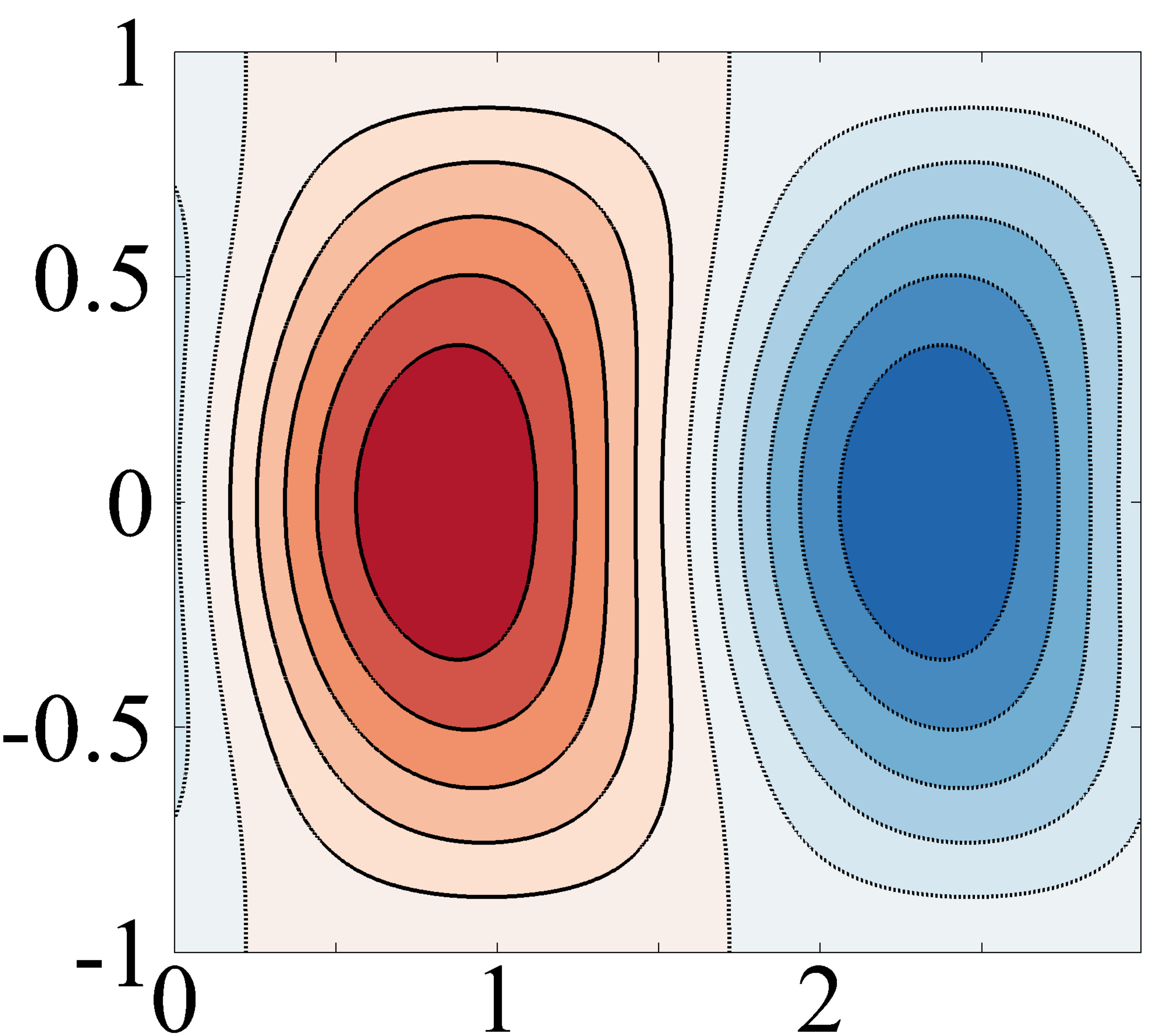}} & 
\makecell{ \\  \vspace{10mm} \rotatebox{90}{\footnotesize{$y$}}} & \makecell{\includegraphics[width=0.21\textwidth]{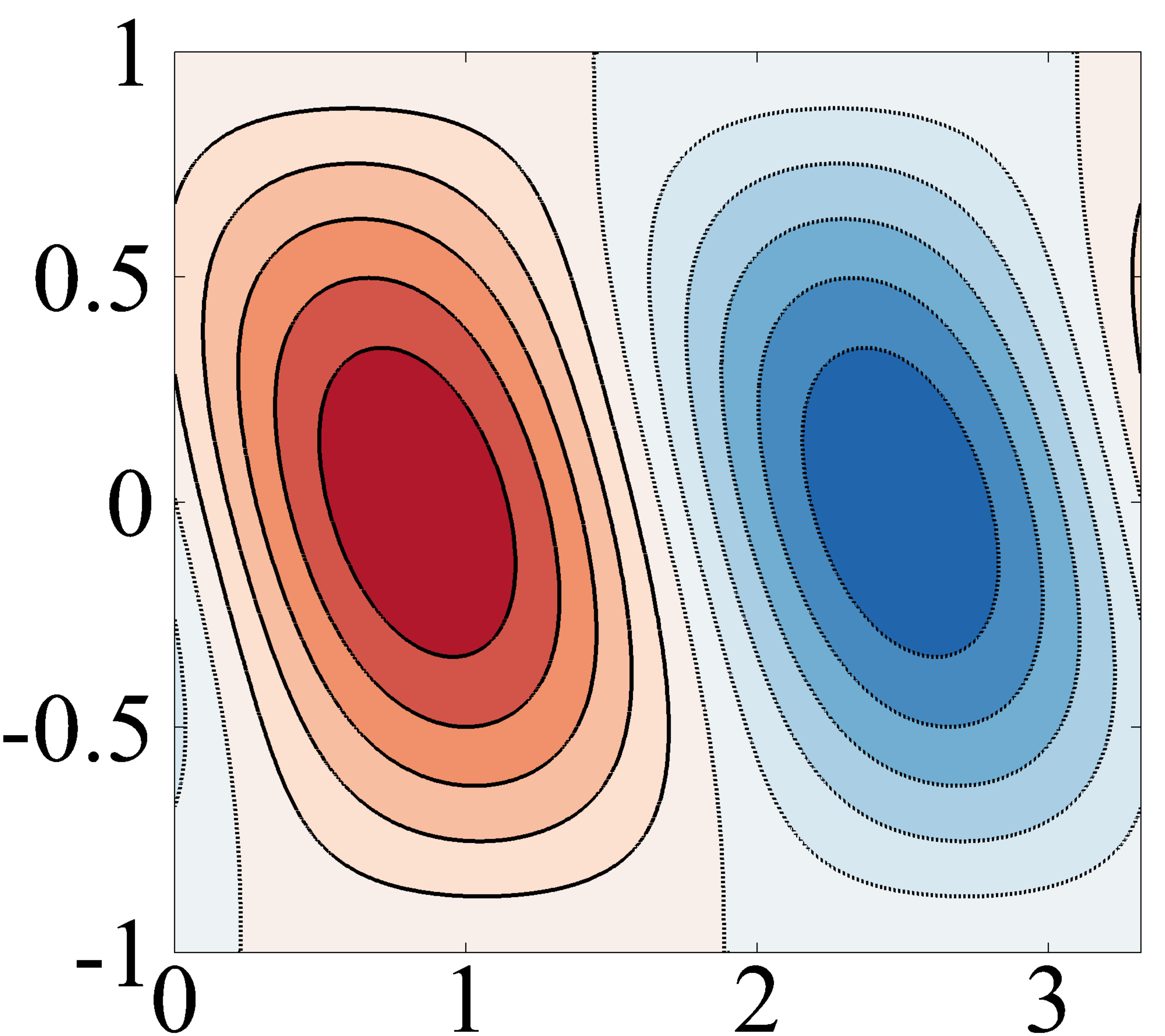}} &
\makecell{ \\  \vspace{10mm} \rotatebox{90}{\footnotesize{$y$}}}  & \makecell{\includegraphics[width=0.21\textwidth]{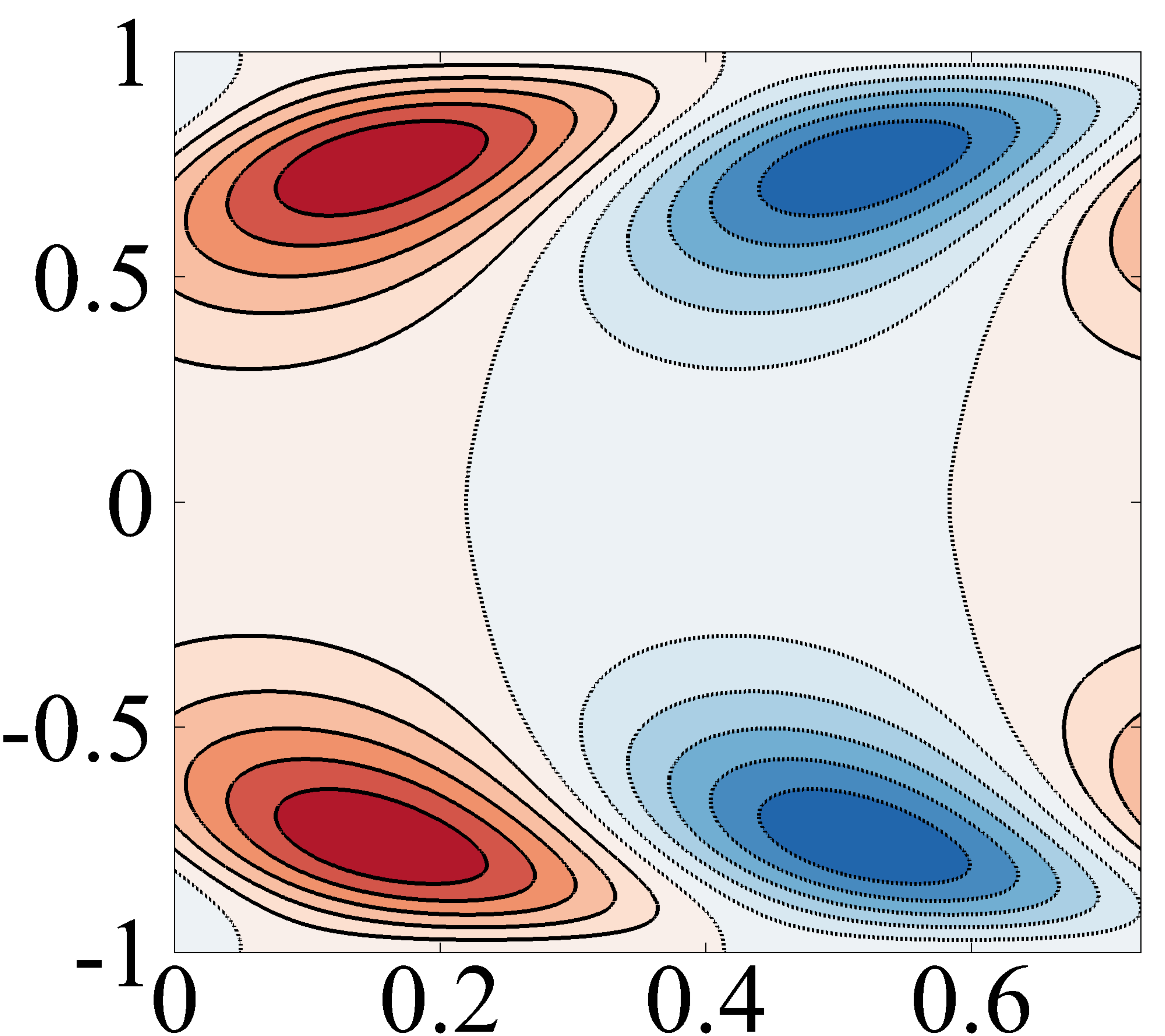}} & 
\makecell{ \\  \vspace{10mm} \rotatebox{90}{\footnotesize{$y$}}} & \makecell{\includegraphics[width=0.21\textwidth]{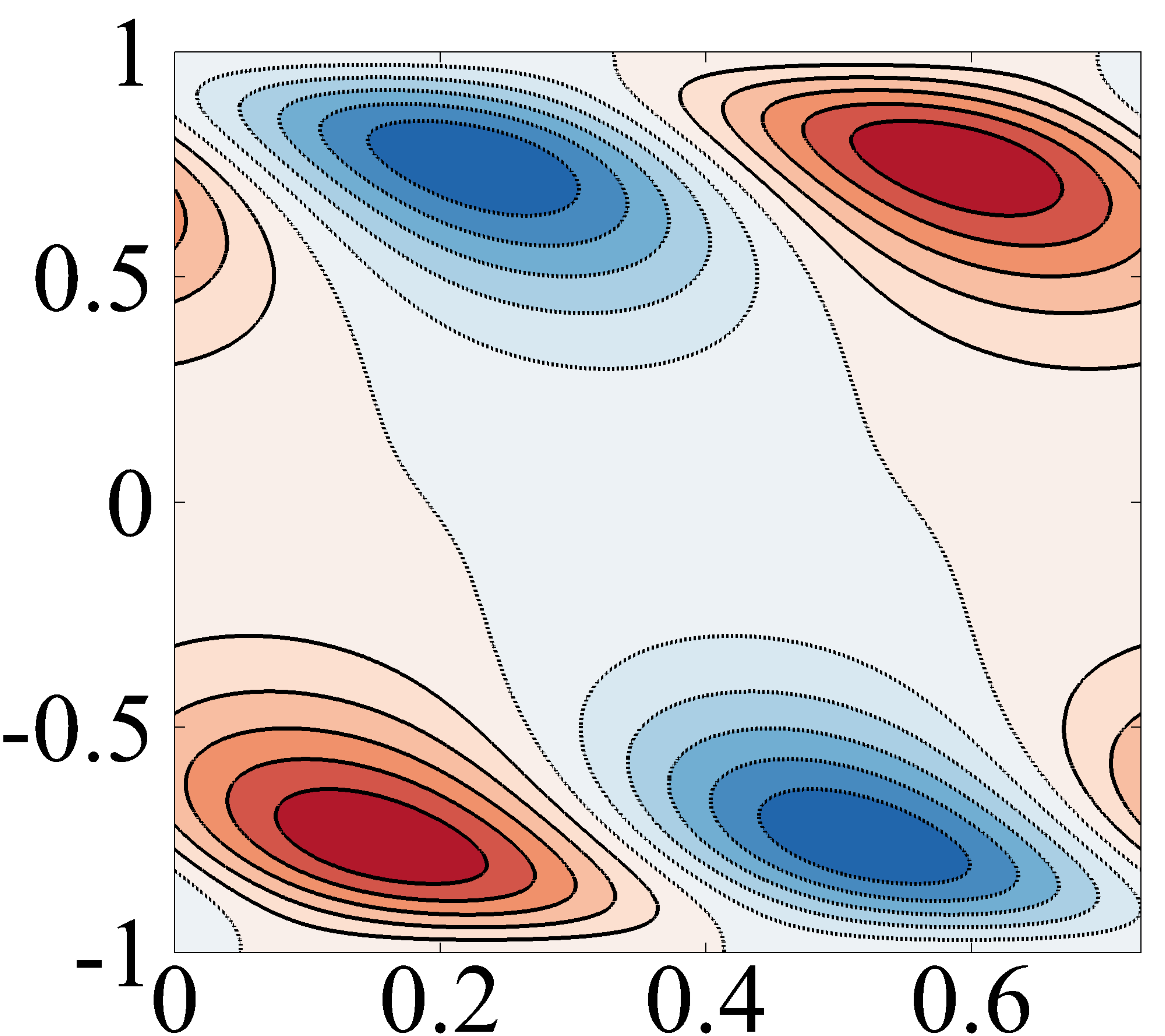}} \\
 & \hspace{16mm} \footnotesize{$x$} &  & \hspace{16mm} \footnotesize{$x$} &  & \hspace{16mm} \footnotesize{$x$} &  & \hspace{16mm} \footnotesize{$x$} \\
\end{tabular}
\addtolength{\tabcolsep}{+2pt}
\addtolength{\extrarowheight}{+10pt}
\end{center}
    \caption{Dominant eigenvectors of the energetic analysis, comparing Shercliff and MHD-Couette flows, $\hat{v}-$velocity contours; solid lines (red flooding) positive; dotted lines (blue flooding) negative.}
    \label{fig:eng_evecs_lowH}
\end{figure}


The eigenvectors from the energetic analysis are provided in \fig\ \ref{fig:eng_evecs_lowH}. Unlike the linear stability analysis, these modes do not directly represent solutions to the SM82 equations \citep{Potherat2007quasi}. The modes are more clearly slanted due to the lower Reynolds numbers. Similar to the linear stability analysis, at higher $H$, a wall mode forms, which again is increasingly compacted toward the wall as $H$ increases. Discounting the irrelevant symmetry or antisymmetry, as in focusing on  $-1<y<0$ in \fig\ \ref{fig:eng_evecs_lowH}, the modes effectively appear identical. Thus, varying the base flow through $\UsubR$ has little effect on the overall dynamics of the dominant modes of the energetic analysis (when comparing the same $H$).

\section{Linear transient growth and pseudospectra}\label{sec:tgp}
\subsection{Formulation}\label{sec:tgp_form}
A lower bound for the Reynolds number at which an instability exponentially grows, and an upper bound on the Reynolds number at which all instabilities monotonically decay, have been derived in the preceding sections. However, non-orthogonality of the linearized evolution operator can lead to the transient growth of a superposition of linearly decaying eigenvectors \citep{Schmid2001stability}. To this end, transient growth analysis is  performed for $\ReyE<Re\lesssim\ReyCrit$. 
The maximum possible transient growth is found by seeking the initial condition for perturbation  
$\hat{\mathbf u}_\tau(t=0)$ that maximizes the gain functional 
$G=|| \hat{\mathbf u}(t=\tau) || / || \hat{\mathbf u}(t=0) ||$ at prescribed time $t=\tau$ of the pertubation's linearized evolution. $G$ represents the gain in perturbation kinetic energy as per Ref.~\cite{Barkley2008direct} under the norm $|| \hat{\vect{u}}  || = \int \hat{\vect{u}} \bcdot \hat{\vect{u}} \,\mathrm{d}\Omega$, where $\Omega$ represents the computational domain. The maximum possible gain $\Gmax$ is found at optimal time $\tau_{\rm opt}$ 
for which the value $G_{\rm max}(\tau_{\rm opt})$ of the optimized functional is maximum. In practice, since $\hat{\mathbf u}$ is a plane wave, $\hat v_\tau(t=0)$ is obtained as the solution of an optimisation 
problem with the linearized evolution equation
\begin{equation} \label{eq:ltg_forward}
\pde{\hat{v}}{t} = (D^2-\alpha^2)^{-1}\left[-\ii\alpha U (D^2-\alpha^2) + \ii\alpha U'' + \frac{1}{\Rey}(D^2 - \alpha^2)^2  - \frac{H}{\Rey}(D^2-\alpha^2) \right]\hat{v},
\end{equation}
as constraint. The optimal is obtained iteratively from a timestepper, set up in MATLAB, which first evolves equation (\ref{eq:ltg_forward}) to time $\tau$, then evolves the adjoint equation  
\begin{equation} \label{eq:ltg_adjoint}
\pde{\hat{\xi}}{t} = (D^2-\alpha^2)^{-1}\left[\ii\alpha U (D^2-\alpha^2) + 2i\alpha U' D + \frac{1}{\Rey}(D^2 - \alpha^2)^2  - \frac{H}{\Rey}(D^2-\alpha^2) \right]\hat{\xi},
\end{equation}
for the Lagrange multiplier of the velocity perturbation $\hat{\xi}$, from $t=\tau$ to $t=0$, until $\hat{\mathbf u}_\tau(t=0)$ has converged to the desired precision. A third-order forward \AdamBash\ scheme  \citep{Hairer1993solving} is used to integrate equations (\ref{eq:ltg_forward}) and (\ref{eq:ltg_adjoint}) in time, subject to $\hat{v}$ and $\hat{\xi}$ satisfying boundary conditions $\hat{v} = D\hat{v} = \hat{\xi} = D\hat{\xi} = 0$ at all walls, and `initial' condition $\hat\xi(\tau)=\hat v(\tau)$. The $j$'th eigenvalue $\lambda_{\mathrm{G},j}$ of the operator representing the action of direct then adjoint evolution is determined with a Krylov subspace scheme \citep{Barkley2008direct, Blackburn2008convective}. With eigenvalues sorted in ascending order by largest real component, the optimized growth $G = \lambda_{\mathrm{G},1}$. The iterative scheme is initialized with random noise for $\hat{v}(t=0)$.

Validation against literature is provided in \tbl\ \ref{tab:trans_Cass_vary_H}. Validation against the rescaled results of Ref.~\cite{Camobreco2020role} is also visible in \fig\ \ref{fig:all_transient}. To maintain six significant figure accuracy in $\Gmax$ requires a timestep of $\Delta t=2\times10^{-5}$, 20 forward-backward iterations and $\Nc=60$, 80 and 100 Chebyshev points for $H \leq 10$, 30, and 100, respectively (for $\Rey \leq 10^5$). $\tau_\mathrm{opt}$ and $\alpha_\mathrm{opt}$ are computed to three significant figures.
\begin{table}
\begin{center}
\begin{tabular}{c|ccc|ccc} 
\hline
       & \multicolumn{3}{c|}{$\Rey=5\times10^3$} & \multicolumn{3}{c}{$\Rey=1.5\times10^4$} \\
\hspace{2mm} $H$ \hspace{2mm}    & \hspace{4mm} Ref.~\cite{Cassels2019from3D} \hspace{4mm} & Present  & \hspace{2mm} $|\%$ error$|$ \hspace{2mm} & \hspace{4mm} Ref.~\cite{Cassels2019from3D} \hspace{4mm} & Present & \hspace{2mm} $|\%$ error$|$ \hspace{2mm} \\
\hline
$10$   & $14.65$ & $14.8272$   & $1.195$ & $27.4$ & $34.0552$       & $19.54$ \\
$30$   & -       & $7.62330$   & -       & $17.7$ & $17.7515$       & $0.290$ \\
$50$   & $6.08$  & $6.13073$   & $0.827$ & $14.2$ & $14.4036$       & $1.414$ \\
$100$  & $4.61$  & $4.60979$   & $0.004$ & $11.0$ & $11.0476$       & $0.431$ \\
$150$  & $3.88$  & $3.89381$   & $0.355$ & $9.43$ & $9.44834$       & $0.194$ \\
$300$  & $2.90$  & $2.90575$   & $0.198$ & $7.11$ & $7.19654$       & $1.203$ \\
$600$  & $2.16$  & $2.16392$   & $0.181$ & $5.43$ & $5.44425$       & $0.262$ \\
$800$  & $1.91$  & $1.91680$   & $0.355$ & $4.83$ & $4.83895$       & $0.185$ \\
\hline
\end{tabular}
\caption{Comparisons of the $\Gmax$ calculated in the present work, and those calculated by Ref.~\cite{Cassels2019from3D} for various $H$, MHD-Poiseuille flow profile, at $\Rey=5\times10^3$ and $1.5\times10^4$. Reference \cite{Cassels2019from3D}'s results, kindly provided from their \fig\ 2, are wave number optimized in a full \threed\ domain, but time optimized at the 3D optimal wave number in a \twod\ domain. The discrepancy at low $H$ reflects the breakdown of the quasi-two-dimensionality assumption, not numerical error.}
\label{tab:trans_Cass_vary_H}
\end{center}
\end{table}

Additionally there was excellent agreement with results obtained with the matrix method 
\cite[provided in Appendix A of Ref.~][]{Schmid2001stability} at low Reynolds and Hartmann numbers. As such, the matrix method is used to further assess the transient growth capability by considering the non-normality of the operator, via the pseudospectrum and condition number of the energy norm weight matrix. A point $z$ on the complex plane 
is within the $\epsilon_p$-pseudospectrum of the SM82-modified Orr--Sommerfeld operator if $||(zI-\mathcal{L}_\mathrm{OS})^{-1}|| \geq \epsilon_p^{-1}$ \citep{Reddy1993pseudospectra}. For a normal operator, a point $z$ on the complex plane will be at most at a distance $\epsilon_p$ from any eigenvalue.  The greater the degree of non-normality, the greater the ratio of the distance between a point $z$ and the nearest eigenvalue, to the bounding value of $\epsilon_p$ at the point $z$. The extent of the pseudospectra into the complex upper half plane forms a lower 
bound on transient growth \citep{Reddy1993energy, Reddy1993pseudospectra, Schmid2001stability, Trefethen1993hydrodynamic, Trefethen2000spectral}. The pseudospectrum is computed by evaluating $||\matbm{W}\,(1/(z\matbm{I}-\lambda)) \matbm{W}^{-1}||_2$, with energy norm weight matrix $\matbm{W}$ \citep{Schmid2001stability}, identity matrix $\matbm{I}$, and diagonalized eigenvalues $\lambda$ of the discretized SM82-modified Orr--Sommerfeld operator. Computations were performed with a discretization of $\Nc=400$ and truncated to the $240$ modes with largest imaginary component.

\subsection{Results: transient growth}\label{sec:tgp_results}
\begin{figure}
\begin{center}
\addtolength{\extrarowheight}{-10pt}
\addtolength{\tabcolsep}{-2pt}
\begin{tabular}{ llll }
\makecell{\vspace{24mm} \footnotesize{(a)} \\  \vspace{31mm} \rotatebox{90}{\footnotesize{$\Gmax$}}} & \makecell{\includegraphics[width=0.458\textwidth]{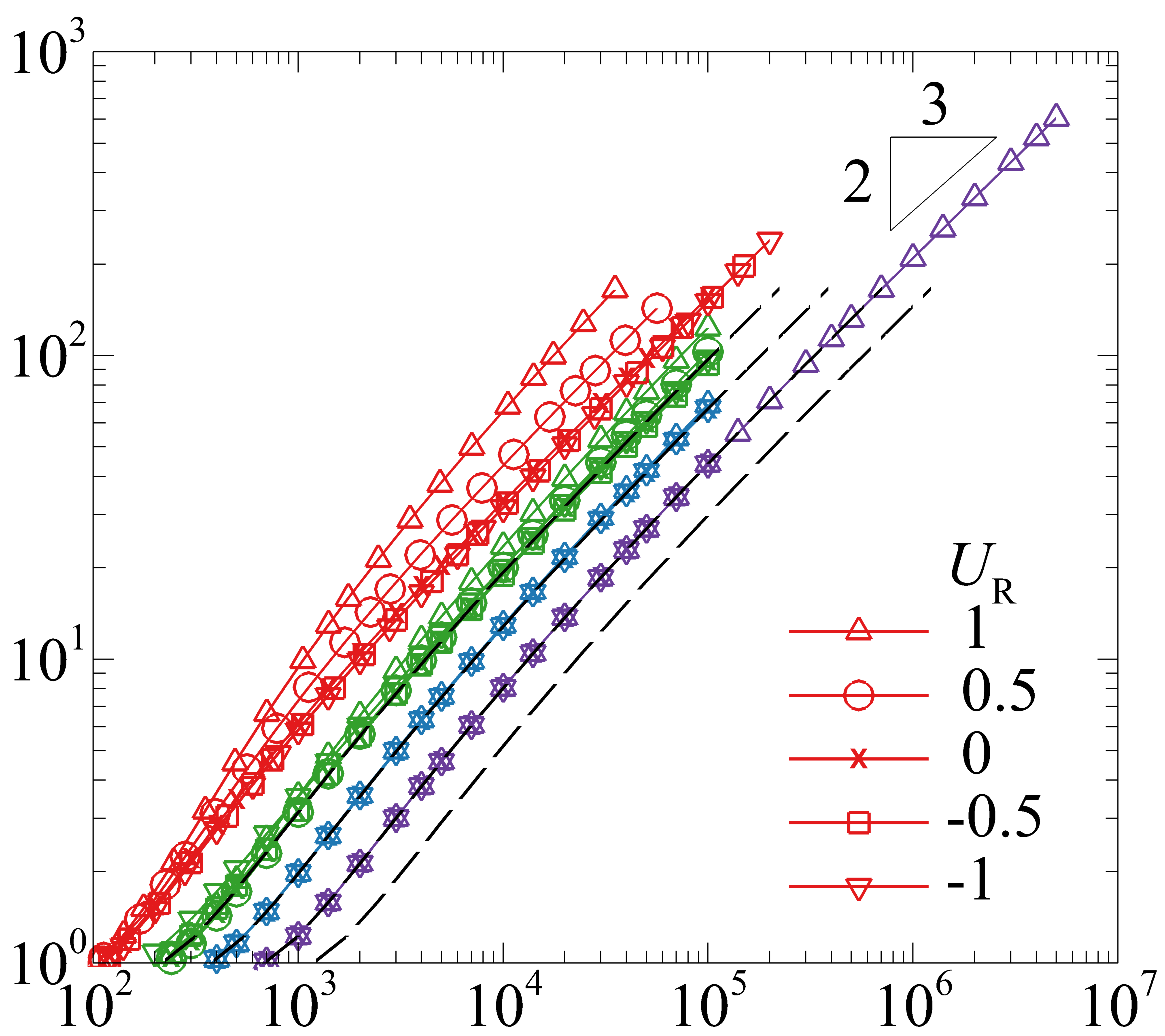}} &
\makecell{\vspace{19.5mm} \footnotesize{(b)} \\  \vspace{28mm} \rotatebox{90}{\footnotesize{$\tauOpt$, $\tauDOpt$}}} & \makecell{\includegraphics[width=0.458\textwidth]{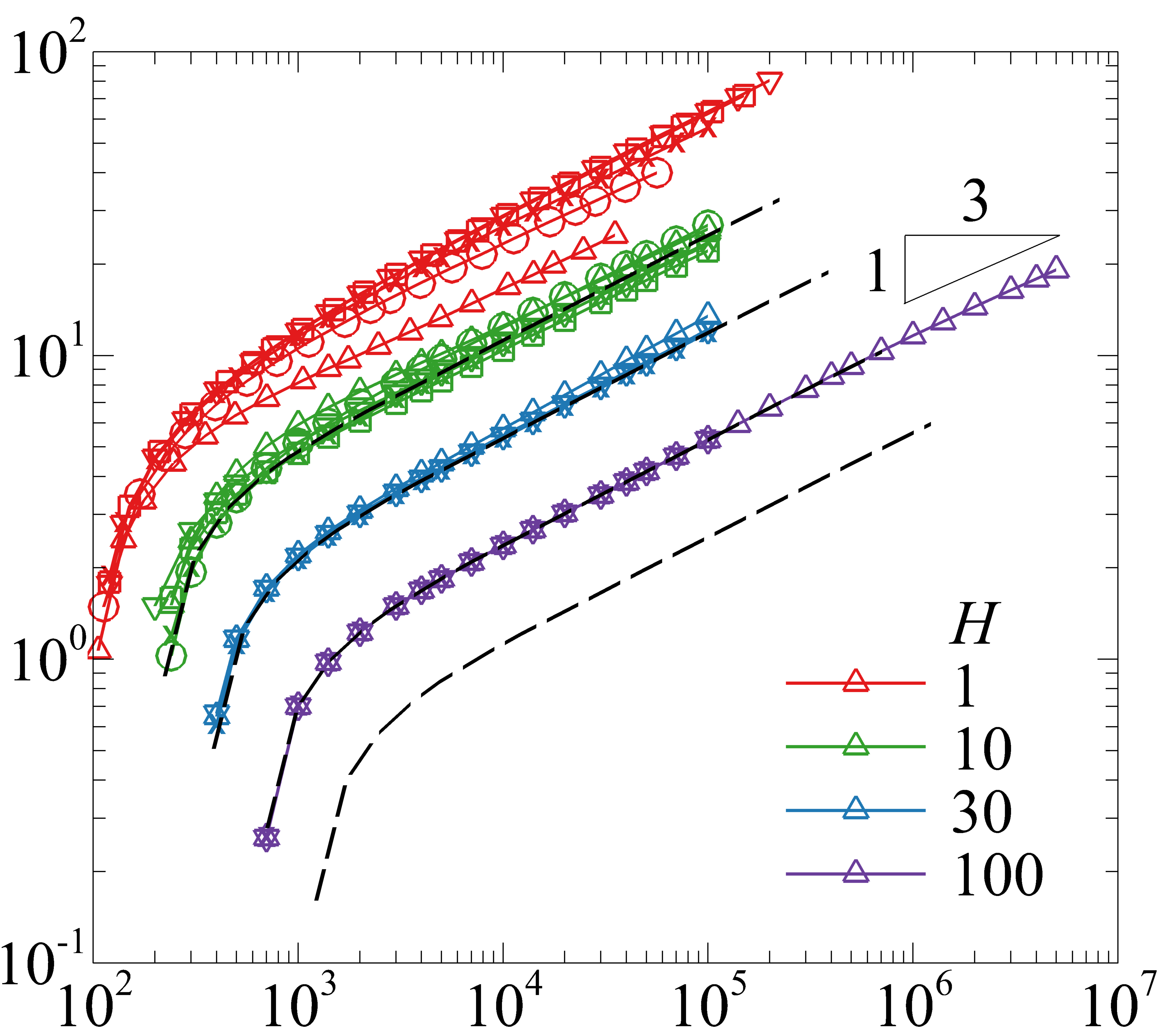}} \\
 & \hspace{33mm} \footnotesize{$\ReyD$, $\Rey$} & & \hspace{33mm} \footnotesize{$\ReyD$, $\Rey$} \\
\end{tabular}
\addtolength{\extrarowheight}{+10pt}
\addtolength{\extrarowheight}{-10pt}
\begin{tabular}{ ll }
\makecell{\vspace{22mm} \footnotesize{(c)} \\  \vspace{31mm} \rotatebox{90}{\footnotesize{$\alphaOpt$}}} & \makecell{\includegraphics[width=0.458\textwidth]{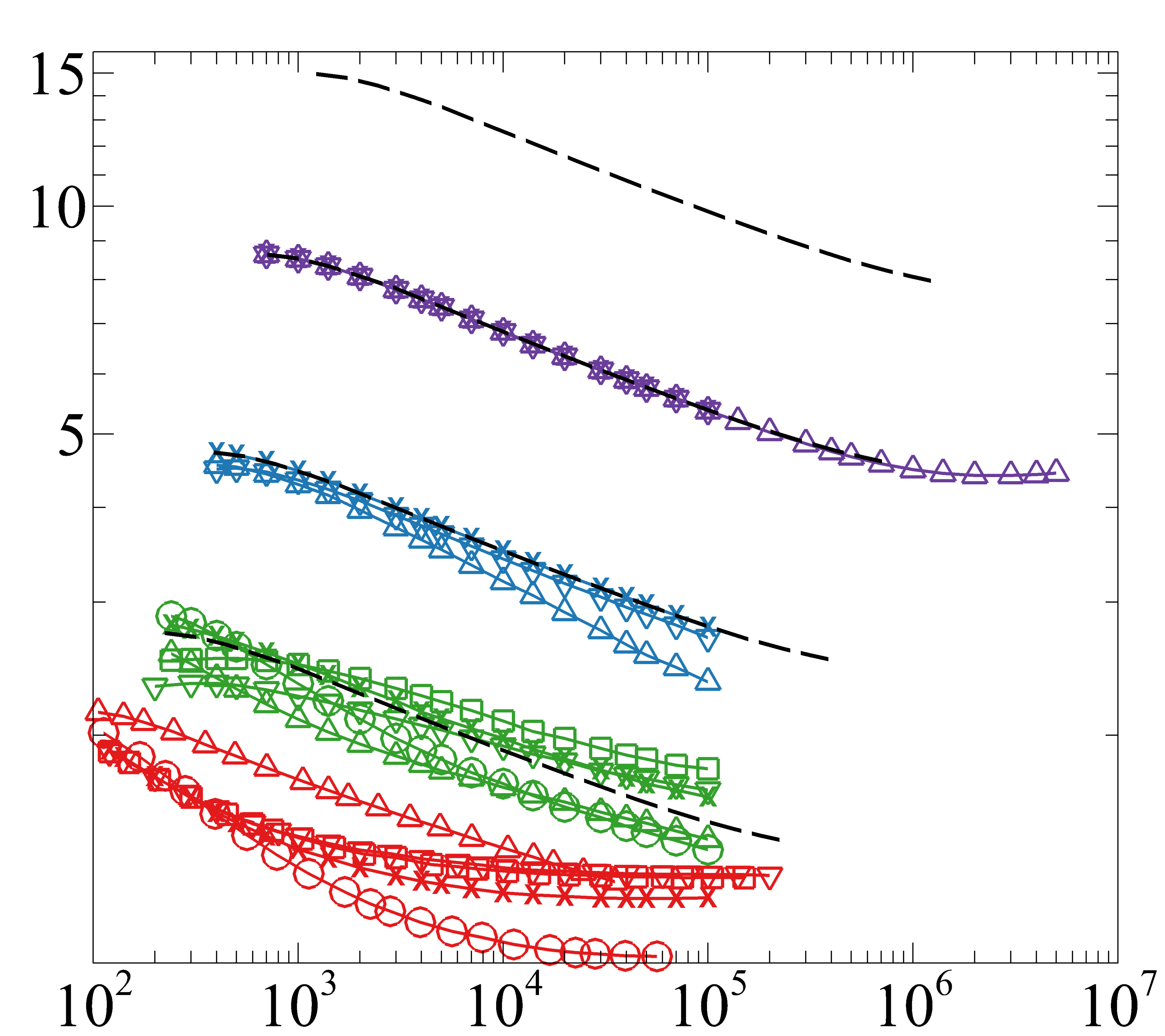}} \\
 & \hspace{33mm} \footnotesize{$\ReyD$, $\Rey$} \\
\end{tabular}
\addtolength{\tabcolsep}{+2pt}
\addtolength{\extrarowheight}{+10pt}
\end{center}
    \caption{
    Transient growth results for various $\UsubR$ and $H$. (a) Time and wave number optimized maximum growth. (b) Optimal time interval. (c) Optimal wave number. $\ReyD$ and $\tauDOpt=\tauOpt/(1-U_\mathrm{min})$ are plotted for $H = 1$, and $Re$ and $\tauOpt$ for $H = 10$, 30 and 100. The black long dashed lines correspond to an isolated Shercliff boundary layer \citep{Camobreco2020role}, at $H=10$, $30$, $100$ and $300$. For these, plotted quantities are $\Rey=\ReyS\,H^{1/2}$, $\alpha_{\mathrm{opt}}=\alpha_\mathrm{opt,S}\,H^{1/2}$ and $\tau_{\mathrm{opt}}=\tau_\mathrm{opt,S}/H^{1/2}$. The $\Rey^{2/3}$ and $\Rey^{1/3}$ power laws for $\Gmax$ and $\tauOpt$ are approximate.}
    \label{fig:all_transient}
\end{figure}

\begin{figure}
\begin{center}
\addtolength{\extrarowheight}{-10pt}
\addtolength{\tabcolsep}{-2pt}
\begin{tabular}{ ll ll ll ll }
\footnotesize{(a)} & \footnotesize{\hspace{3mm} $H=1$, $\UsubR=1$}  &
\footnotesize{(b)} & \footnotesize{\hspace{3mm} $H=1$, $\UsubR=-1$} & 
\footnotesize{(c)} & \footnotesize{\hspace{3mm} $H=100$, $\UsubR=1$} &
\footnotesize{(d)} & \footnotesize{\hspace{3mm} $H=100$, $\UsubR=-1$} \\
\makecell{ \\  \vspace{10mm} \rotatebox{90}{\footnotesize{$y$}}} & \makecell{\includegraphics[width=0.208\textwidth]{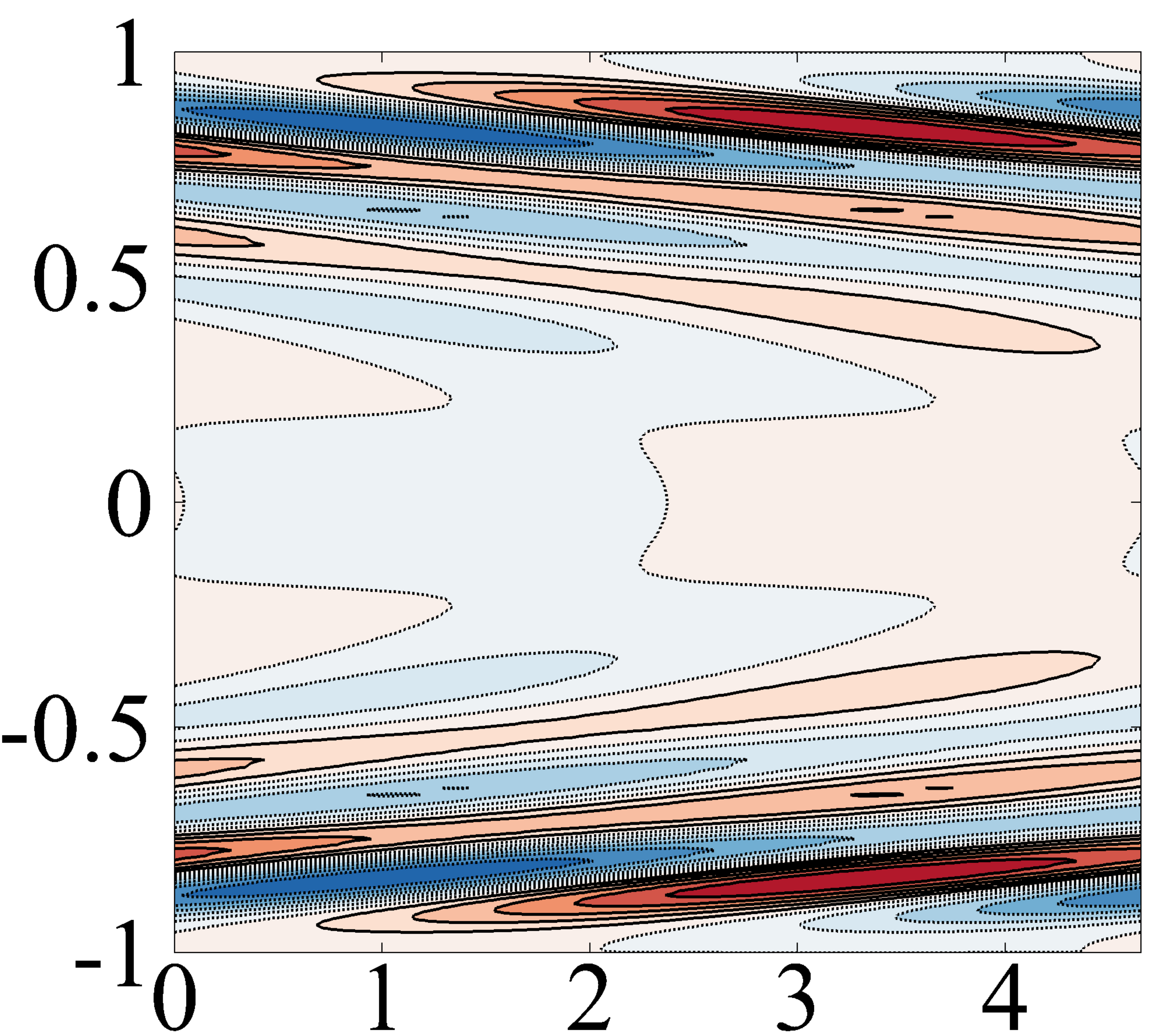}} & 
\makecell{ \\  \vspace{10mm} \rotatebox{90}{\footnotesize{$y$}}} & \makecell{\includegraphics[width=0.208\textwidth]{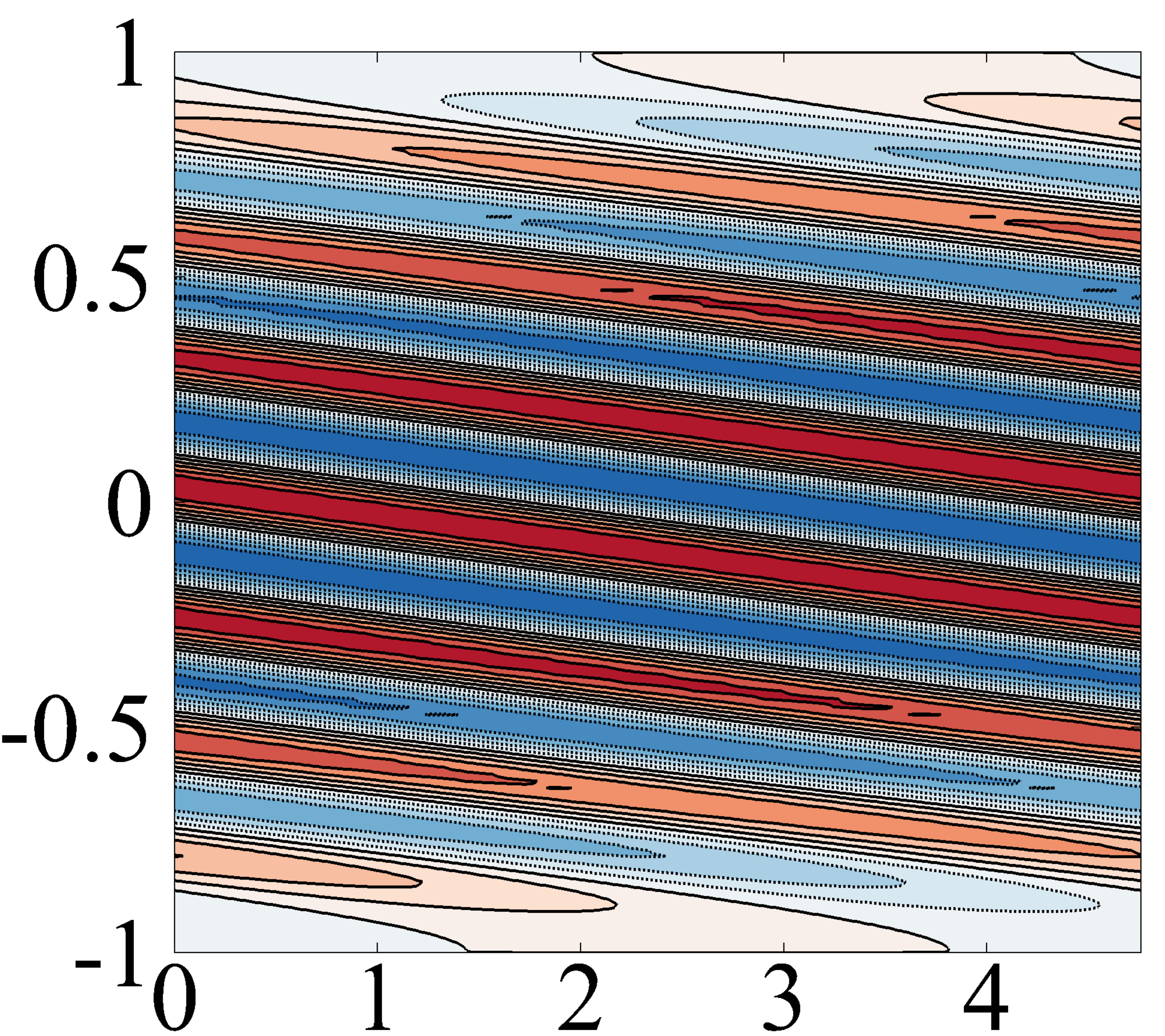}} &
\makecell{ \\  \vspace{10mm} \rotatebox{90}{\footnotesize{$y$}}}  & \makecell{\includegraphics[width=0.208\textwidth]{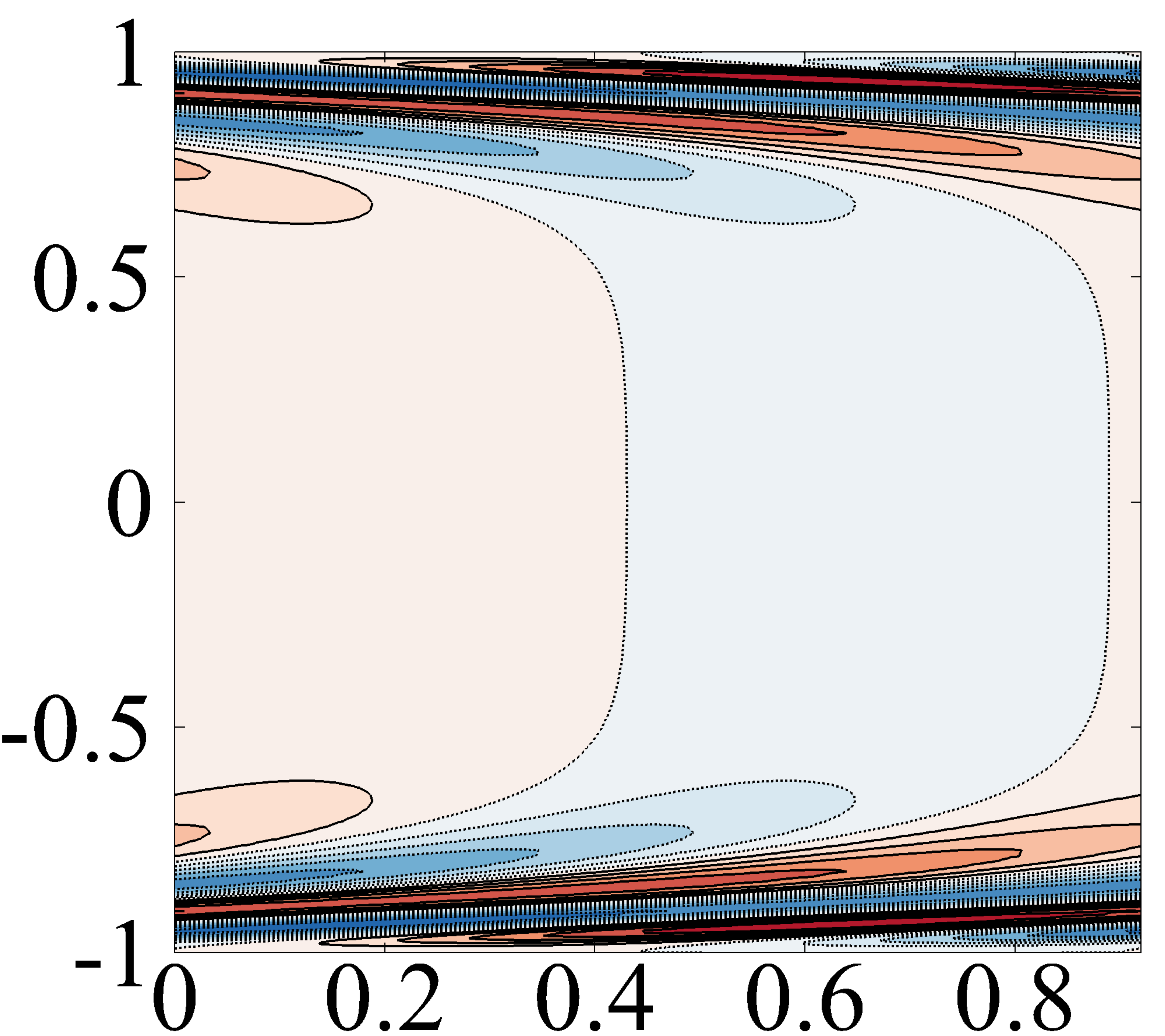}} & 
\makecell{ \\  \vspace{10mm} \rotatebox{90}{\footnotesize{$y$}}} & \makecell{\includegraphics[width=0.208\textwidth]{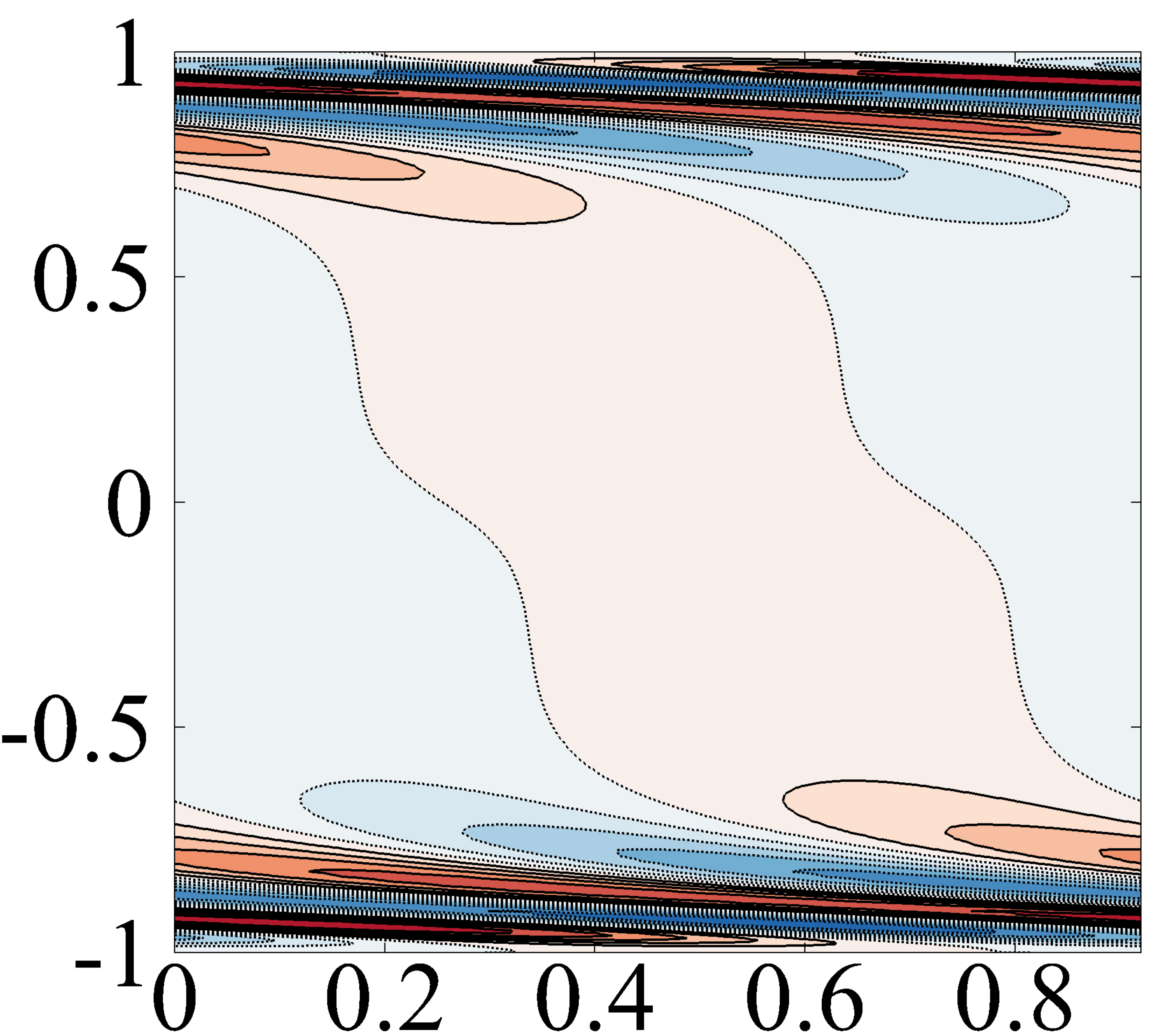}} \\
 & \hspace{16mm} \footnotesize{$x$} &  & \hspace{16mm} \footnotesize{$x$} &  & \hspace{16mm} \footnotesize{$x$} &  & \hspace{16mm} \footnotesize{$x$} \\
\makecell{ \\  \vspace{10mm} \rotatebox{90}{\footnotesize{$y$}}} & \makecell{\includegraphics[width=0.208\textwidth]{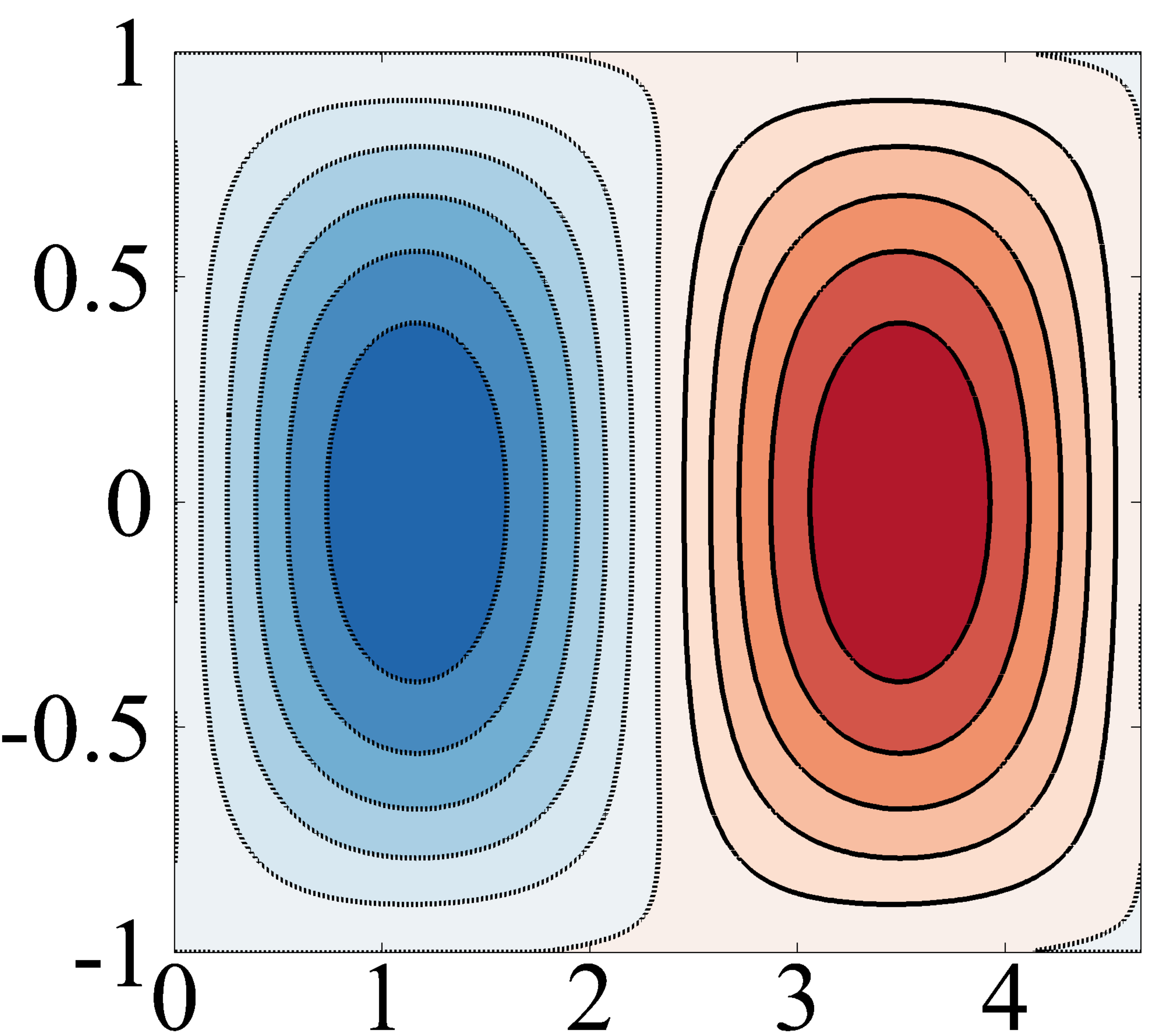}} & 
\makecell{ \\  \vspace{10mm} \rotatebox{90}{\footnotesize{$y$}}} & \makecell{\includegraphics[width=0.208\textwidth]{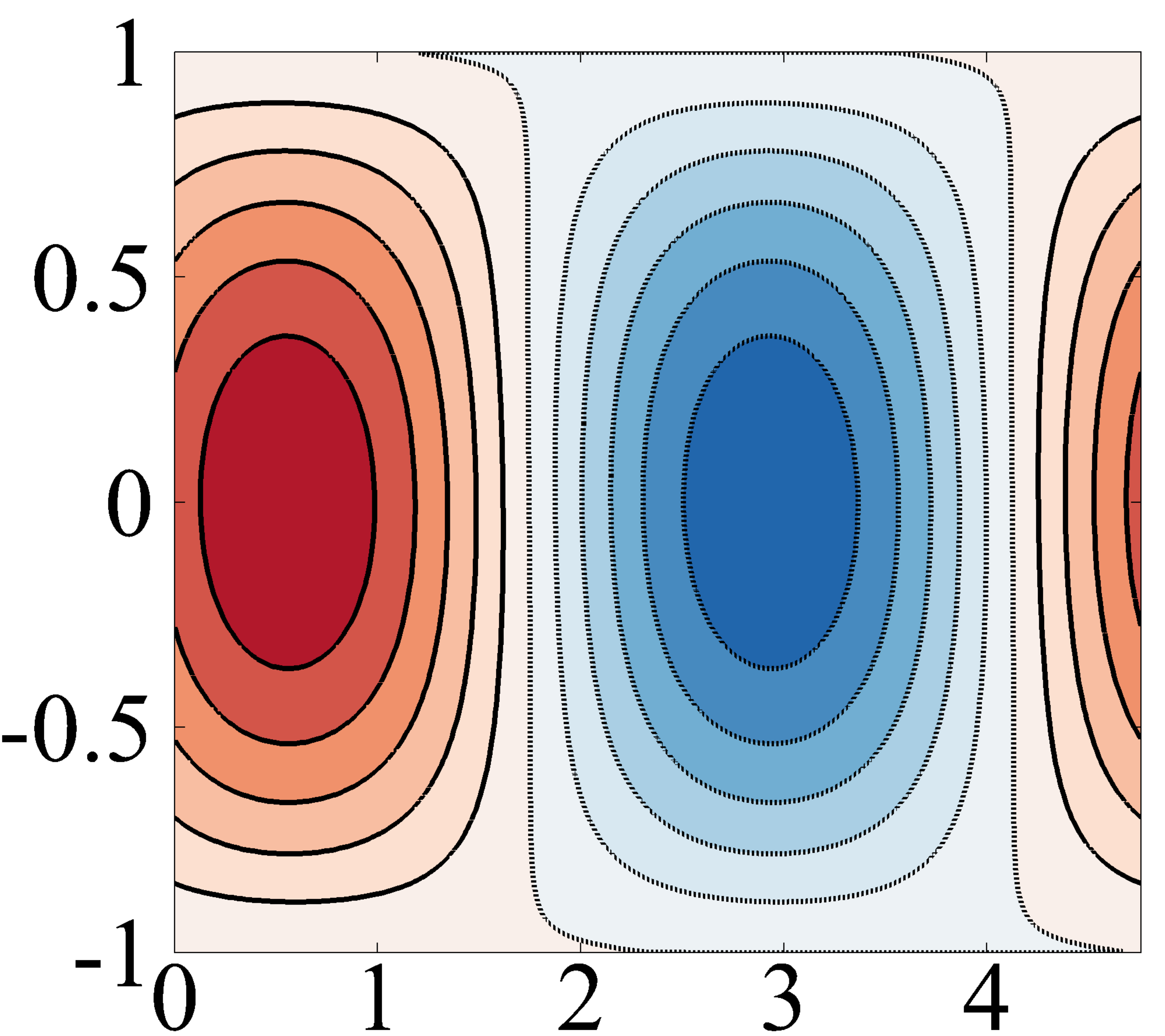}} &
\makecell{ \\  \vspace{10mm} \rotatebox{90}{\footnotesize{$y$}}}  & \makecell{\includegraphics[width=0.208\textwidth]{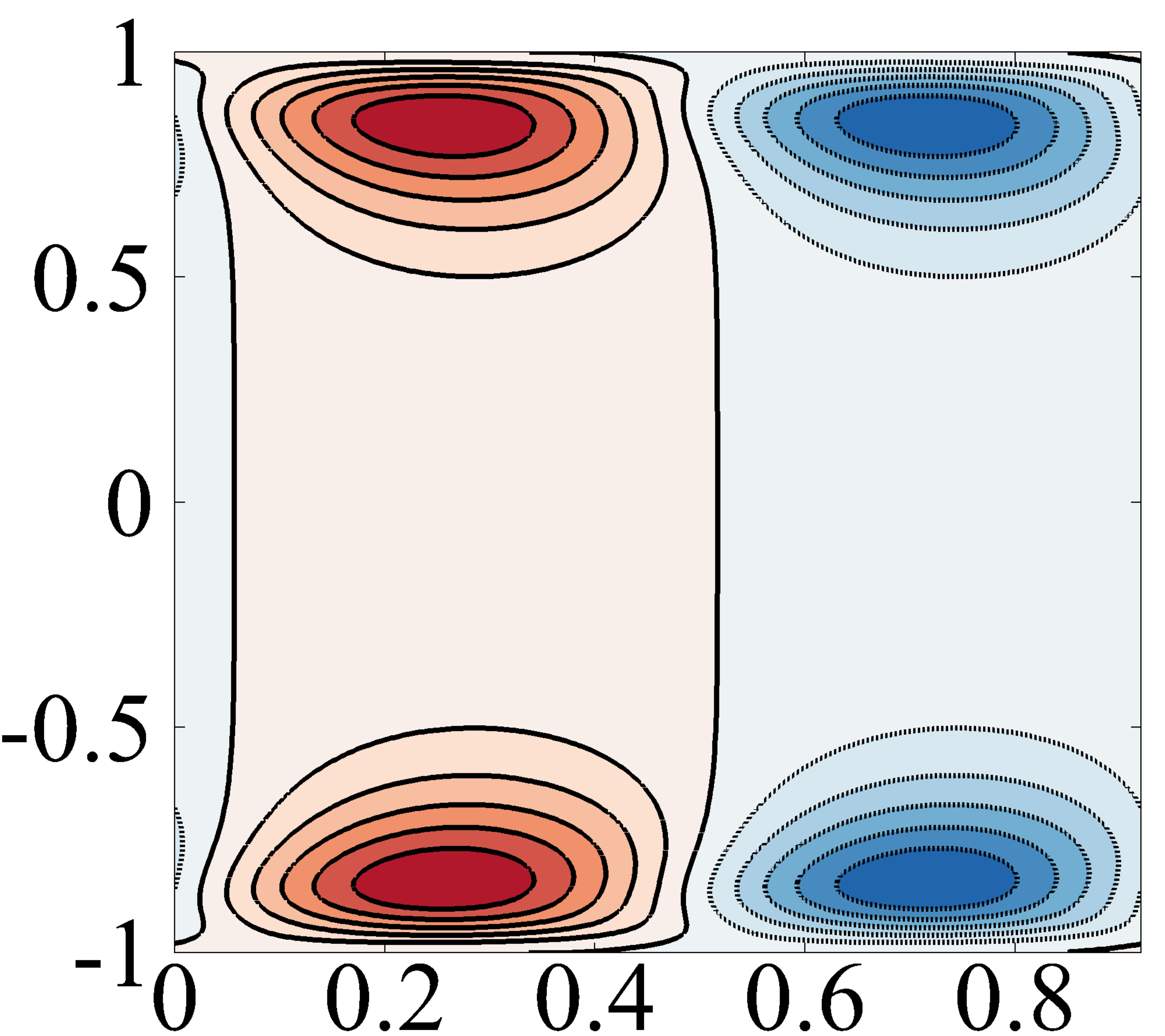}} & 
\makecell{ \\  \vspace{10mm} \rotatebox{90}{\footnotesize{$y$}}} & \makecell{\includegraphics[width=0.208\textwidth]{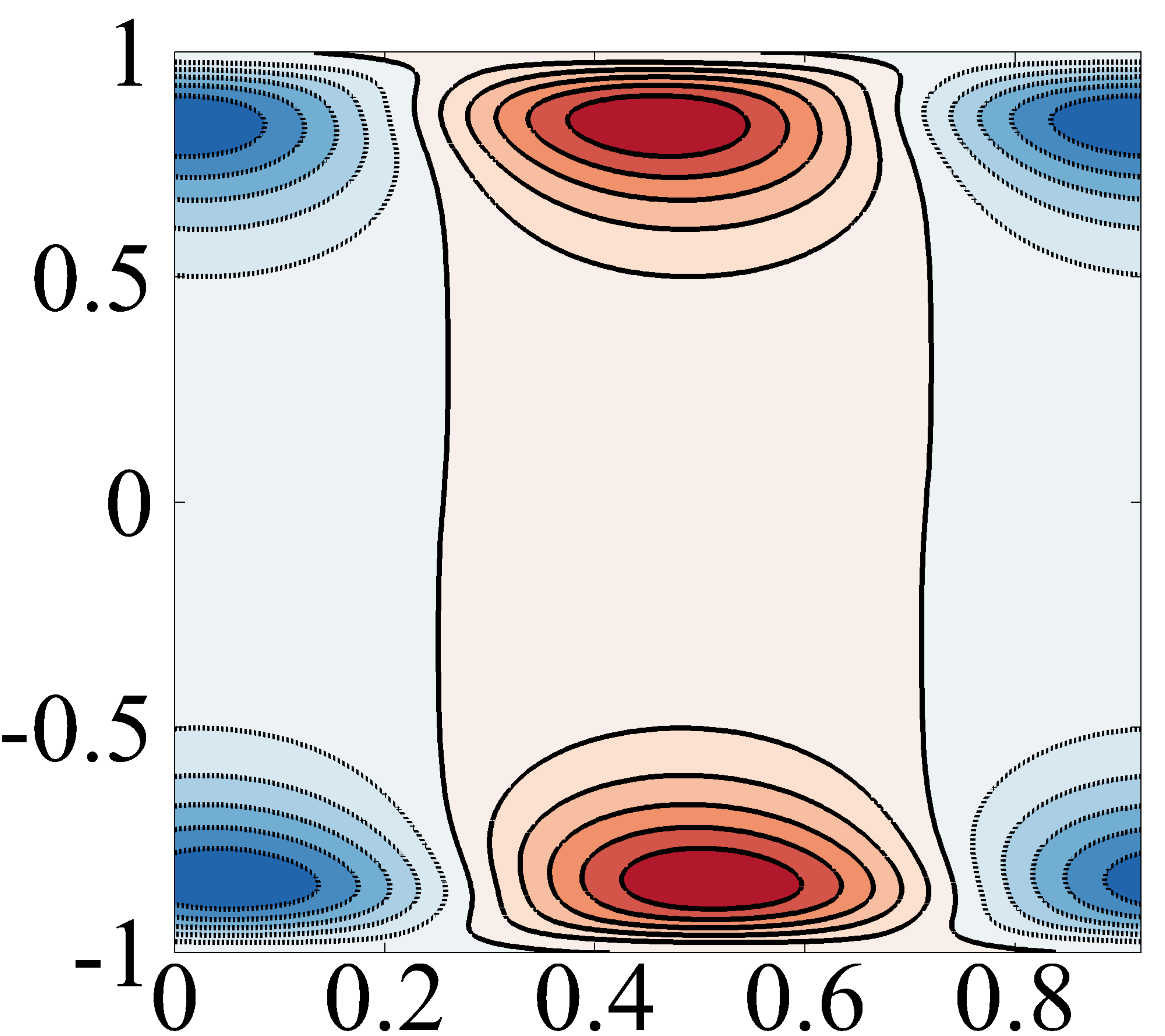}} \\
 & \hspace{16mm} \footnotesize{$x$} &  & \hspace{16mm} \footnotesize{$x$} &  & \hspace{16mm} \footnotesize{$x$} &  & \hspace{16mm} \footnotesize{$x$} \\
\end{tabular}
\addtolength{\tabcolsep}{+2pt}
\addtolength{\extrarowheight}{+10pt}
\end{center}
    \caption{Optimized perturbations at $\ReyD \approx 10^4$, at $t=0$ (top row) and linearly evolved to $t=\tauOpt$ (bottom row), comparing Shercliff and MHD-Couette flows. $\hat{v}-$velocity contours, solid lines (red flooding) positive; dotted lines (blue flooding) negative.}
    \label{fig:trans_lvecs}
\end{figure}

The optimized growth for various base flows, over a range of $H$ values, is depicted in \fig\ \ref{fig:all_transient}. Unlike in 3D flows where the lift-up mechanism incites significant growth 
\citep{Butler1992optimal, Schmid2001stability}, 2D transient growth is driven by the less efficient Orr mechanism. The maximum transient growth found in the present study is accordingly lower, scaling as $\Gmax \sim \Rey^{2/3}$, with magnitudes of only $\Gmax\simeq 10^2$ for Reynolds numbers of ${10^4}$ to ${10^5}$, depending on $H$ and $\UsubR$.
At $H = 30$ the transient growth already closely matches that of an isolated exponential boundary layer (long dashed lines in \fig\ \ref{fig:all_transient}) for all $\UsubR$. By $H=100$, $\Gmax$, $\alphaOpt$ and $\tau_\mathrm{opt}$ all respectively collapse to that limit.
As in the energetic analysis, this collapse occurs at far lower $H$ than the linear stability analysis.
This could be due to the much larger wave numbers at which the transient growth and the energetic analysis optimals occur. The \TS\ waves thereby penetrate a shorter distance into the bulk (see \fig\ \ref{fig:trans_lvecs}) and therefore become isolated at a smaller friction parameter.
The \TS\ wave optimals otherwise have the same general appearance as the linear stability eigenmodes (\fig\ \ref{fig:lin_evecs_vary_H_sym}), except that both MHD-Couette and Shercliff flows have wave structures at both walls, which thereby require similar friction parameters to isolate. This leads to the overall difference in transient growth across the family of profiles to be negligible even at relatively low $H \geq 30$. Constructive interference between modes at the top and bottom walls may be the cause of the slightly larger growth observed for symmetric base flows at smaller $H$. However, even this is not large, such that the base flow does not make a significant difference in generating Q2D linear transient growth. Furthermore, the degree of symmetry in the base flow is not relevant once $H$ is sufficiently large to flatten the central region, and isolate the boundary layers, after which all growth values collapse to those of an isolated exponential boundary layer.


\subsection{Results: pseudospectra}\label{sec:psd_results}

\begin{figure}
\begin{center}
\addtolength{\extrarowheight}{-10pt}
\addtolength{\tabcolsep}{-2pt}
\begin{tabular}{ llll }
  \footnotesize{(a)} & \footnotesize{\hspace{4mm} $H=10$ ($\alphaOpt=1.71$), $\UsubR = 1$} 
& \footnotesize{(b)} & \footnotesize{\hspace{4mm} $H=10$ ($\alphaOpt=1.93$), $\UsubR = -1$} \\
\makecell{\vspace{24mm}  \\  \vspace{31mm} \rotatebox{90}{\footnotesize{$\Im$}}} & \makecell{\includegraphics[width=0.458\textwidth]{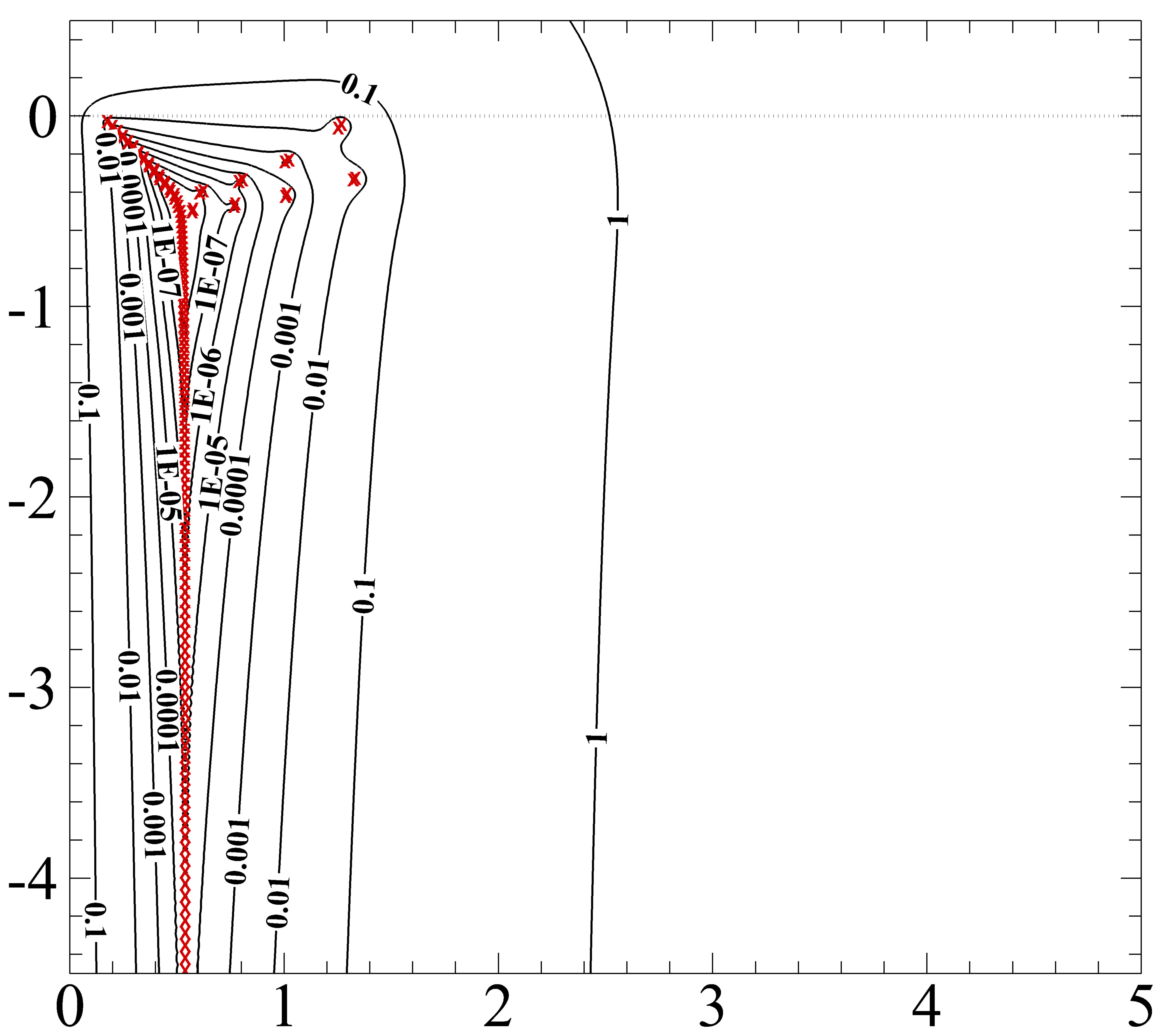}} &
\makecell{\vspace{24mm} \\  \vspace{31mm} \rotatebox{90}{\footnotesize{$\Im$}}} & \makecell{\includegraphics[width=0.458\textwidth]{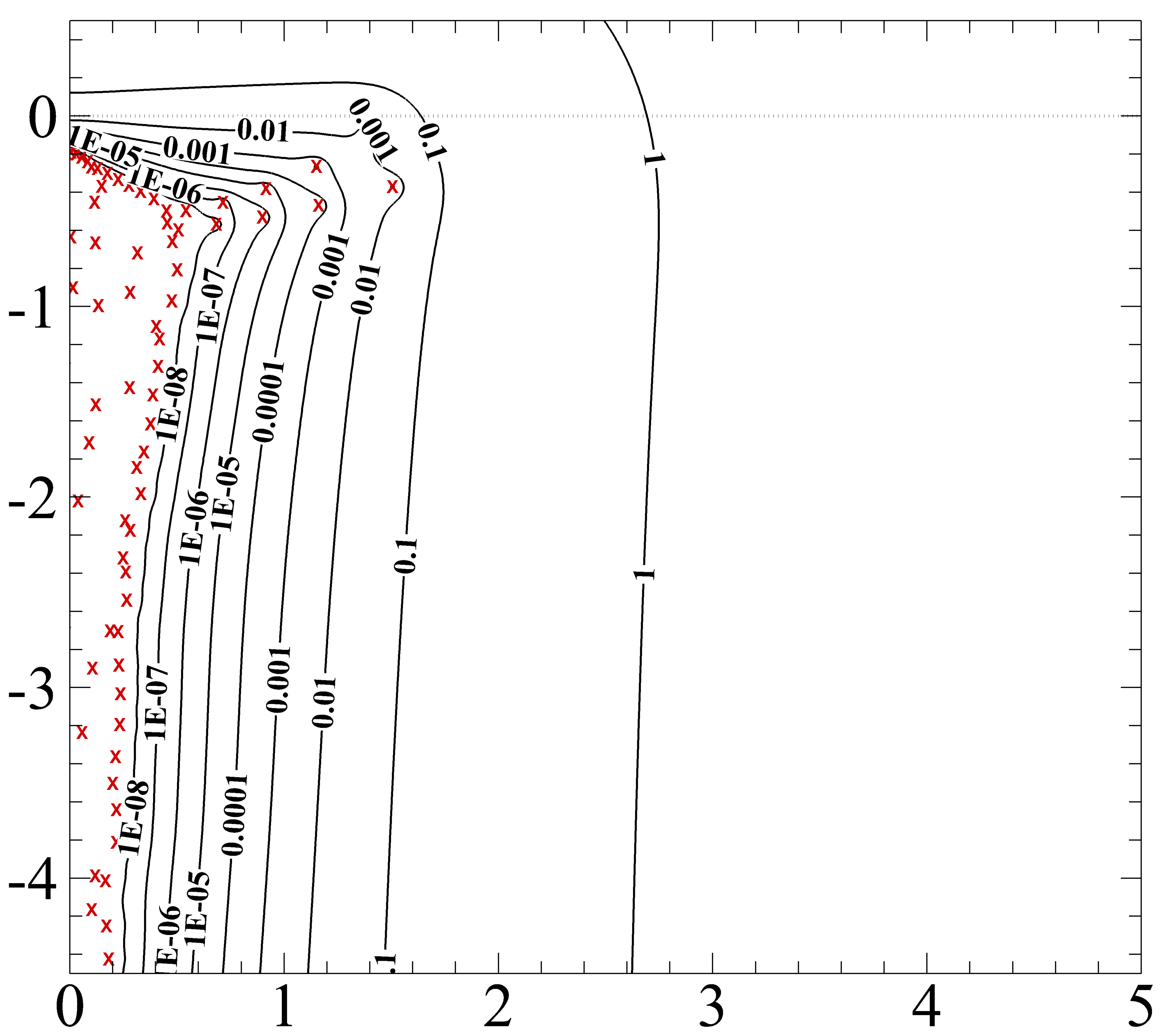}} \\
& \hspace{37mm} \footnotesize{$\Re$} & & \hspace{37mm} \footnotesize{$\Re$} \\
  \footnotesize{(c)} & \footnotesize{\hspace{4mm} $H=100$ ($\alphaOpt=6.83$), $\UsubR = 1$} 
& \footnotesize{(d)} & \footnotesize{\hspace{4mm} $H=100$ ($\alphaOpt=6.83$), $\UsubR = -1$} \\
\makecell{\vspace{24mm}  \\  \vspace{31mm} \rotatebox{90}{\footnotesize{$\Im$}}} & \makecell{\includegraphics[width=0.458\textwidth]{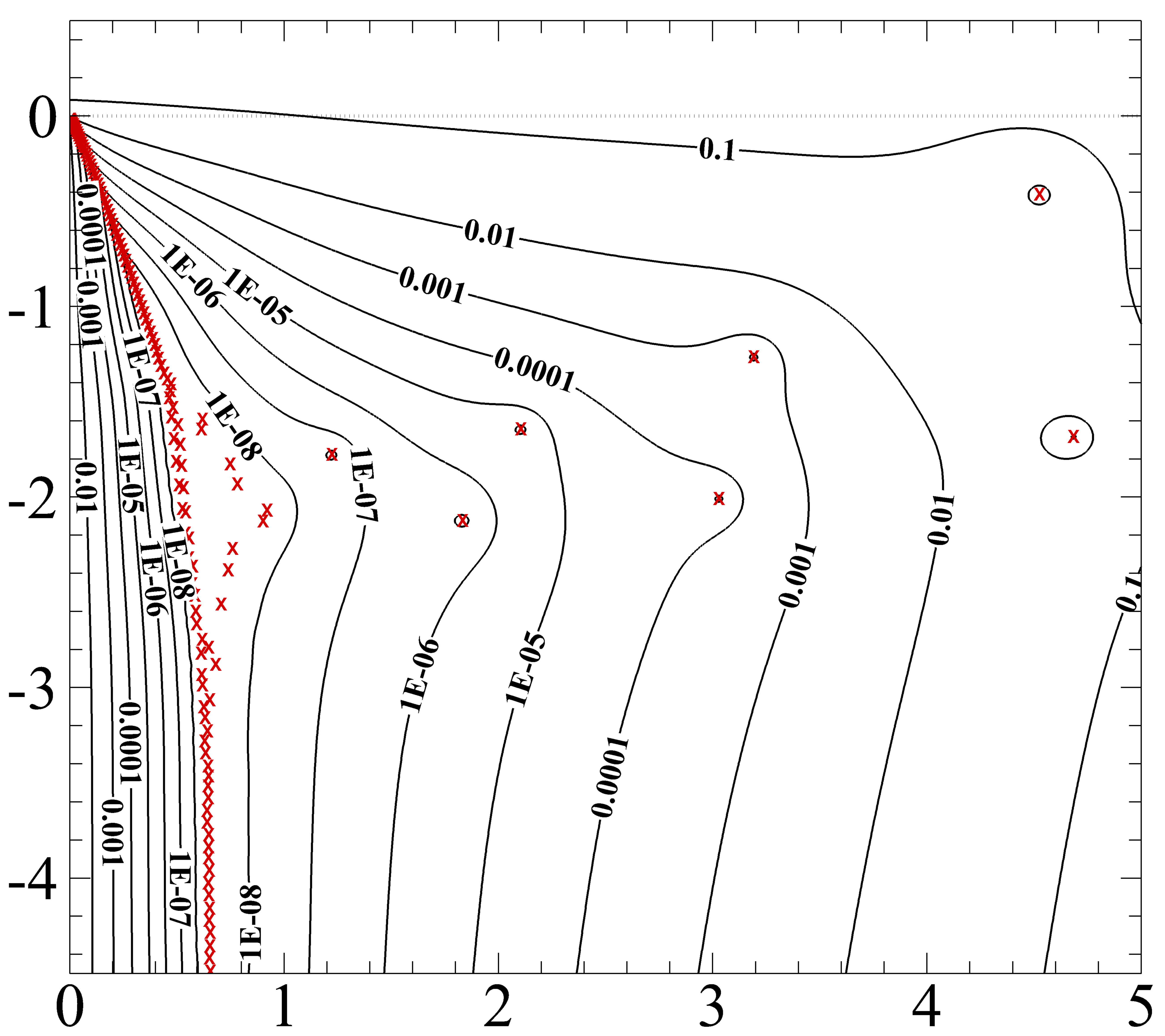}} &
\makecell{\vspace{24mm}  \\  \vspace{31mm} \rotatebox{90}{\footnotesize{$\Im$}}} & \makecell{\includegraphics[width=0.458\textwidth]{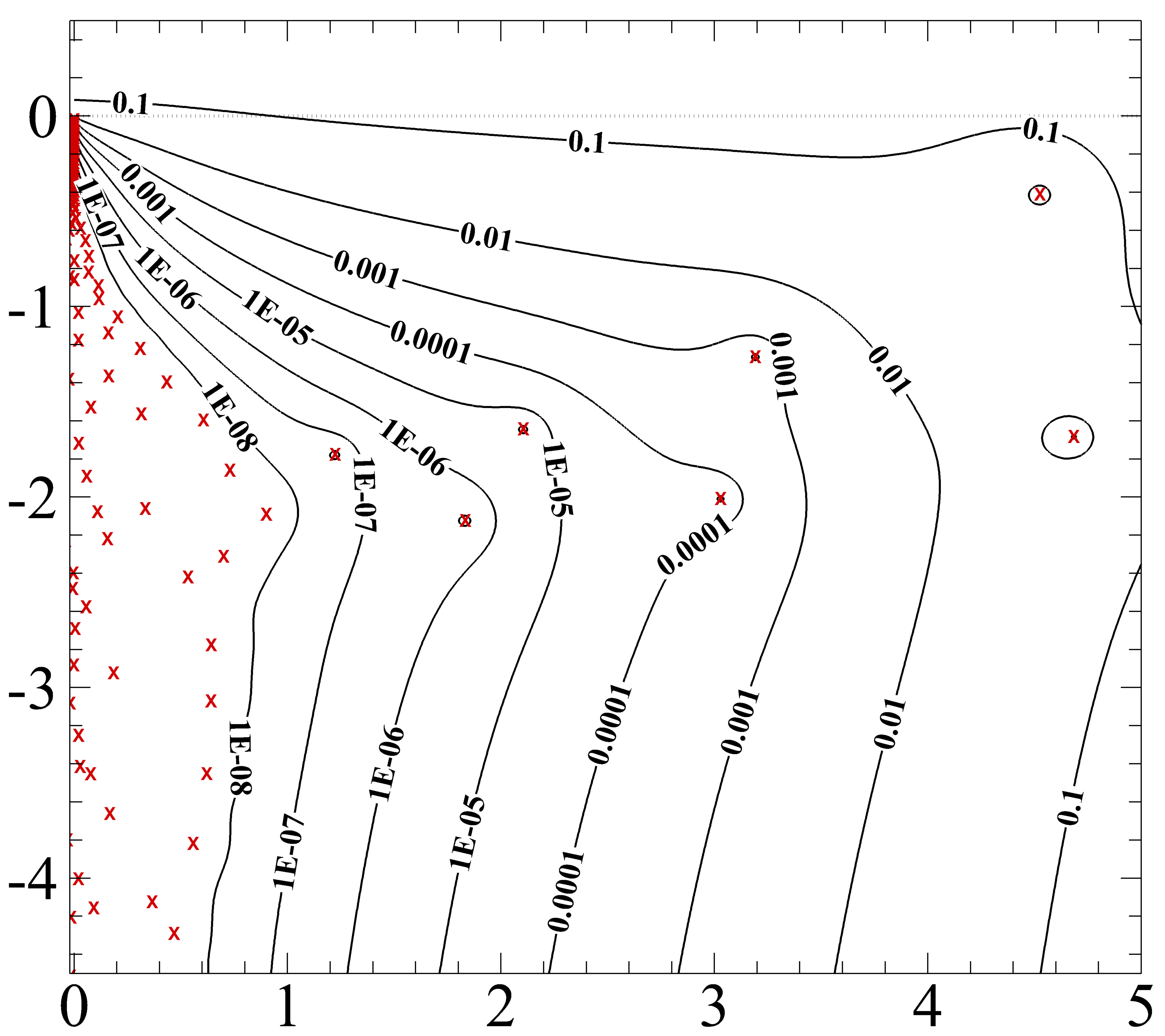}} \\
& \hspace{37mm} \footnotesize{$\Re$} & & \hspace{37mm} \footnotesize{$\Re$} \\
\end{tabular}
\addtolength{\tabcolsep}{+2pt}
\addtolength{\extrarowheight}{+10pt}
\end{center}
    \caption{Pseudospectra (contours from $10^{-8}$ to $1$) at $\ReyD \approx 10^4$, comparing Shercliff and MHD-Couette flows (the latter symmetric about the imaginary axis). Red crosses denote eigenvalues from the linear stability analysis $\lambda$, computed at the same parameters as the linear transient optima.}
    \label{fig:psd_H100_1e4}
\end{figure}

The transient growth results are also supported by the pseudospectra. \Fig\ \ref{fig:psd_H100_1e4} depicts pseudospectra obtained at $\ReyD \approx 10^4$ for both the Shercliff and MHD-Couette base flows, at $H=10$ and $H=100$. Increasing the Reynolds number directly brings more eigenvalues close to the real axis, allowing smaller perturbations to cross to the positive imaginary half plane, thereby generating more transient growth \citep{Reddy1993pseudospectra}.
However, as demonstrated in \fig\ \ref{fig:psd_H100_1e4}, increasing the Hartmann friction parameter mainly stretches the pseudospectra along the real axis, with the further separation of the eigenvalues appearing to lead to reduced transient growth for a given Reynolds number. 
This is supported by determining the condition number of the basis, $\kappa=||\matbm{W}||_2||\matbm{W}^{-1}||_2$ \citep{Reddy1993pseudospectra, Schmid2001stability}, recalling that a normal operator has a condition number of unity. At or near hydrodynamic conditions, the condition number of MHD-Couette flow is much higher than for Shercliff flow. This was observed in  
3D non-MHD Couette and Poiseuille flows \citep{Reddy1993pseudospectra} and remains unexplained. For example at $H=1$, $\ReyD \approx 10^3$ the condition numbers for Shercliff and MHD-Couette flows are $1.9\times 10^3$ and $1.2\times 10^8$, respectively (at $\alphaOpt$). However, at $H=100$, $\ReyD \approx 10^3$ the condition numbers are respectively $1.0 \times 10^4$ and $1.3 \times 10^6$. Hence, an increasing Hartmann friction parameter acts to re-orient eigenvectors such that they are more normal for MHD-Couette flow, and less normal for Shercliff flow. It also indicates the increasing similarity between these base flow profiles with increasing $H$. 

\section{Weakly nonlinear stability}\label{sec:wnl}
\subsection{Formulation}\label{sec:wnl_form}
By assuming a small perturbation amplitude $\bigO{\epsilon}$, to allow linearization, linear stability analysis becomes amplitude independent. However, if amplitude dependence is maintained, a weakly nonlinear analysis can be performed. To remain accurate, the weakly nonlinear analysis is concerned only with expansion about a leading perturbation which is close to neutrally stable. This ensures only one mode is unstable \citep{Drazin2004hydrodyamic}. Linearly, a single unstable mode would either slowly grow or decay exponentially. However, if weakly nonlinear self-interaction occurs, the overall growth rate will increase or decrease, depending on whether the leading nonlinear growth term is positive or negative. A positive nonlinear growth can outweigh a negative linear growth rate if the linear growth is sufficiently small (close to the neutral curve), such that growth occurs at $\Rey<\ReyCrit$ until a saturation amplitude, or a turbulent state, is reached (in which case the bifurcation is subcritical). If the nonlinear term is negative, $\Rey>\ReyCrit$ is required for non-transient growth (the bifurcation is supercritical). The amplitude dependence of the plane-wave mode $\hat{w}_n(y) = w(y)e^{i\alpha n x}$ is expanded as
\begin{equation}\label{eq:amp_exp}
\hat{w}_n = \sum_{m=0}^{\infty}\epsilon^{|n|+2m}\tilde{A}^{|n|}|\tilde{A}|^{2m}\hat{w}_{n,|n|+2m},
\end{equation}
where $\hat{w}_{n,|n|+2m}$ now denotes a perturbation (the first subscript is the harmonic, the second the amplitude), in line with Ref.~\cite{Hagan2013weakly}, and $\tilde{A} = A/\epsilon$ is the normalized amplitude. The wave frequency $\omega$ is also expanded as $\omega = \omega_0 + \epsilon^2 \tilde{\omega}_2 + \cdots$, where the normalized amplitude $\tilde{\omega}_2 = \omega_2/\epsilon^2$. The linearly unstable mode $\hat{w}_{1,1}$ (which is $\hat{v}$ under rescaling) of $\bigO{\epsilon}$ excites via self-interaction through the nonlinear term a second harmonic $\hat{w}_{2,2}$ and a modification to the base flow $\hat{u}_{0,2}$ (zeroth harmonic), which both have amplitude of $\bigO{\epsilon^2}$ \citep{Hagan2013weakly}. These harmonics also interact with the original perturbation, resulting in another harmonic $\hat{h}_{1,3}^w$ with amplitude of $\bigO{\epsilon^3}$ \citep{Hagan2013weakly}. Higher order terms are neglected, as they have a rapidly increasing radius of convergence \citep{Schmid2001stability}. However, such an expansion is sufficient to define the bifurcation type as sub- or supercritical and determine whether the system is sensitive to subcritical perturbations of finite amplitude. 

The weakly non-linear stability is calculated following the method outlined in Ref.~\cite{Hagan2013weakly}, where the key equations are provided here. 
Denoting $U=\hat{u}_0$ in line with Ref.~\cite{Hagan2013weakly}, the equations governing higher-order harmonics of the base flow and the perturbation are
\begin{equation}\label{eq:L_b}
D^2\hat{u}_{0,2m} - H\hat{u}_{0,2m} = \hat{g}_{0,2m},
\end{equation}
%
\begin{equation}\label{eq:L_n}
\mathcal{L}_n\hat{w}_n = [(D_n^2 - \ii\lambda n)D_n^2 - H D_n^2 - \ii\alpha n(\hat{u}_0''-u_{0,0}D_n^2)]\hat{w}_{n} = \hat{h}_{n,|n|+2m}^w,
\end{equation}
respectively, where $D_n = \partial/\partial y \vect{e}_y + in\alpha \vect{e}_x$ and where the RHS's, representing the curl of the nonlinear term, are
\begin{equation}
\hat{g}_{0,2m}=\ii\sum_{m\neq 0}^{\infty}(\alpha m)^{-1}\hat{w}_m^*D^2\hat{w}_m,
\end{equation}
\begin{equation}\label{eq:h_eq}
\hat{h}_{n,|n|+2m}^w = n\sum_{m\neq 0}^{\infty}m^{-1}(\hat{w}_{n-m}D_m^2D\hat{w}_m - D(\hat{w}_m)D_{n-m}^2\hat{w}_{n-m}).
\end{equation}
Here $*$ denotes complex conjugation. Equations (\ref{eq:L_b}) and (\ref{eq:L_n}) are identical to those used to determine the base flows in Sec.~\ref{sec:base_flows} and the linear stability results in Sec.~\ref{sec:lin}, respectively, if the RHS's are set to zero (taking $m=0$, $n=1$). Equations (\ref{eq:L_b}) through (\ref{eq:h_eq}) are discretized into matrix operators and solved as follows, noting that after determining the RHS's of equations (\ref{eq:L_b}) and 
(\ref{eq:L_n}), the amplitude expansion for $\hat{w}_n$ should be substituted in. First, the SM82-modified Orr--Sommerfeld eigenvalue problem
\begin{equation}\label{eq:lin_HP}
[\matbm{A}_1^{-1}\matbm{M}_1(u_{0,0}) - \lambda \matbm{I}]w_{1} = 0,
\end{equation}
is solved in the standard form, which provides the leading eigenvalue $\lambda_1$, with frequency $\omega_0 = \Imz(\lambda_1)$, and the corresponding right and left eigenvectors, $w_{1,1}$ and $w_{1,1}^\dag$, respectively. $\ReyCrit$ and $\alphaCrit$ are determined from the linear stability problem, equation~(\ref{eq:lin_HP}), with neutral conditions satisfying $\Rez(\lambda_1)=0$ in this formulation. The following are then solved,
\begin{equation}
u_{0,2}  = -2\alpha^{-1}(\matbm{D}^2 - H\matbm{I})^{-1}\Imz(w_{1,1}^*\matbm{F}_0 \matbm{D}^2w_{1,1}),
\end{equation}
\begin{equation}\label{eq:w22}
w_{2,2}  = (\matbm{M}_2 - 2i\omega_0\matbm{A}_2)^{-1}[2\matbm{F}_2\matbm{D}(\matbm{D}(w_{1,1}\matbm{D}w_{1,1}) - 2(\matbm{D}w_{1,1})^2)],
\end{equation}              
\begin{equation}
h_{1,3}^w = 0.5(w_{1,1}^*\matbm{A}_2\matbm{D}w_{2,2} - \matbm{D}(w_{2,2})\matbm{A}_1w_{1,1}^*) - (w_{2,2}\matbm{A}_1\matbm{D}w_{1,1}^* - \matbm{D}(w_{1,1}^*)\matbm{A}_2w_{2,2}),
\end{equation}
where,
\begin{equation}
\matbm{A}_n = \matbm{D}^2 - n^2\alpha^2\matbm{I}, \, \matbm{N}_n(u_{0,0}) = \ii\alpha(u_{0,0}''\matbm{I}-u_{0,0}\matbm{A}_n), \, M_n = \matbm{F}_n[\matbm{A}_n^2 - H\matbm{A}_n + \Rey\matbm{N}_n],
\end{equation}
with boundary condition matrix $\matbm{F}_n$ as given in \cite{Hagan2013capacitance, Hagan2013weakly}. $w_{1,1}$ is normalized such that $D^2w_{1,1}(1)=1$, 
and $w_{1,1}^\dag$ such that $w_{1,1}\cdot w_{1,1}^\dag=1$. 

The $n=3$ harmonic of equation (\ref{eq:L_n}) only has a solution when the RHS is not proportional to $w_{1,1}$. Thus the RHS must be orthogonal to the adjoint eigenfunction $w_{1,1}^\dag$ \cite{Hagan2013weakly}. The $n=3$ harmonic is
\begin{equation}
(\matbm{A}_1^{-1}\matbm{M}_1-i\omega_0\matbm{I})w_{1,3} = \matbm{A}_1^{-1}[\matbm{F}_1h_{1,3}^w + 
|A|^{-2}(\matbm{F}_1\matbm{N}_1[(\Rey-\ReyCrit)u_0 + |A|^2 u_{0,2}] + i\omega_2\matbm{A}_1)w_{1,1}].
\end{equation}
The RHS will be zero once orthogonal to $w_{1,1}^\dag$ if the frequency perturbation satisfies
\begin{equation}\label{eq:io2}
\ii\omega_2 = \mu_1(\Rey-\ReyCrit) + \mu_2 |A|^2,
\end{equation} 
where
\begin{equation}
\mu_1 = w_{1,1}^\dag \cdot \matbm{A}_1^{-1}\matbm{F}_1\matbm{N}_1w_{1,1},
\end{equation}
\begin{equation}
\mu_2 = w_{1,1}^\dag \cdot \matbm{A}_1^{-1}\matbm{F}_1(\matbm{N}_1(u_{0,2})w_{1,1}-h_{1,3}^w).
\end{equation}
The linear growth rate correction is then $\mu_1(\Rey-\ReyCrit)$  and the first Landau coefficient $\mu_2$  \citep{Schmid2001stability,Drazin2004hydrodyamic}. Rearranging the real part of equation (\ref{eq:io2}) yields $|A|^2 = -(\Rey-\ReyCrit)\Rez(\mu_1)/\Rez(\mu_2)$. Thus, $\Rez(\mu_1)>0$ is required for the existence of a finite amplitude state, while $\Rez(\mu_2)>0$ (resp.\ $\Rez(\mu_2)<0$) defines a subcritical  (resp.\ supercritical) bifurcation. Note that all coefficients quoted in this paper are rescaled by $\alpha^2\Rey$, following Ref.~\cite{Hagan2013weakly} and Ref.~\cite{Sen1983stability}.

Weakly non-linear analysis is valid only near the neutral curve, such that only one mode is ever unstable. However, MHD-Couette flow yields a conjugate pair of equally unstable modes.
This issue has been circumvented by taking $\UsubR = -1+10^{-10}$ to approximate MHD-Couette flow, which breaks antisymmetry above machine precision. This ensures that there is only one unstable eigenvalue, while having a negligible effect on the linear computations.

Extensive literature comparisons \cite{Reynolds1967finite, Sen1983stability, Stewartson1971nonlinear} were performed when Ref. \cite{Hagan2013weakly} validated their method, for the $H=0$ Posieuille flow problem. Testing the present formulation against this benchmark recovered the values for $\mu_1$ and $\mu_2$ to all 6 significant figures provided in Ref. \cite{Hagan2013weakly}. The resolution required for higher $H$, \tbl\ \ref{tab:val_weakly}, demonstrates that the discretization for the linear stability problem yields acceptable results for the additional weakly nonlinear computations. 
%
\begin{table}
\begin{center}
\begin{tabular}{ c|ccccc } 
\hline
$\Nc$; $H=100$ & $\Rez(\mu_1)\times10^{-4}$ & $\Imz(\mu_1)\times10^{-5}$ & $\Rez(\mu_2)\times10^{3}$ & $\Imz(\mu_2)\times10^{3}$ \\
\hline
$100$ & $1.26052772556414$ & $-2.83996157965910$ & $3.26503843783969$ & $-2.7618383295094$ \\
$\mathbf{200}$ & $1.24061408024751$ & $-2.83961314112167$ & $3.26170229761087$ & $-2.7527141776193$ \\
$300$  & $1.24061417438959$ & $-2.83961313972489$ & $3.26170132992112$ & $-2.7527132574776$ \\
$400$  & $1.24061417435937$ & $-2.83961313972252$ & $3.26170134197776$ & $-2.7527132674919$ \\
\hline
$\Nc$; $H=1000$ & $\Rez(\mu_1)\times10^{-6}$ & $\Imz(\mu_1)\times10^{-8}$ & $\Rez(\mu_2)\times10^{3}$ & $\Imz(\mu_2)\times10^{4}$ \\
\hline
$400$ & $1.02741326199010$ & $-1.78689676323013$ & $1.08671495177896$ & $5.88065081880331$ \\
$\mathbf{500}$ & $1.02741326027113$ & $-1.78689676325672$ & $1.08671494954617$ & $5.88065083277250$ \\
$600$ & $1.02741326031740$ & $-1.78689676325717$ & $1.08671494979183$ & $5.88065083423307$ \\
$700$ & $1.02741326030699$ & $-1.78689676325563$ & $1.08671494908700$ & $5.88065083086592$ \\
\hline
\end{tabular}
\caption{Resolution testing for the weakly nonlinear analysis, Shercliff flow ($\UsubR=1$) at $H=100$, $\ReyCrit = 4.40223\times 10^{5}$, $\alphaCrit = 1.73897$ and MHD-Couette flow ($\UsubR=-1+10^{-10}$), at $H=1000$, $\ReyCrit = 1.52886 \times 10^{6}$, $\alphaCrit = 5.10748$. The bold resolutions are chosen, identical to those for the linear stability analysis.}
\label{tab:val_weakly}
\end{center}
\end{table}

\subsection{Results}\label{sec:wnl_results}
\begin{figure}
\begin{center}
\addtolength{\extrarowheight}{-10pt}
\addtolength{\tabcolsep}{-2pt}
\begin{tabular}{ llll }
\makecell{\vspace{23mm} \footnotesize{(a)} \\  \vspace{31mm} \rotatebox{90}{\footnotesize{$\Re(\mu_1)$}}} & \makecell{\includegraphics[width=0.458\textwidth]{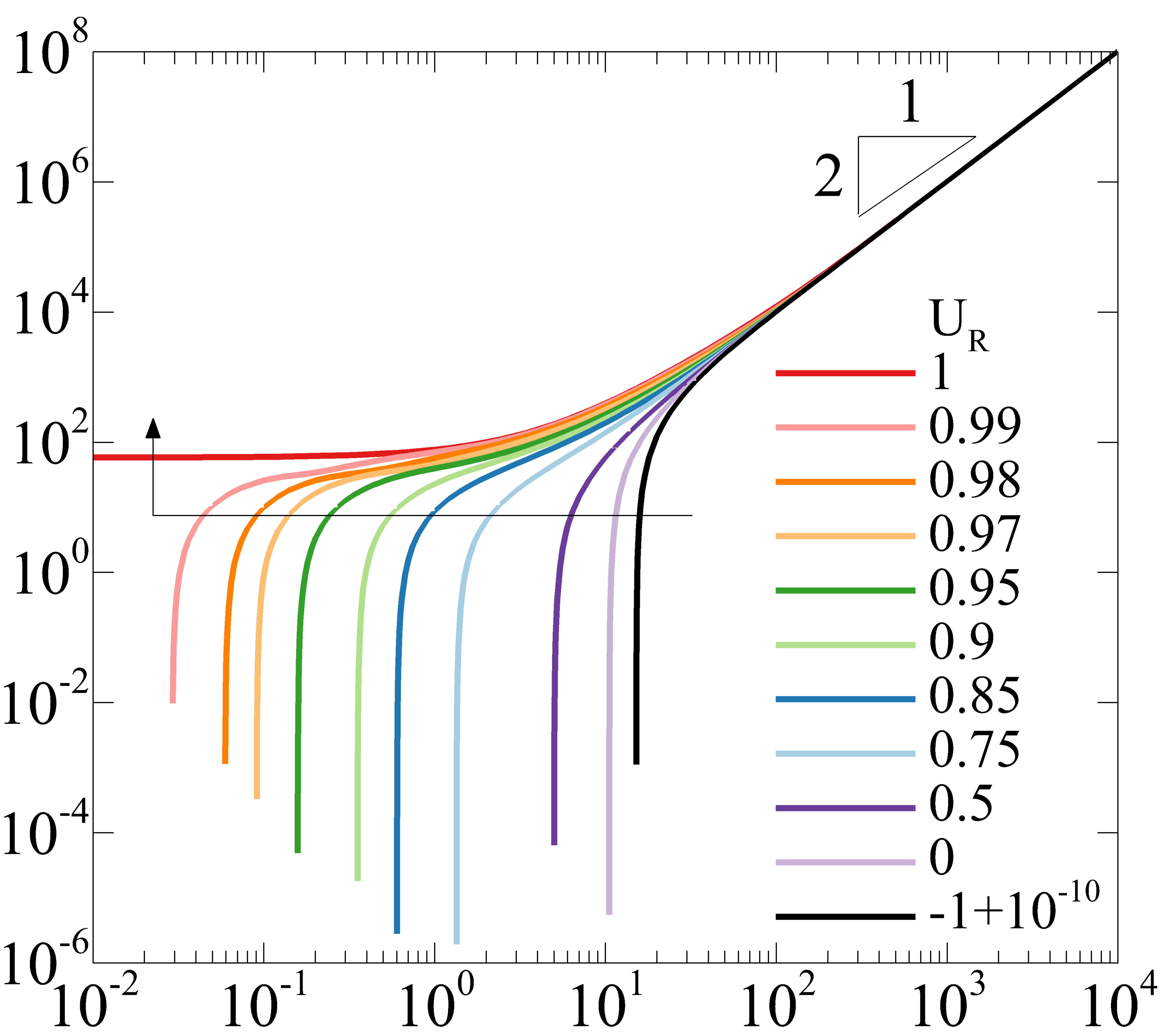}} &
\makecell{\vspace{22mm} \footnotesize{(b)} \\  \vspace{32mm} \rotatebox{90}{\footnotesize{$\Re(\mu_2)$}}} & \makecell{\includegraphics[width=0.458\textwidth]{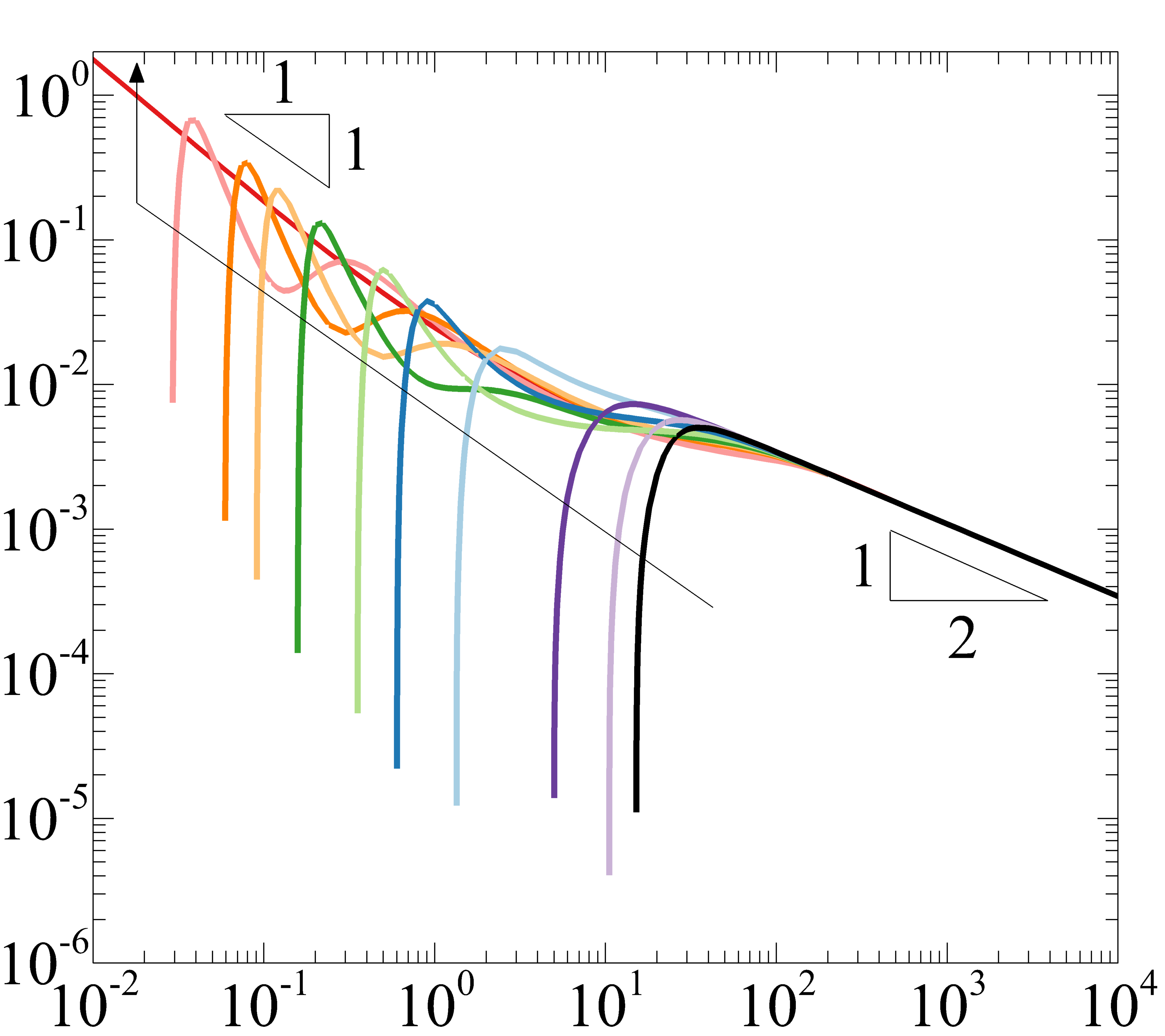}} \\
 & \hspace{40mm} \footnotesize{$H$} & & \hspace{40mm} \footnotesize{$H$} \\
\end{tabular}
\begin{tabular}{ llll }
\makecell{\vspace{12mm} \footnotesize{(c)} \\  \vspace{25mm} \rotatebox{90}{\footnotesize{$|A|^2/(\ReyCrit-\Rey)$}}} & \makecell{\includegraphics[width=0.458\textwidth]{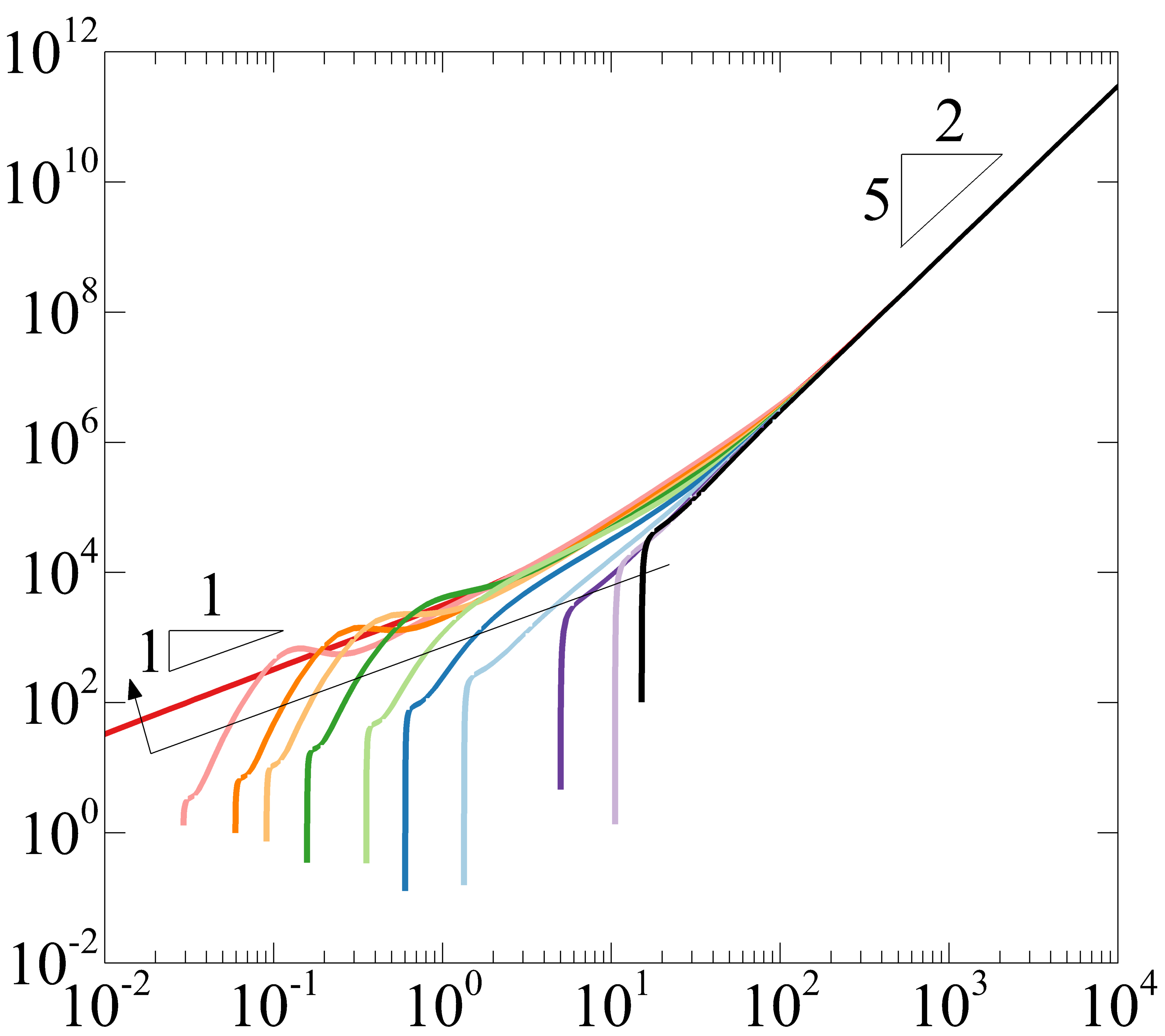}} \\
 & \hspace{40mm} \footnotesize{$H$} \\
\end{tabular}
\addtolength{\tabcolsep}{+2pt}
\addtolength{\extrarowheight}{+10pt}
\end{center}
    \caption{Weakly nonlinear stability for various $\UsubR$ at the critical points (recall \fig\ \ref{fig:lin_and_eng}). (a) Real part of linear growth rate correction coefficient. (b) Real part of first Landau coefficient. (c) Normalized amplitude. Arrows indicate increasing $\UsubR$. As $H \rightarrow \infty$, $|A|^2 = 29.8970H^{5/2}(\ReyCrit-\Rey)$.}
    \label{fig:wnl_all}
\end{figure}
%
\begin{figure}
\begin{center}
\addtolength{\extrarowheight}{-10pt}
\addtolength{\tabcolsep}{-2pt}
\begin{tabular}{ llll }
\makecell{\vspace{26mm} \footnotesize{(a)} \\  \vspace{31mm} \rotatebox{90}{\footnotesize{$\alpha$}}} & \makecell{\includegraphics[width=0.458\textwidth]{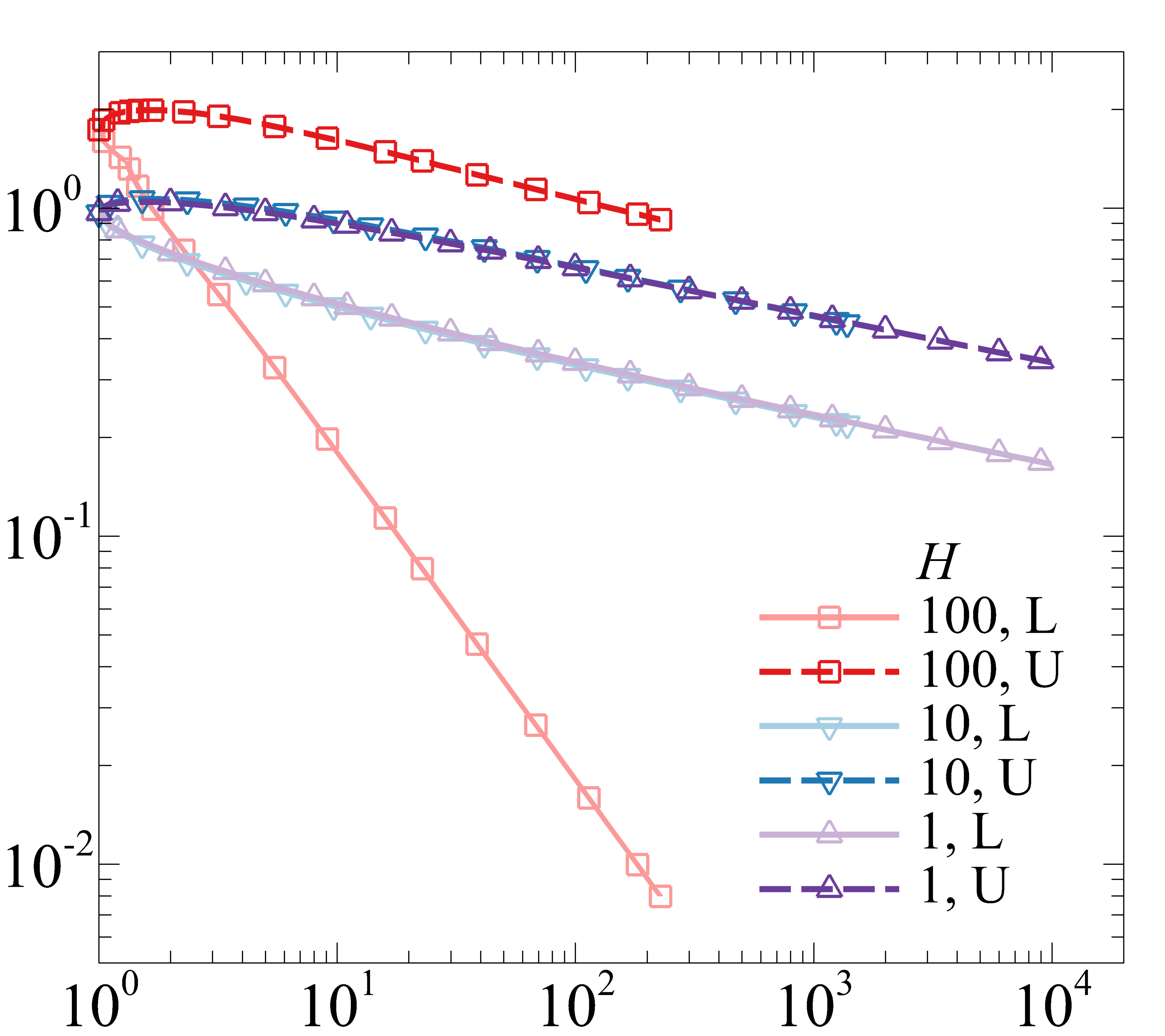}} &
\makecell{\vspace{21mm} \footnotesize{(b)} \\  \vspace{29mm} \rotatebox{90}{\footnotesize{$|\Re(\mu_2)|$}}} & \makecell{\includegraphics[width=0.458\textwidth]{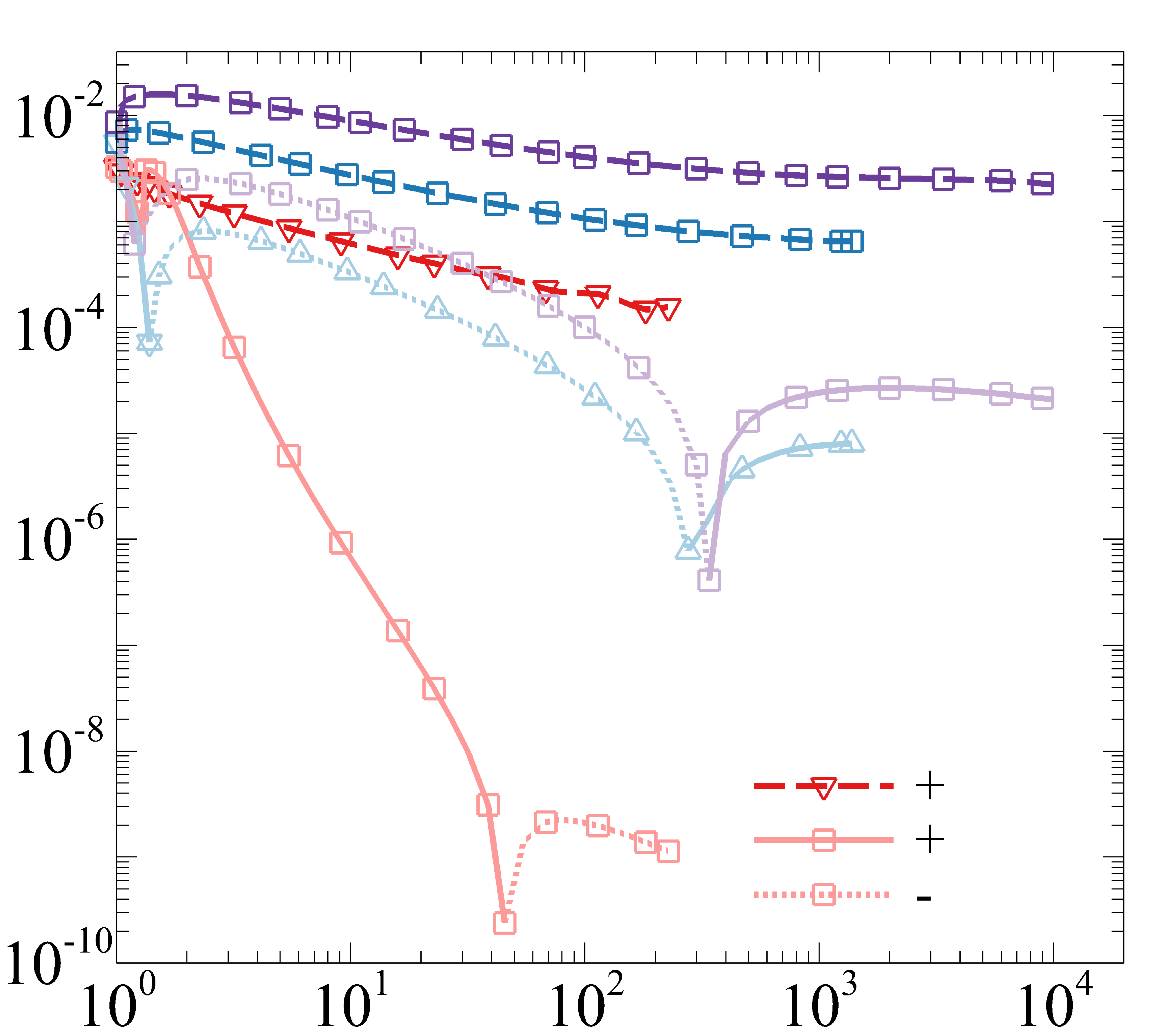}} \\
 & \hspace{32mm} \footnotesize{$\Rey/\ReyCrit$} & & \hspace{32mm} \footnotesize{$\Rey/\ReyCrit$} \\
\makecell{\vspace{26mm} \footnotesize{(c)} \\  \vspace{31mm} \rotatebox{90}{\footnotesize{$\alpha$}}} & \makecell{\includegraphics[width=0.458\textwidth]{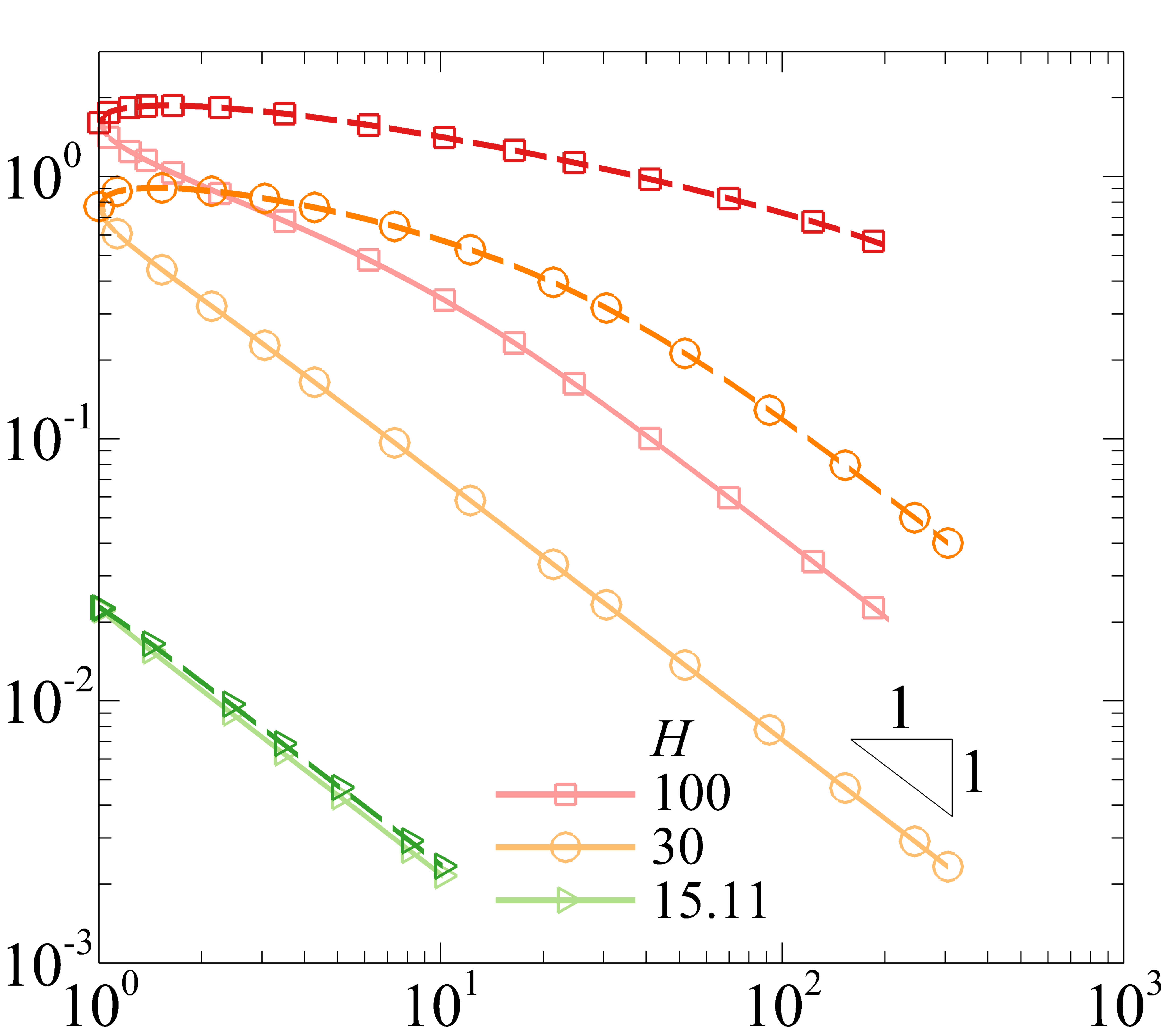}} &
\makecell{\vspace{21mm} \footnotesize{(d)} \\  \vspace{29mm} \rotatebox{90}{\footnotesize{$|\Re(\mu_2)|$}}} & \makecell{\includegraphics[width=0.458\textwidth]{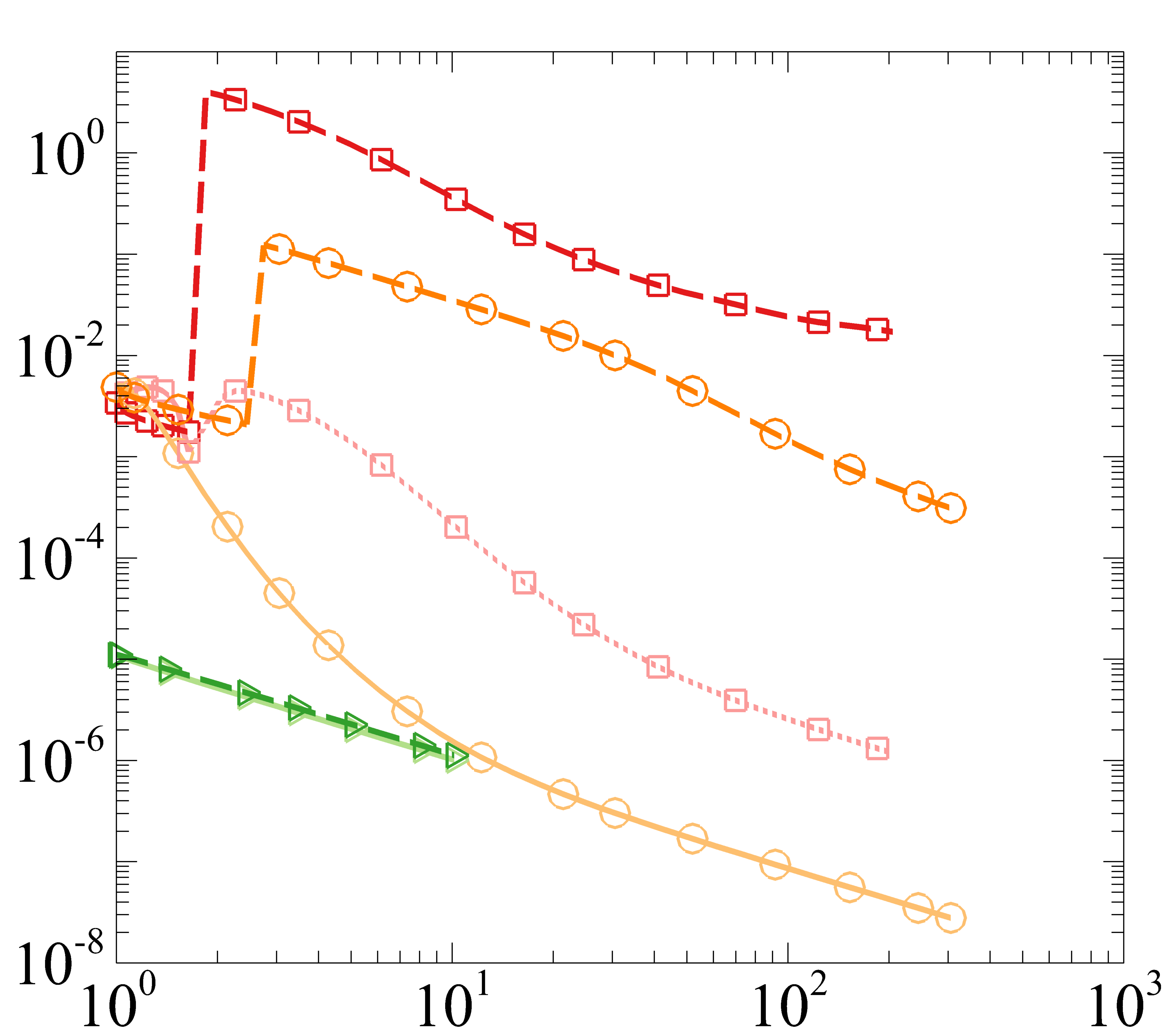}} \\
 & \hspace{32mm} \footnotesize{$\Rey/\ReyCrit$} & & \hspace{32mm} \footnotesize{$\Rey/\ReyCrit$} \\
\makecell{\vspace{26mm} \footnotesize{(e)} \\  \vspace{31mm} \rotatebox{90}{\footnotesize{$\alpha$}}} & \makecell{\includegraphics[width=0.458\textwidth]{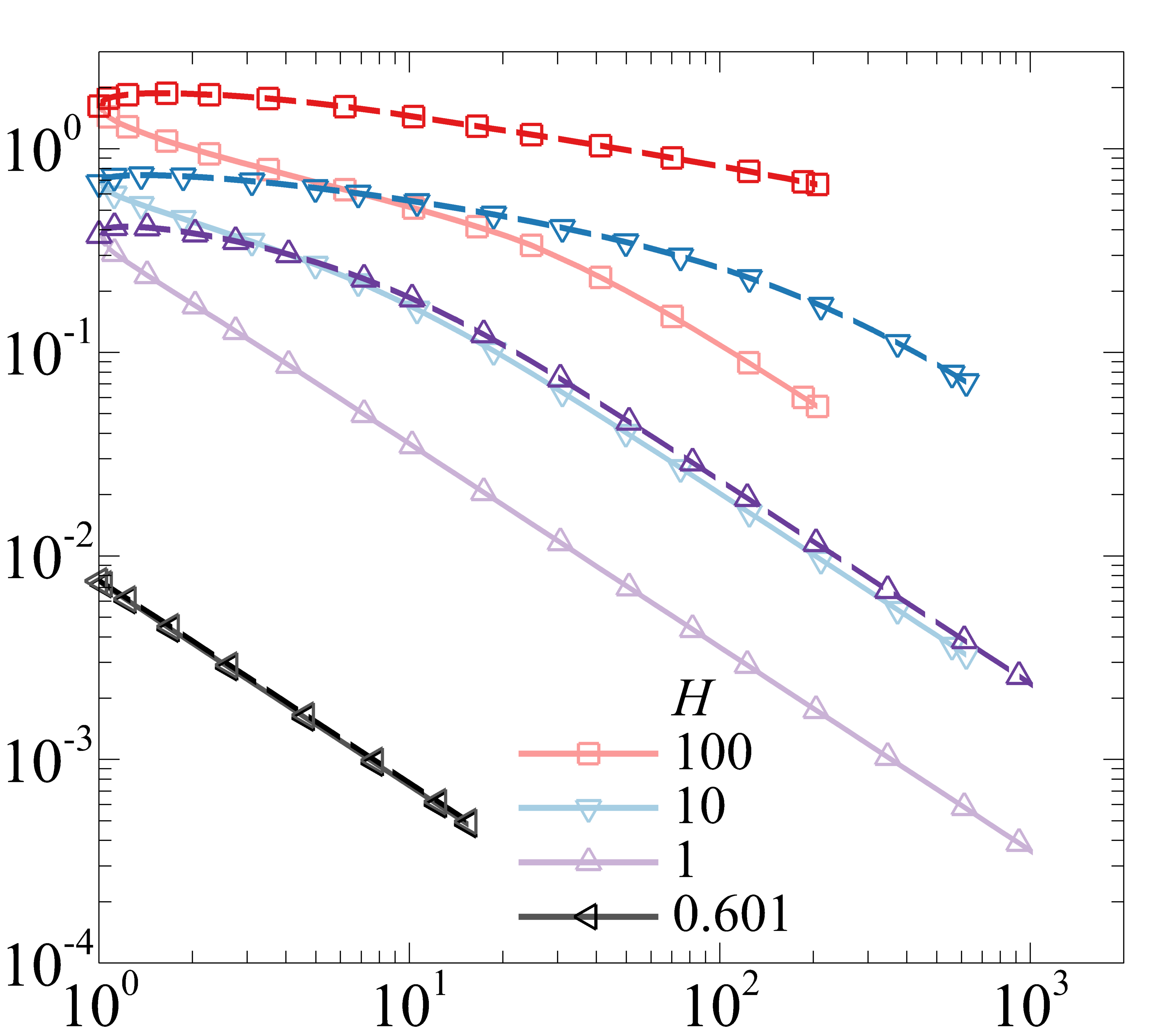}} &
\makecell{\vspace{21mm} \footnotesize{(f)} \\  \vspace{29mm} \rotatebox{90}{\footnotesize{$|\Re(\mu_2)|$}}} & \makecell{\includegraphics[width=0.458\textwidth]{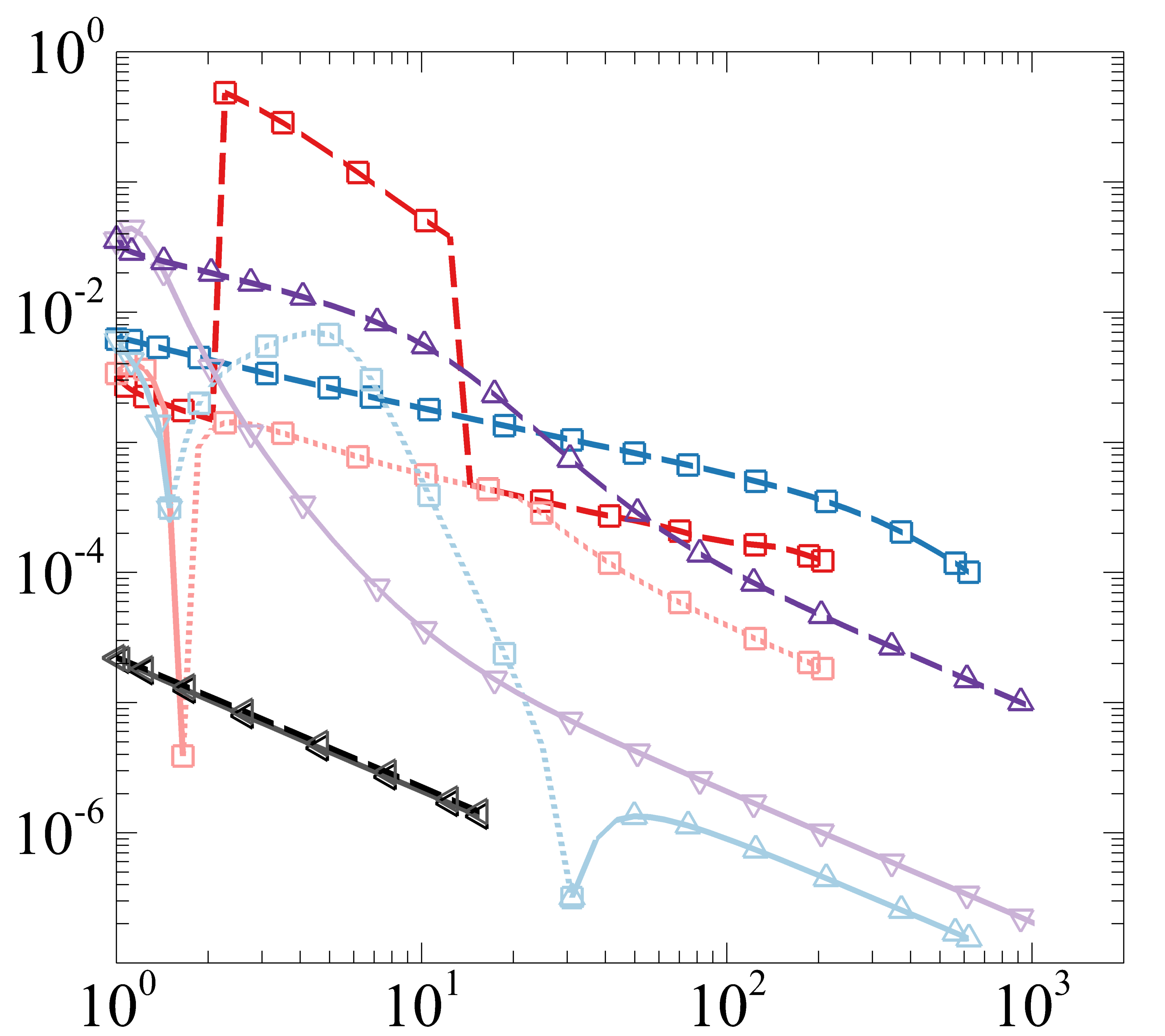}} \\
 & \hspace{32mm} \footnotesize{$\Rey/\ReyCrit$} & & \hspace{32mm} \footnotesize{$\Rey/\ReyCrit$} \\
\end{tabular}
\addtolength{\tabcolsep}{+2pt}
\addtolength{\extrarowheight}{+10pt}
\end{center}
    \caption{Weakly nonlinear stability along the neutral curve. (a--b) $\UsubR=1$. (c--d) $\UsubR=-1+10^{-10}$. (e--f) $\UsubR=0.85$. Neutrally stable wave numbers (left column) and real part of the first Landau coefficient (right column). Dotted lines denote where $\Re(\mu_2)$ was negative, indicating supercritical bifurcation. `L' and `U' denote the lower and upper branches of the neutral curve, respectively.}
    \label{fig:wnl_all3}
\end{figure}
\Fig\ \ref{fig:wnl_all} depicts the weakly nonlinear behavior solely at the critical points (\fig\ \ref{fig:lin_and_eng}). Locally, the transition is subcritical ($\mu_2>0$) and the finite amplitude state can be reached ($\mu_1>0$) at all critical points, including along the $\ReyCrit \rightarrow \infty$ asymptotes. However, the magnitude of $\mu_2$ directly quantifies the level of subcriticality of the transition. The variations of $\Rez(\mu_2(H))$ are opposite to those of $\ReyCrit(H)$. As such, for the larger values of $H$, $\Rez(\mu_2)$ scales with the Shercliff layer thickness and decreases as $\Rez(\mu_2)\sim H^{-1/2}$. Near asymptotes where $\ReyCrit$ diverges, on the other hand, $\Rez(\mu_2)$ increases sharply with $H$ from  $-\infty$. This is expected, since in this limit, any growth of finite amplitude takes place at a 
vanishingly small critical parameter $\Rey/\ReyCrit-1$.
The byproduct of this is that the saturation amplitude at which the perturbation is `large' enough for non-linear effects to be important increases with $H$, as $|A|^2 \sim H^{5/2}$ for large $H$, see \fig\ \ref{fig:wnl_all}(c). However, to compare between $H$, a constant $\Rey/\ReyCrit$ scaling of $|A|^2 \sim H^2$ is more appropriate, as $\ReyCrit \sim H^{1/2}$ for large $H$.
Since, at constant $\Rey$, linear transient growth decreases at least as $H^{-1/2}$ \citep{Cassels2019from3D}, it is unlikely to provide a mechanism to support the growth of perturbations at large $H$. Thus, although subcritical bifurcations exist, they are unlikely to be obtained, given the lack of transient growth at subcritical $\Rey$.


$\Rez(\mu_2)$ is depicted along neutral curves from the linear stability analysis 
in \fig\ \ref{fig:wnl_all3}, for Shercliff, MHD-Couette and mixed ($\UsubR=0.85$) base flows. For Shercliff flow, at $H=1$ and $10$, $\Rez(\mu_2)$ changes sign twice along the lower branch, so the bifurcation associated with modes on this branch is supercritical between these two points, and subcritical elsewhere.
At $H=100$, the bifurcation becomes supercritical at a much higher Reynolds number (likely because the \TS\ mode does not define the edge of the neutral curve there), and remains supercritical to the computed extent of the lower branch (to $\Rey=10^7$). Comparatively, for MHD-Couette flow, at $H=15.11$ and $30$, there is no supercritical region. At $H=100$, a supercritical bifurcation appears along the lower branch, as at higher $H$, there is less sensitivity to the exact base flow profile. The mixed  flow displays a clearer transition from subcritical to supercritical bifurcation with increasing $H$. At $H$ near $H^\infty$ ($0.601$ and $1$), the bifurcation is everywhere subcritical. At $H=10$, a small region of supercritical bifurcation exists along the lower branch. By $H=100$, this region of supercritical bifurcation is much larger, and does not switch back to subcritical to the 
computed extent of the neutral curve ($\Rey=10^7$). However, the top branch always 
remains open to subcritical bifurcation.

At large Reynolds numbers, the scalings $\alpha\sim\Rey^{-1}$, $\Rez(\mu_1)\sim\Rey^{-1}$ and $\Rez(\mu_2)\sim\Rey^{-1}$ hold, and the phase speed asymptotes to a constant. Furthermore, as $\Rez(\mu_1)(\ReyMarg-\Rey)$ always remains positive, the finite amplitude state can always be reached ($\ReyMarg$ is the Reynolds number `on' the neutral curve). 

\section{Direct Numerical Simulations}\label{sec:dns}
\subsection{Formulation}\label{sec:dns_form}

Finally, we shall now assess whether transition to quasi-two-dimensional turbulence may actually take place under the full nonlinear dynamics, by performing direct numerical simulations (DNS) of equations (\ref{eq:SM82_continuity}) and (\ref{eq:SM82_momentum}). `Natural' conditions are reproduced with white noise added in varying fraction $E_{0}(t=0)=\int_{-1}^1\hat{u}^2+\hat{v}^2\,\dUP\Omega/\int_{-1}^1U^2\,\dUP\Omega$, where $\Omega$ represents the computational domain. Periodic boundary conditions, $\vect{u}(x=0)=\vect{u}(x=W)$ and $p(x=0)=p(x=W)$, are applied at the downstream and upstream boundaries of a domain with length $W=2\pi/\alpha_{\rm max}$ set to match the wave number that achieved maximal linear growth. 
The simulations typically exhibited a rapid drop in disturbance energy, followed by a linear phase of exponential growth, which is finally superseded by nonlinear effects. The exponential growth rate $\sigma_{\rm max}$ from the linear growth regime is obtained by fitting the natural logarithm of $\int |v|\,\dUP\Omega$ data over a few thousand time units. As $\timeD = t/(1-U_{\rm min})$, the rescaled growth rate $\sigma_{\Delta \rm max}=\sigma_{\rm max}/(1-U_{\rm min})$.


The random noise seeds (perturbations) are evolved with an in-house spectral element solver, which employs a third order backward differencing scheme, with operator splitting, for time integration \citep{Karniadakis2005spectral}.  
High-order Neumann pressure boundary conditions are imposed on impermeable walls to maintain third order time accuracy \citep{Karniadakis2005spectral}. 
The Cartesian domain is discretized with quadrilateral elements over which Gauss--Legendre--Lobatto nodes are placed ($\Np=19$ nodes per element to take advantage of spectral convergence). 
Elements are uniformly distributed in both streamwise and transverse directions, with greater element compression in the wall-normal direction. At the highest $H$ value simulated, at least 20 nodes reside within the Shercliff boundary layer. The solver, incorporating the SM82 friction term, has been previously introduced and validated \cite{Cassels2016heat, Cassels2019from3D, Hussam2012optimal, Sheard2009cylinders}.
%
\begin{table}
\begin{center}
\begin{tabular}{ cc|cc|cc } 
\hline
$E_x$ & $E_y$ &  $\sigma_{\Delta \rm max}$, $\UsubR=1$  &  $|$\% error$|$ & $\sigma_{\Delta \rm max}$, $\UsubR=-1$  & $|$\% error$|$ \\
\hline
$3$          & $12$          & $5.13119051\times10^{-3}$ & \hspace{1mm}  $2.236\times10^{-1}$ \hspace{1mm} & $3.92816302\times10^{-3}$ & $1.367\times10^{-1}$ \\
$6$          & $12$          & $5.13138001\times10^{-3}$ & $2.273\times10^{-1}$ & $3.92831007\times10^{-3}$ & $1.394\times10^{-1}$ \\
$\mathbf{3}$ & $\mathbf{24}$ & $5.11994480\times10^{-3}$ & $3.959\times10^{-3}$ & $3.93378788\times10^{-3}$ & $6.248\times10^{-3}$ \\
$6$          & $24$          & $5.11998768\times10^{-3}$ & $4.797\times10^{-3}$ & $3.93158203\times10^{-3}$ & \hspace{1mm}  $4.983\times10^{-2}$  \hspace{1mm}  \\
LSA & LSA  & \hspace{1mm} $5.11974211\times10^{-3}$  \hspace{1mm} & - & \hspace{1mm} $3.93354213\times10^{-3}$ \hspace{1mm} & - \\
\hline
\end{tabular}
\caption{DNS mesh resolution testing at $H=100$ and $\Rey=10^6$, for Shercliff ($\ReyCrit=4.40223 \times 10^5$; $\alphaMax=1.52813$ at $\Rey=10^6$) and MHD-Couette ($\ReyCrit=4.87187\times 10^5$; $\alphaMax=1.39883$ at $\Rey=10^6$) flow. The interpolant for each macro-element is order $\Np = 19$ and the timestep is $\Delta t = 1.25\times 10^{-3}$, as validated in Ref.~\cite{Camobreco2020role}. The final resolution choices are in bold, where $E_x$ and $E_y$ represent the number of elements per unit length in $x$ and $y$, respectively. The chosen resolution has roughly 650 elements and $6.25\times10^{5}$ degrees of freedom for these wave numbers.}
\label{tab:val_DNS}
\end{center}
\end{table}

Mesh validation results are provided in \tbl\ \ref{tab:val_DNS}, comparing growth rates measured in the DNS against linear stability analysis (LSA) predictions. The agreement is excellent in the exponential growth (or decay) stages. Some additional comparisons at the chosen resolution (3 elements per unit length in $x$, 24 elements per unit height in $y$) can be found in \tbl\ \ref{tab:subc_vary}. 

Fourier analysis is also performed at select instants in time, exploiting the streamwise periodicity of the domain. The absolute values of the Fourier coefficients $c_\kappa = \lvert (1/\Nf)\sum_{n=0}^{n=\Nf-1}[\hat{u}^2(x_n)+\hat{v}^2(x_n)] e^{-2\pi i \kappa n/\Nf}\rvert$ were obtained using the discrete Fourier transform in MATLAB, where $x_n$ represents the $n$'th $x$-location linearly spaced between $x_0=0$ and $x_{\Nf}=W$, interpolating in the discretized domain when necessary, and taking $\Nf=10000$. A mean Fourier coefficient $\meanfoco$ is obtained by averaging the coefficients obtained at 21 $y-$values. A time averaged mean Fourier coefficient $\meanfocoti$ is also determined by averaging over approximately 20 time instants, for stages after the initial linear and nonlinear growth. Note that although the number of sample points $\Nf$ is high only $\kappa$ up to about 100 to 200 are well resolved by the spatial discretization, depending on $H$. 
%
\subsection{Subcritical regime}\label{sec:dns_results}

The results for DNS at subcritical Reynolds numbers are collated in \tbl\ \ref{tab:subc_vary}. All initial seeds decayed exponentially, with excellent agreement to the LSA decay rates, and without any observable linear transient growth. It appears that supercritical Reynolds numbers are required to induce nonlinear behavior and transitions to turbulence, if the initial field is random noise, more in line with a supercritical bifurcation. Thus, subcritical transitions may only be attainable for a small range of Reynolds numbers near $\ReyCrit$.  Subcritical tests of MHD-Couette flow for $H<H^{\infty}$ (for which $\alphaMax \rightarrow 0$) were also simulated, with $W=20\pi$ arbitrarily chosen; only monotonic decay was observed. Note that in all these cases, the transient growth optimals have wavelengths shorter than the domain length required to maximize linear growth, and are hence not excluded.

\begin{table}
\begin{center}
\begin{tabular}{ ccc|ccc } 
\hline
$\Rey/\ReyCrit$ & $\sigma_{\Delta \rm max} \times 10^{4}$ &  $|$\% error$|$ &
 $E_0$          & $\sigma_{\Delta \rm max} \times 10^{4}$ &  $|$\% error$|$ \\
 & \hspace{1mm} LSA \hspace{14mm} DNS  &  &
 & \hspace{1mm} LSA \hspace{14mm} DNS  &  \\
 \hline
$0.5$ & $-76.4704$       \hspace{5mm} $-76.4726$  & \hspace{1mm} $2.9\times10^{-3}$ \hspace{1mm}  &
 $10^{2}$ & $-8.97961$  \hspace{5mm} $-8.98230$ & \hspace{1mm} $3.0\times10^{-2}$ \hspace{1mm} \\
$0.6$ & $-52.0623$       \hspace{5mm} $-52.0586$  & $7.1\times10^{-3}$ &
 $10^{-0}$ & $-8.97961$ \hspace{5mm} $-8.98277$ & $3.5\times10^{-2}$ \\
$0.7$ & $-33.9759$       \hspace{5mm} $-33.9774$  & $4.4\times10^{-3}$ &
 $10^{-2}$ & $-8.97961$ \hspace{5mm} $-8.98232$ & $3.0\times10^{-2}$ \\
$0.8$ & $-20.0392$       \hspace{5mm} $-20.0420$  & $1.4\times10^{-2}$ &
 $10^{-4}$ & $-8.97961$ \hspace{5mm} $-8.98240$ & $3.1\times10^{-2}$ \\
$0.9$ &  $-8.97961$     \hspace{5mm} $-8.98232$ & $3.0\times10^{-2}$ &
 $10^{-6}$ & $-8.97961$ \hspace{5mm} $-8.94802$ & $3.5\times10^{-1}$  \\
 \hline
 \end{tabular}
\caption{Subcritical test cases for Shercliff flow at $H=1$, with different levels of criticality at $E_0=10^{-2}$ (left) and with different values of $E_0$ at $\Rey/\ReyCrit=0.9$ (right).}
\label{tab:subc_vary}
\end{center}
\end{table}


\subsection{Supercritical regime}\label{sec:dns_results2}

\begin{figure}
\begin{center}
\addtolength{\extrarowheight}{-10pt}
\addtolength{\tabcolsep}{-2pt}
\begin{tabular}{ llll }
\makecell{\vspace{19mm} \footnotesize{(a)} \\  \vspace{26mm} \rotatebox{90}{\footnotesize{$\int \hat{u}^2+\hat{v}^2 \mathrm{d}\Omega$}}} & \makecell{\includegraphics[width=0.458\textwidth]{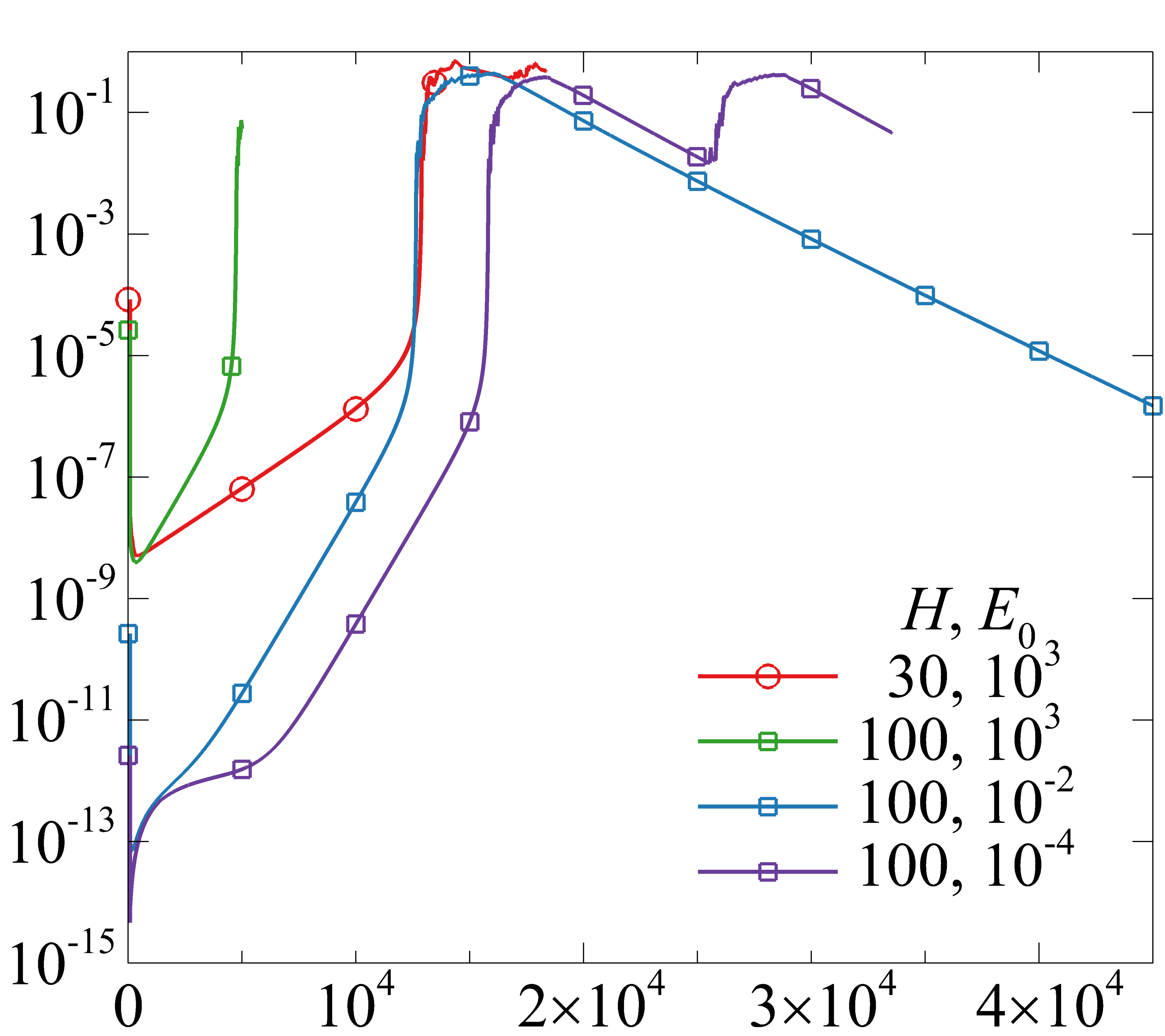}} &
\makecell{\vspace{23mm} \footnotesize{(b)} \\  \vspace{31mm} \rotatebox{90}{\footnotesize{$\int \hat{v}^2 \mathrm{d}\Omega$}}} & \makecell{\includegraphics[width=0.458\textwidth]{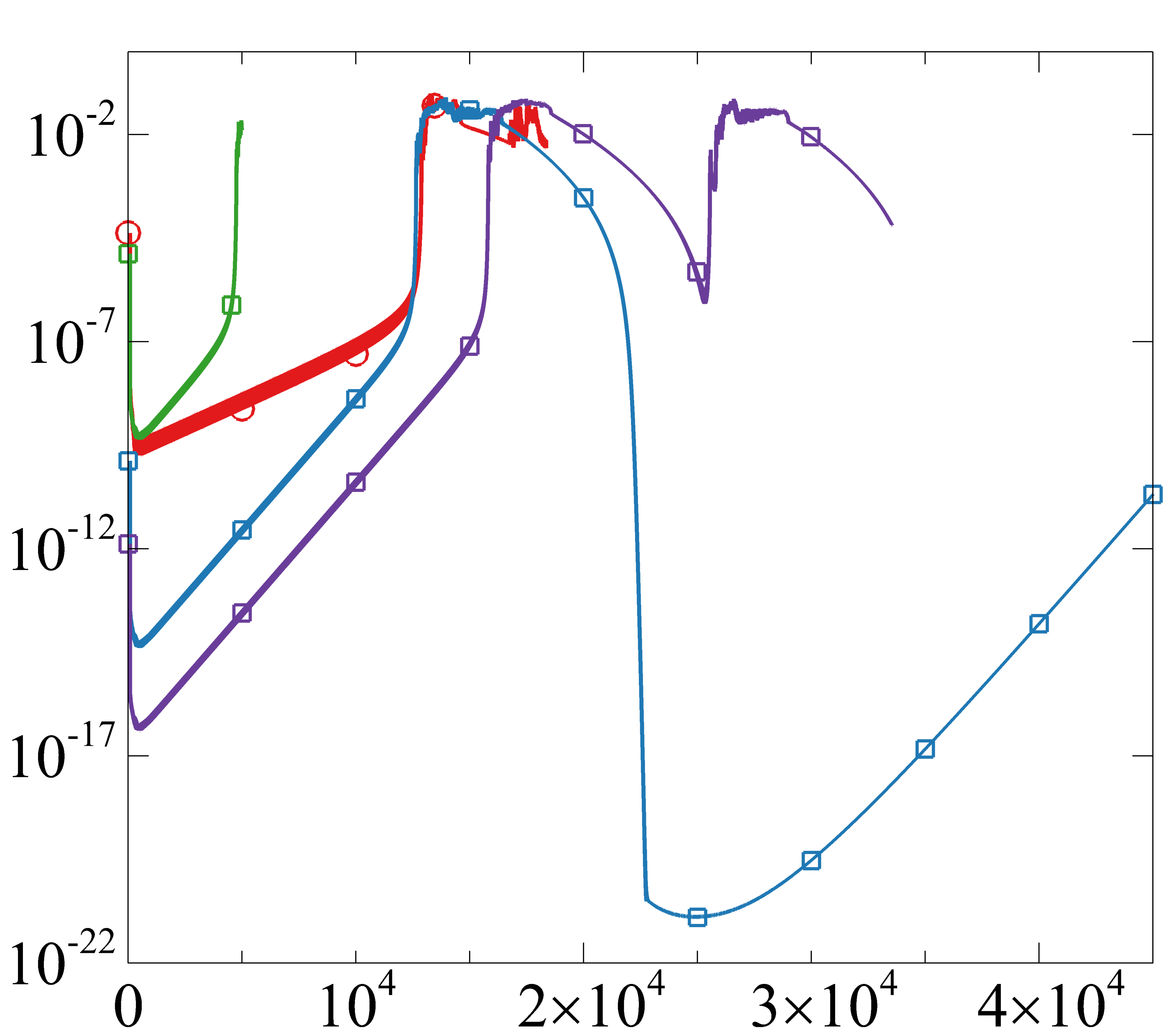}} \\
 & \hspace{40mm} \footnotesize{$t$} & & \hspace{40mm} \footnotesize{$t$} \\
\end{tabular}
\addtolength{\tabcolsep}{+2pt}
\addtolength{\extrarowheight}{+10pt}
\end{center}
    \caption{Temporal evolution of perturbations initiated with random noise for MHD-Couette flow ($\UsubR=-1$) at $\Rey/\ReyCrit=1.1$, and various $H>H^\infty$ and $E_0$. (a) $\int \hat{u}^2+\hat{v}^2\,\dUP\Omega$. (b) $\int \hat{v}^2\,\dUP\Omega$.}
    \label{fig:Couette_DNS}
\end{figure}
\begin{figure}
\begin{center}
\addtolength{\extrarowheight}{-10pt}
\addtolength{\tabcolsep}{-2pt}
\begin{tabular}{ llll }
\makecell{\vspace{19mm} \footnotesize{(a)} \\  \vspace{26mm} \rotatebox{90}{\footnotesize{$\int \hat{u}^2+\hat{v}^2 \mathrm{d}\Omega$}}} & \makecell{\includegraphics[width=0.458\textwidth]{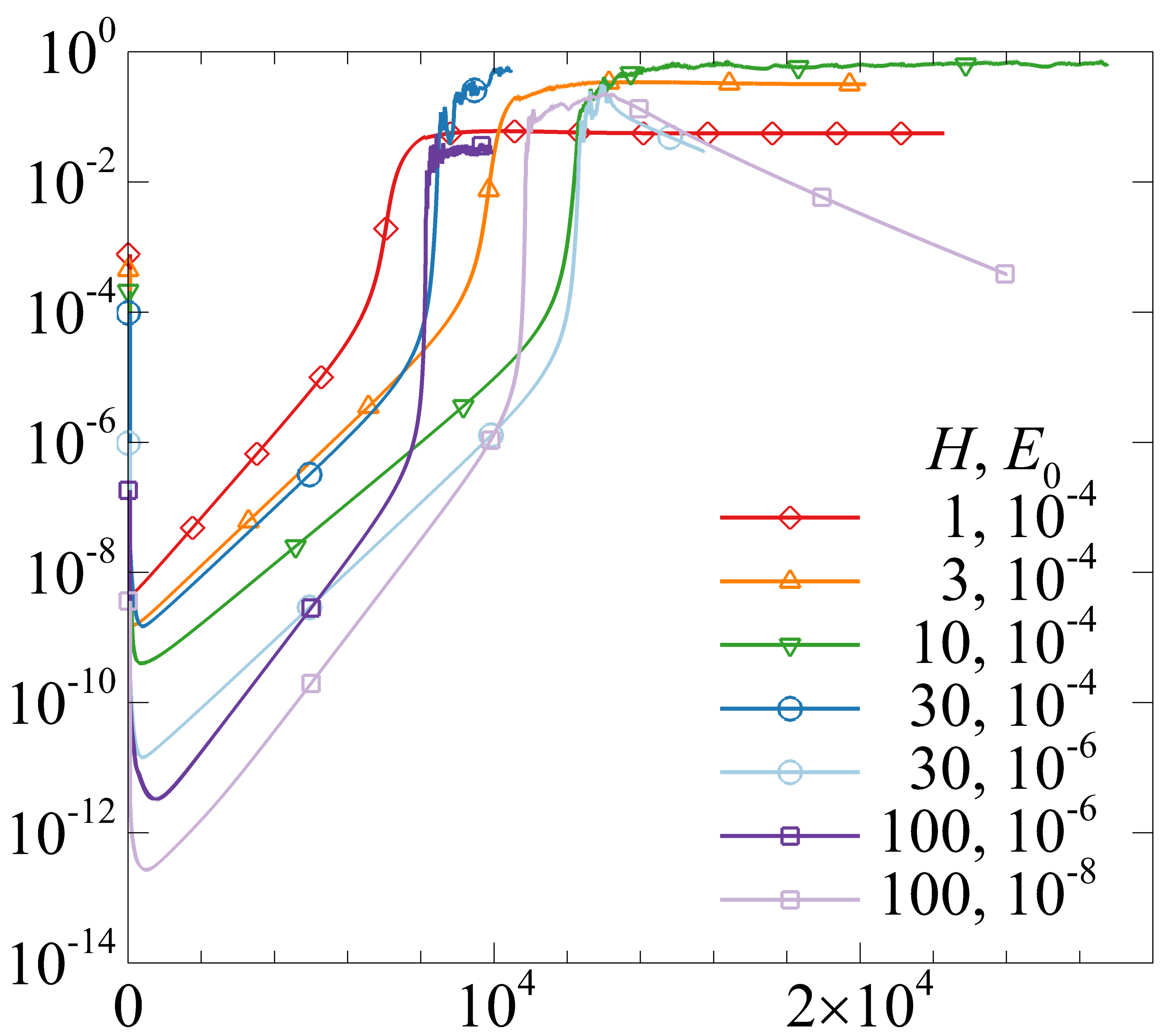}} &
\makecell{\vspace{23mm} \footnotesize{(b)} \\  \vspace{31mm} \rotatebox{90}{\footnotesize{$\int \hat{v}^2 \mathrm{d}\Omega$}}}  & \makecell{\includegraphics[width=0.458\textwidth]{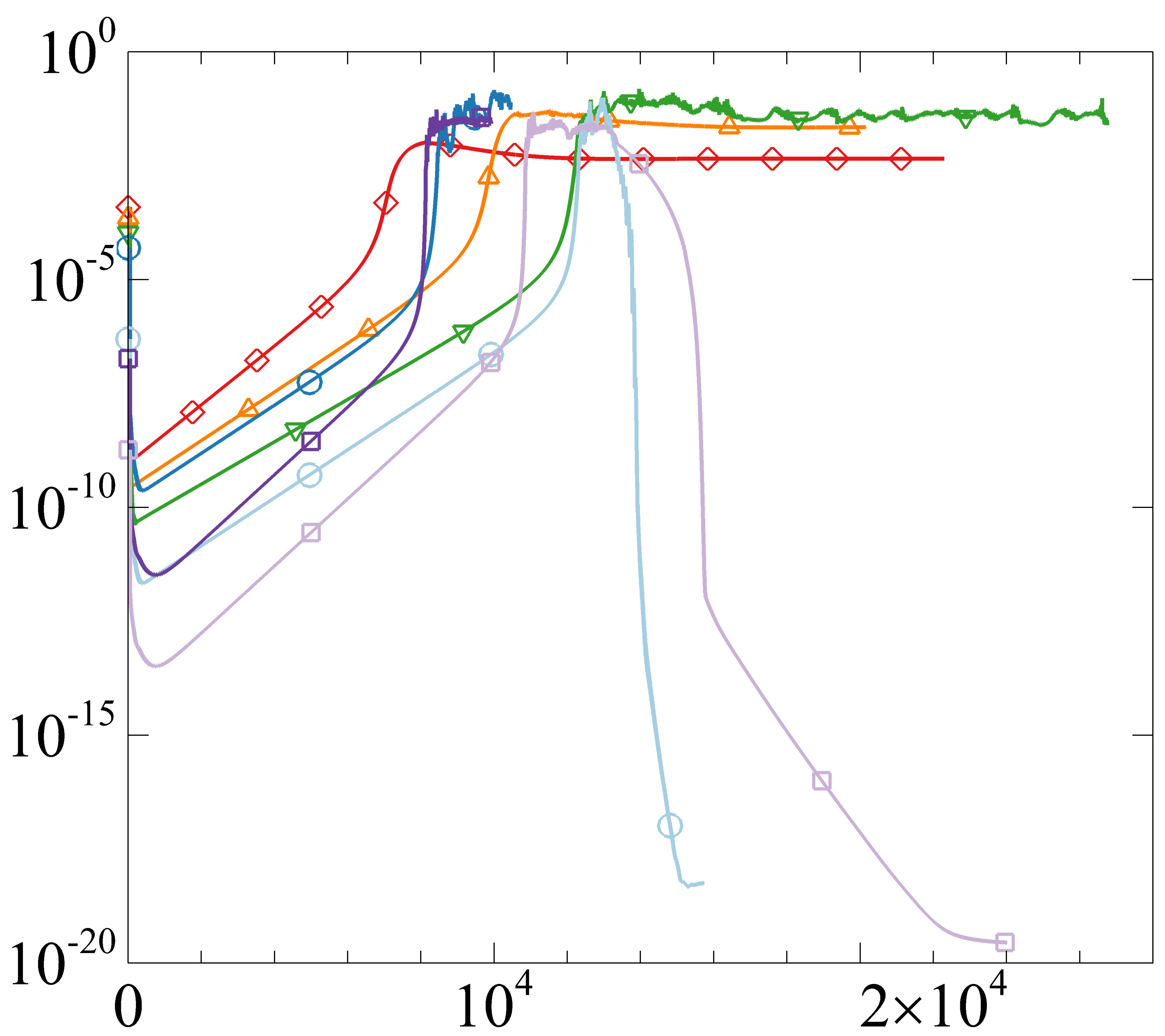}} \\
 & \hspace{40mm} \footnotesize{$\timeD$} & & \hspace{40mm} \footnotesize{$\timeD$} \\
\end{tabular}
\addtolength{\tabcolsep}{+2pt}
\addtolength{\extrarowheight}{+10pt}
\end{center}
    \caption{Temporal evolution (rescaled time) of perturbations initiated with random noise for Shercliff flow ($\UsubR=1$) at $\Rey/\ReyCrit=1.1$, and various $H$ and $E_0$. (a) $\int \hat{u}^2+\hat{v}^2\,\dUP\Omega$. (b) $\int \hat{v}^2\,\dUP\Omega$.}
    \label{fig:Poiseuille_DNS}
\end{figure}

The energy growth for supercritical MHD-Couette flow (for $H>15.102$) is shown in \fig\ \ref{fig:Couette_DNS} and for supercritical Shercliff flow in \fig\ \ref{fig:Poiseuille_DNS}. These are separated into growth in $\int \hat{u}^2+\hat{v}^2 \mathrm{d}\Omega$, to represent the total perturbation kinetic energy and to highlight the formation of streamwise independent structures, and growth in $\int \hat{v}^2 \mathrm{d}\Omega$, which better isolates the growth or decay of the perturbation. The linear growth, nonlinear growth, and initial turbulent stages are very similar between the MHD-Couette and Shercliff flows. However, the relaminarization and decay stages are quite different. For MHD-Couette flow (\fig\ \ref{fig:Couette_DNS}), $\int \hat{v}^2 \mathrm{d}\Omega$ displays clear re-excitations. The $H=30$ and $H=100$, $E_0=10^{-4}$ cases relaminarize, but are both quickly re-excited (very rapidly in the $H=30$ case) while at high amplitudes, when nonlinearity still plays a role. The $H=100$, $E_0=10^{-2}$ case cleanly decays to the floor, after which growth begins again, via the linear mechanism. This was not observed for Shercliff flow, with both (smaller $E_0$) $H=30$ and $H=100$ cases relaminarizing and rapidly monotonically decaying (the larger $E_0$ cases require exceedingly small time steps and as such their final behaviors remain unknown). The smaller $E_0$ case at $H=30$ also relaminarizes and decays more rapidly than at $H=100$, in spite of less Hartmann friction. Note that the energy in the Shercliff and MHD-Couette base flows at the same $H$ differ, so it is not necessarily appropriate to compare the same $E_0$ between different base flows.


Computations of Shercliff flow reveal two interesting changes in behavior with decreasing $H$. Unlike the relaminarization and monotonic decay for $H \geq 30$, the $H=3$ and $H=10$ cases (with $E_0=10^{-4}$) maintain turbulent states. At $H=3$ relaminarization again occurs, but the perturbation saturates to a stable finite amplitude state, rather than decaying. However, $H=10$ maintains the turbulent state for the computed extent of the simulation, excepting two brief attempts at relaminarization, which are not stable, resulting in a return to turbulence. With further decreasing $H \leq 1$, no turbulent state is triggered by the linear and nonlinear growth, with only an eventual saturation to a stable finite amplitude state. This behavior echoes that discussed in Ref.~\cite{Krasnov2008optimal}, which observe that for all $\Har \geq 0$ a purely two-dimensional finite-amplitude state can be reach via evolution of an Orr mode formed of purely spanwise vortices (recall Sec.~\ref{sec:tgp_results} indicating that the transient Orr optimal was almost identical to the linear optimal). However, the addition of three-dimensional noise to the finite amplitude state triggers (3D) turbulence at low $\Har$, but destabilizes the finite amplitude state at high $\Har$ such that the solution decays back to the laminar base state, with only short lived turbulence. It is presumed by Ref.~\cite{Krasnov2008optimal} that this is due to nonlinear interactions feeding energy from 2D modes to 3D modes, which are then more rapidly dissipated at high $\Har$. Since this could not occur in these purely Q2D simulations, a different mechanism may be at play. Reference \cite{Kuhnen2018destabilizing} and Ref.~\cite{Budanur2020upper} argue that in hydrodynamic pipe flows, the flattening of the mean profile reduces turbulence production in the bulk, such that turbulence cannot be sustained. In the present configuration production is almost solely due to $\hat{u}\hat{v}\partial U/\partial y$. The vanishing of this term in the core flow for $H\gtrsim30$ may therefore explain why turbulence collapses in this parameter regime. Turbulence can still be re-excited as $\partial U/\partial y$ remains large near the wall. A possible explanation of the lack of transition at lower $H$ then follows, as with reducing $H$, $\partial U/\partial y$ near the wall reduces, and so too production. 



\begin{figure}
\begin{center}
\addtolength{\extrarowheight}{-10pt}
\addtolength{\tabcolsep}{-2pt}
\begin{tabular}{ llll }
\footnotesize{(a)} & \hspace{6mm} \footnotesize{$H=1$} &  \footnotesize{(b)} & \hspace{6mm} \footnotesize{$H=3$}\\
\makecell{\vspace{26mm} \\  \vspace{31mm} \rotatebox{90}{\footnotesize{$\meanfoco$}}} & \makecell{\includegraphics[width=0.458\textwidth]{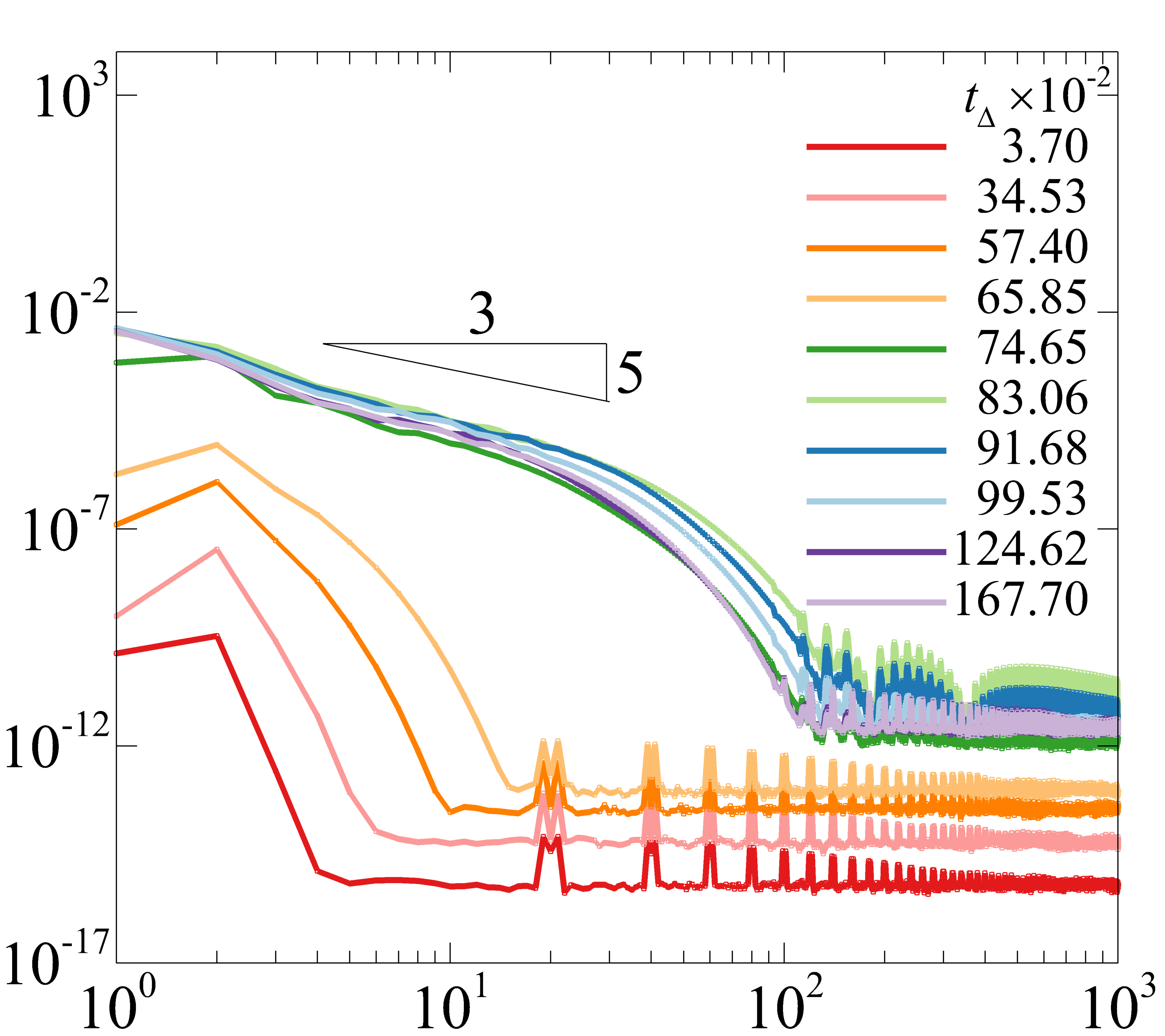}} &
\makecell{\vspace{26mm} \\  \vspace{31mm} \rotatebox{90}{\footnotesize{$\meanfoco$}}} & \makecell{\includegraphics[width=0.458\textwidth]{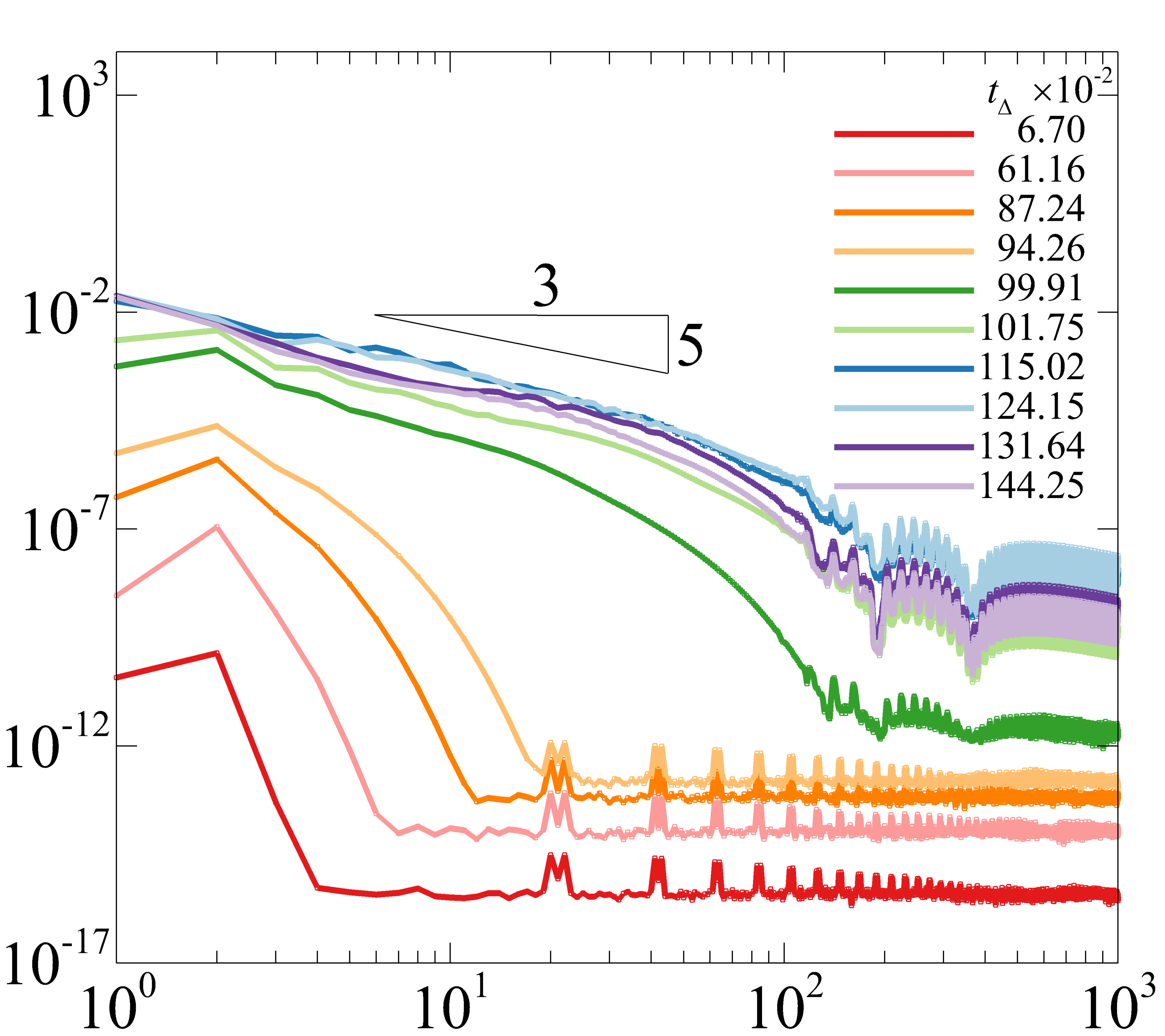}} \\
 & \hspace{38mm} \footnotesize{$\kappa$} & & \hspace{38mm} \footnotesize{$\kappa$} \\
\footnotesize{(c)} & \hspace{6mm} \footnotesize{$H=10$} &  \footnotesize{(d)} & \hspace{6mm} \footnotesize{$H=30$}\\
 \makecell{\vspace{26mm}  \\  \vspace{31mm} \rotatebox{90}{\footnotesize{$\meanfoco$}}} & \makecell{\includegraphics[width=0.458\textwidth]{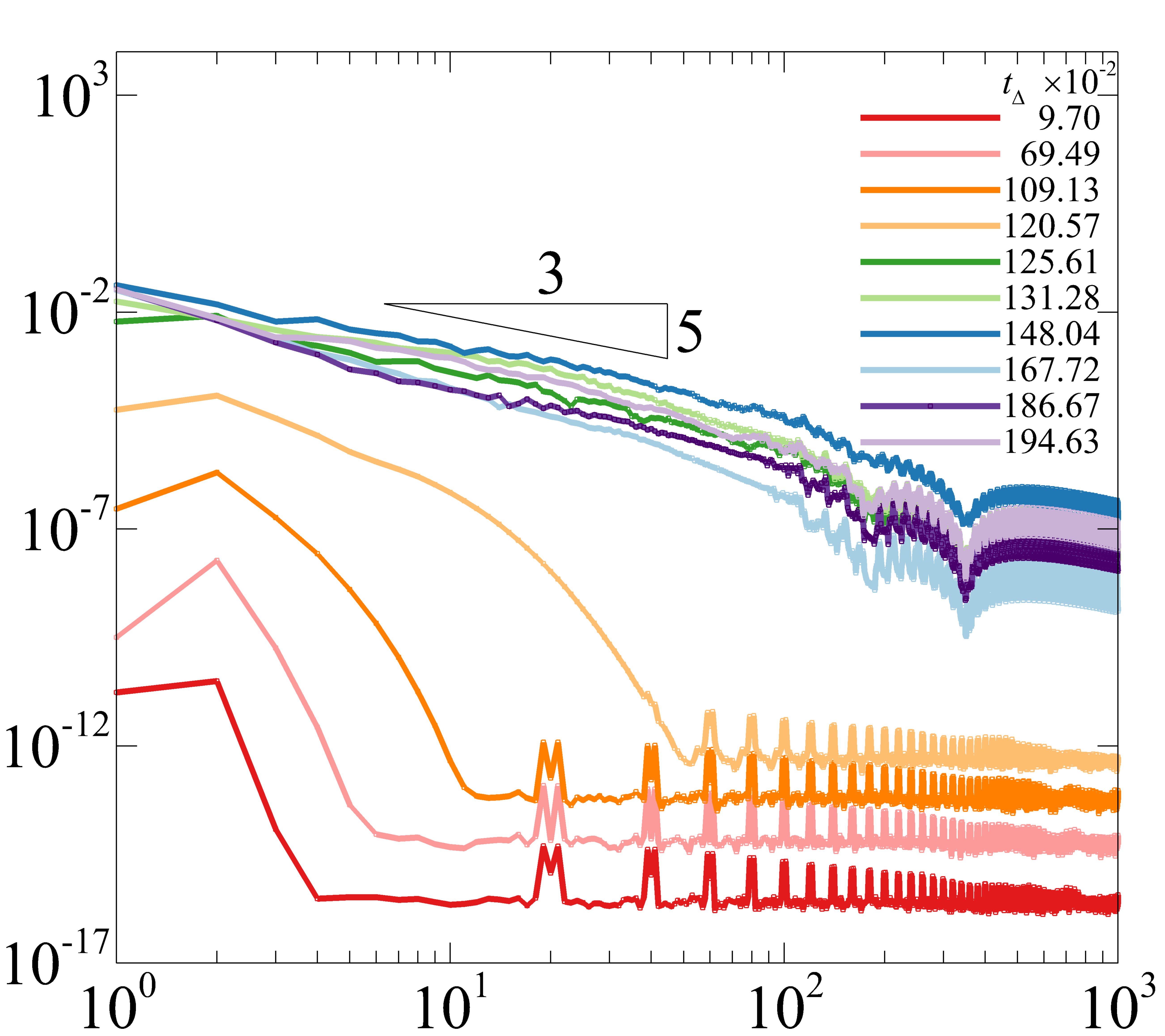}} &
\makecell{\vspace{26mm}  \\  \vspace{31mm} \rotatebox{90}{\footnotesize{$\meanfoco$}}} & \makecell{\includegraphics[width=0.458\textwidth]{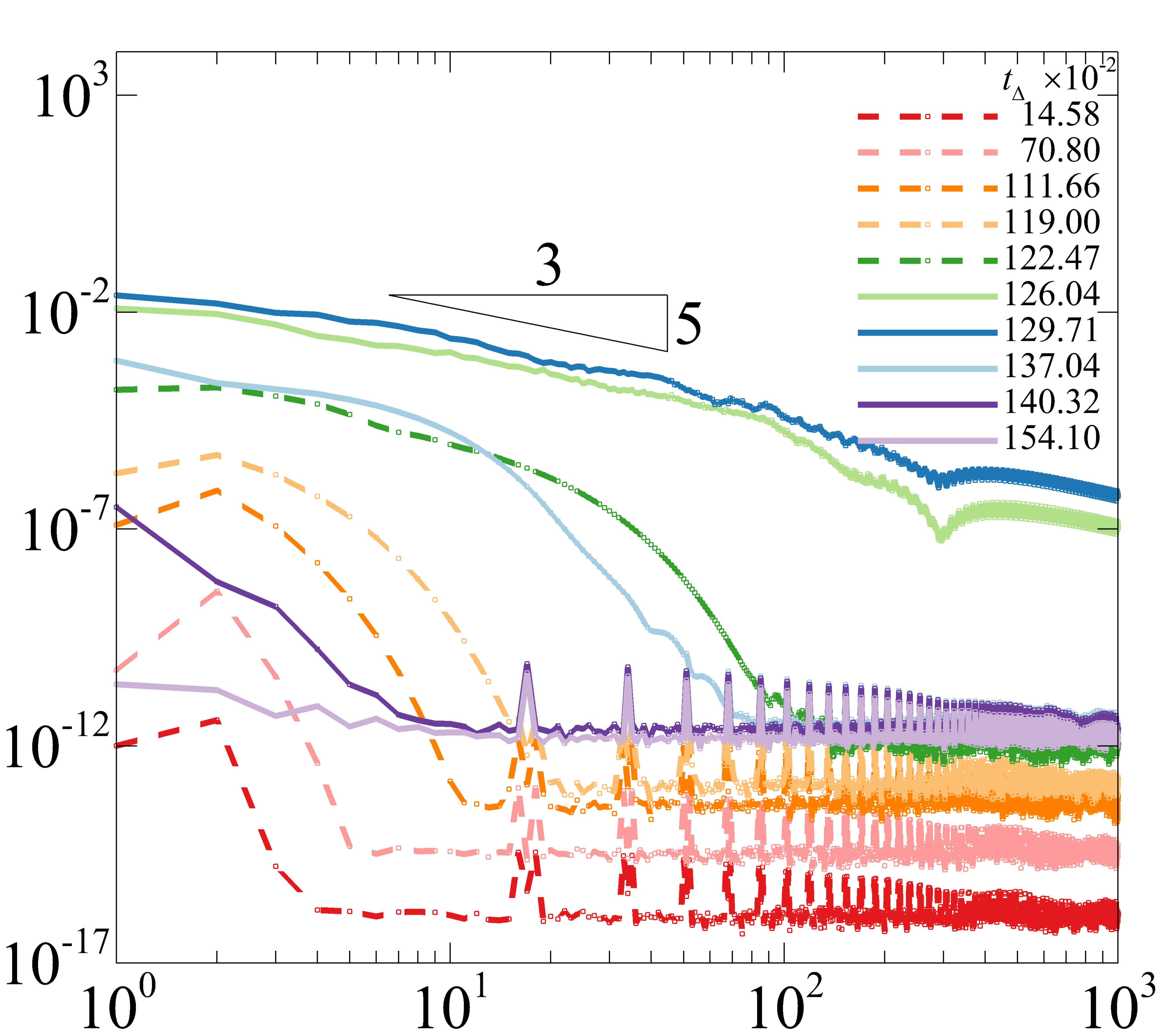}} \\
 & \hspace{38mm} \footnotesize{$\kappa$} & & \hspace{38mm} \footnotesize{$\kappa$} \\
\end{tabular}
\addtolength{\tabcolsep}{+2pt}
\addtolength{\extrarowheight}{+10pt}
\end{center}
    \caption{Instantaneous (rescaled time) values of the $y-$averaged Fourier coefficients for  $1<\kappa<10^3$, for Shercliff flow ($\UsubR=1$) at $\Rey/\ReyCrit=1.1$; $E_0=10^{-4}$ for $H\leq10$, $E_0=10^{-6}$ for $H=30$.}
    \label{fig:H4times10}
\end{figure}

\begin{figure}
\begin{center}
\addtolength{\extrarowheight}{-10pt}
\addtolength{\tabcolsep}{-2pt}
\begin{tabular}{ llll }
\makecell{\vspace{26mm} \footnotesize{(a)} \\  \vspace{30mm} \rotatebox{90}{\footnotesize{$\meanfocoti$}}} & \makecell{\includegraphics[width=0.458\textwidth]{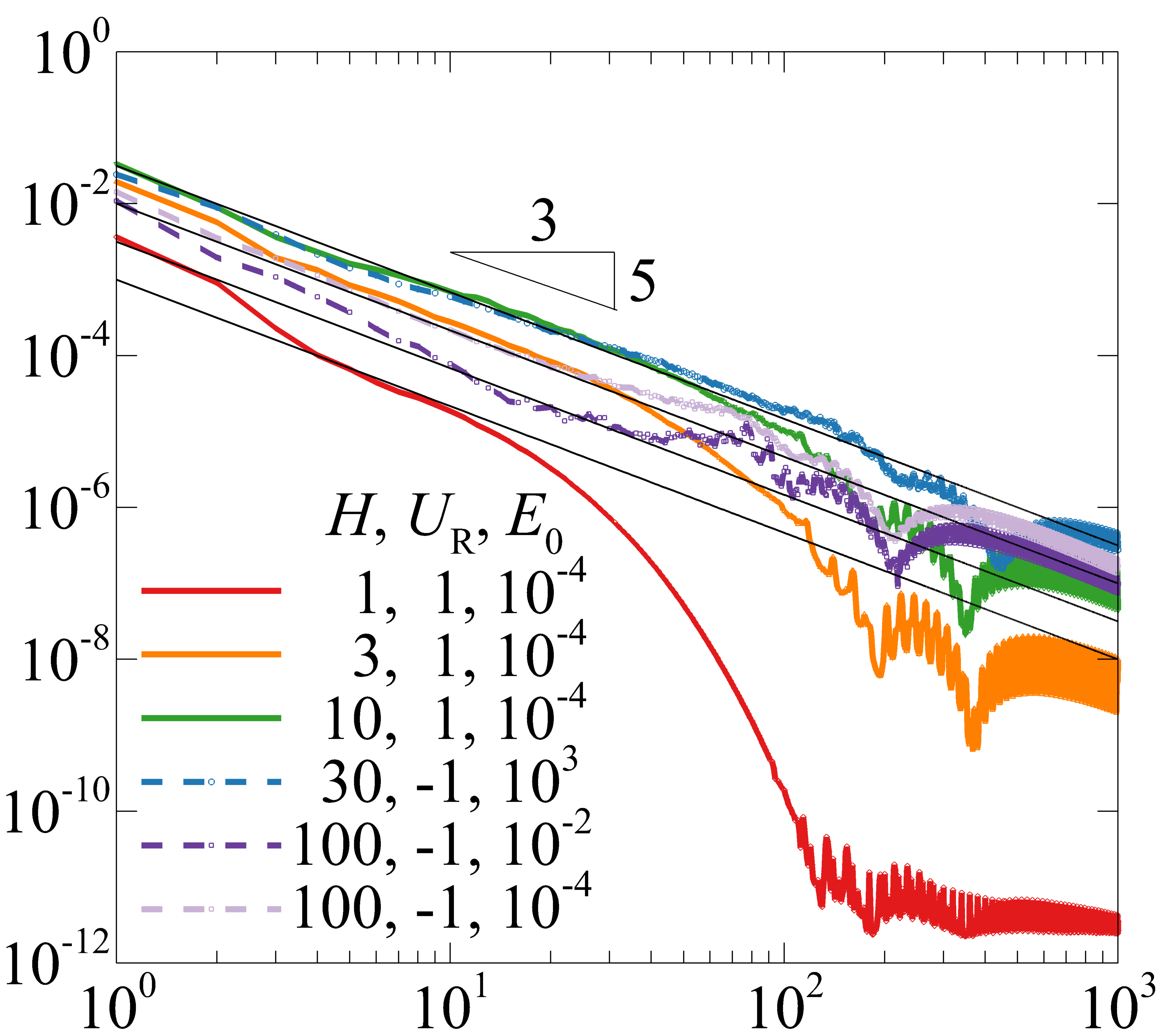}} &
\makecell{\vspace{17mm} \footnotesize{(b)} \\  \vspace{22mm} \rotatebox{90}{\footnotesize{$\meanfoco \, ,\, \int \hat{u}^2+\hat{v}^2 \mathrm{d}\Omega$}}} & \makecell{\includegraphics[width=0.458\textwidth]{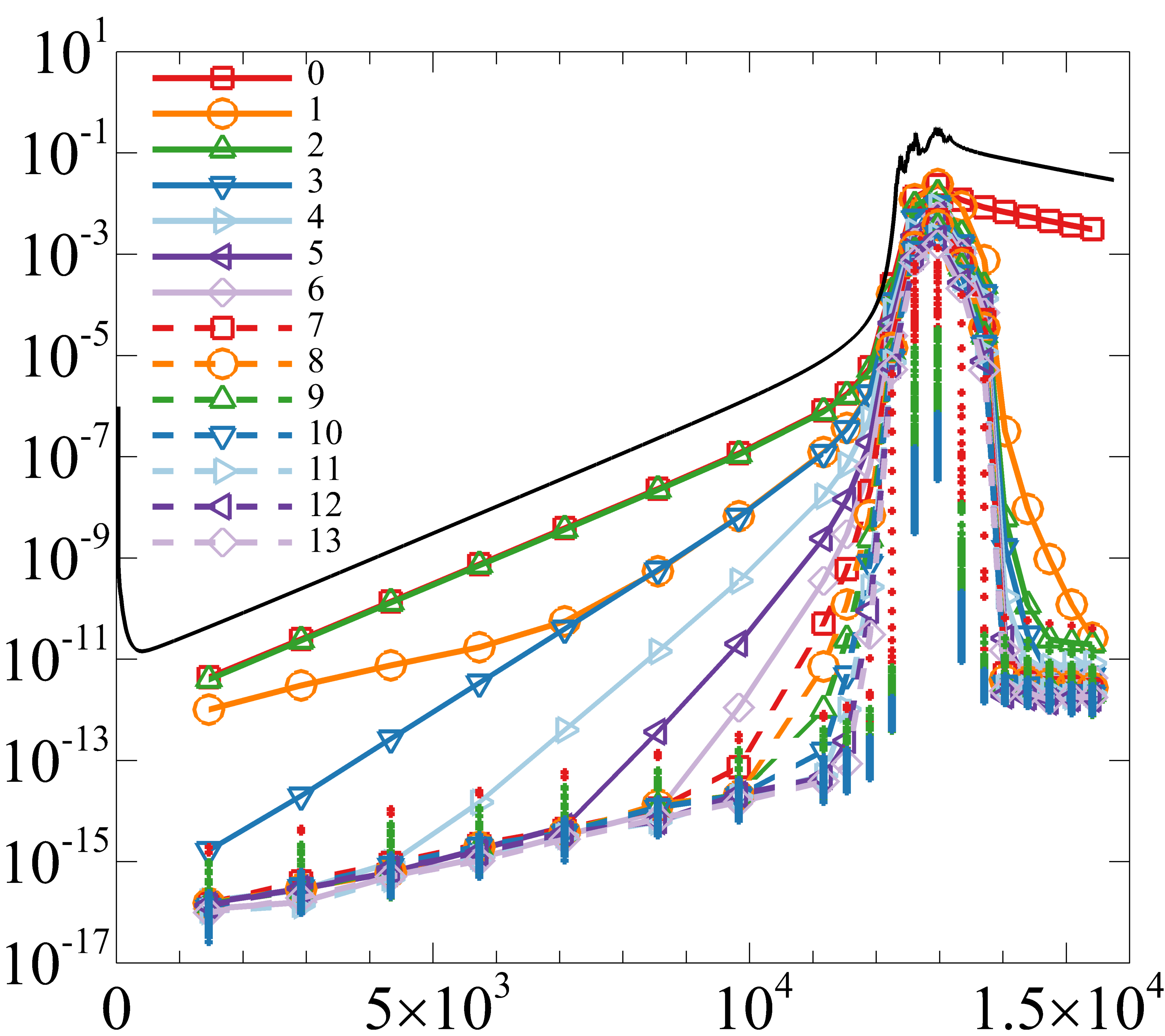}} \\
 & \hspace{38mm} \footnotesize{$\kappa$} &  & \hspace{38mm} \footnotesize{$\timeD$}\\
\end{tabular}
\addtolength{\tabcolsep}{+2pt}
\addtolength{\extrarowheight}{+10pt}
\end{center}
    \caption{(a) Time and $y-$averaged Fourier coefficients for $1<\kappa<10^3$, for both Shercliff and MHD-Couette flows at $\Rey/\ReyCrit=1.1$, for various $H$ and $E_0$. The thin black lines show  $\kappa^{-5/3}$ trends. $H \geq 30$ for Shercliff flow are not shown as the turbulence is short lived. (b) $y-$averaged Fourier coefficients as a function of rescaled time $\timeD$ for Shercliff flow at $\Rey/\ReyCrit=1.1$, $H=30$ and $E_0=10^{-6}$. Modes $0<\kappa<13$ are as defined in the legend. Thereafter every 5th mode is plotted, with $15<\kappa<100$ in red, $105<\kappa<1000$ in green and $1005<\kappa<5000$ in blue. The black solid line is twice the perturbation energy, identical to that from \fig\ \ref{fig:Poiseuille_DNS}(a).}
    \label{fig:fit53}
\end{figure}

\Figs\ \ref{fig:H4times10} and \ref{fig:fit53} depict the $y-$averaged Fourier coefficients $\meanfoco$ to compare the three different behaviors observed when $H \lesssim 1$ (high amplitude non-turbulent), $1<H<30$ (possibly long-lived turbulence) or $H \gtrsim 30$ (short-lived turbulence). Very few modes are energized throughout the linear region, with a rapid jump in the number of modes energized in the nonlinear growth phase. This is shown in any one of \fig\ \ref{fig:H4times10}(a-d), by comparing the fourth and fifth curves, which are closely spaced in time but which exhibit approximately an order of magnitude increase in the number of noticeably (relative to the floor) energized modes. For the $H=1$ case, there is then no further change in the general form of the $\meanfoco$ curves. However, for $H \geq 3$ even more modes continue to be energized, until the spectra are contaminated by under resolution for $\kappa \gtrsim 200$. This is also shown in \fig\ \ref{fig:fit53}(a) by comparing the time averaged $\meanfoco$, averaged only after the initial nonlinear growth.  Only the cases with $H \geq 3$, for either Shercliff or MHD-Couette flow, demonstrate a range of wave numbers with perturbation energy with a $\kappa^{-5/3}$ dependence, which suggests the formation of an inertial subrange. There is also a sudden jump in the spectral floor for cases with $H \geq 3$ (also shown in \fig\ \ref{fig:H4times10}, particularly at $H=10$, comparing the curves at times $\timeD = 1.2057\times 10^4$ and $\timeD = 1.2561 \times 10^4$). This is a good indication of a transition to turbulence, as the chaotic state with a limited number of excited modes becomes a turbulent state, where all available modes are excited. Conversely, the $H=1$ data do not hold to the $\kappa^{-5/3}$ dependence for any distinct inertial subrange of $\kappa$, and a floor of low energy modes is always observed, such that the low-$H$ state never becomes turbulent. The $H=30$ case in \fig\ \ref{fig:H4times10} also shows the resulting decay of the perturbation at larger times, with a rapid reduction in the number of energized modes, until the energy in all modes reaches the floor (the $-5/3$ trend holds briefly before this occurs). \Fig\ \ref{fig:fit53}(b) further supports the temporary turbulent nature of the flow in this $H=30$ case, with the clear energization of all modes at $\timeD \approx 1.25 \times 10^4$, and the rapid decay of all but the zeroth mode (the streamwise independent structure) shortly thereafter, at $\timeD \approx 1.4 \times 10^4$. It also provides a different means of viewing the energization of an increasing number of modes before noticeable nonlinear growth is achieved. 


\begin{figure}
\begin{center}
\addtolength{\extrarowheight}{-10pt}
\addtolength{\tabcolsep}{-2pt}
\begin{tabular}{ llll }
\footnotesize{(a)} & \hspace{4mm} \footnotesize{$t = 1.17\times10^4$, $(-2.17 < \hat{v} < 1.86)\times10^{-4}$} &  \footnotesize{(b)} & \hspace{4mm} \footnotesize{$(-1.93 < \hat{u} < 2.21)\times10^{-3}$}\\
\makecell{ \\  \vspace{7mm} \rotatebox{90}{\footnotesize{$y$}}} & \makecell{\includegraphics[width=0.458\textwidth]{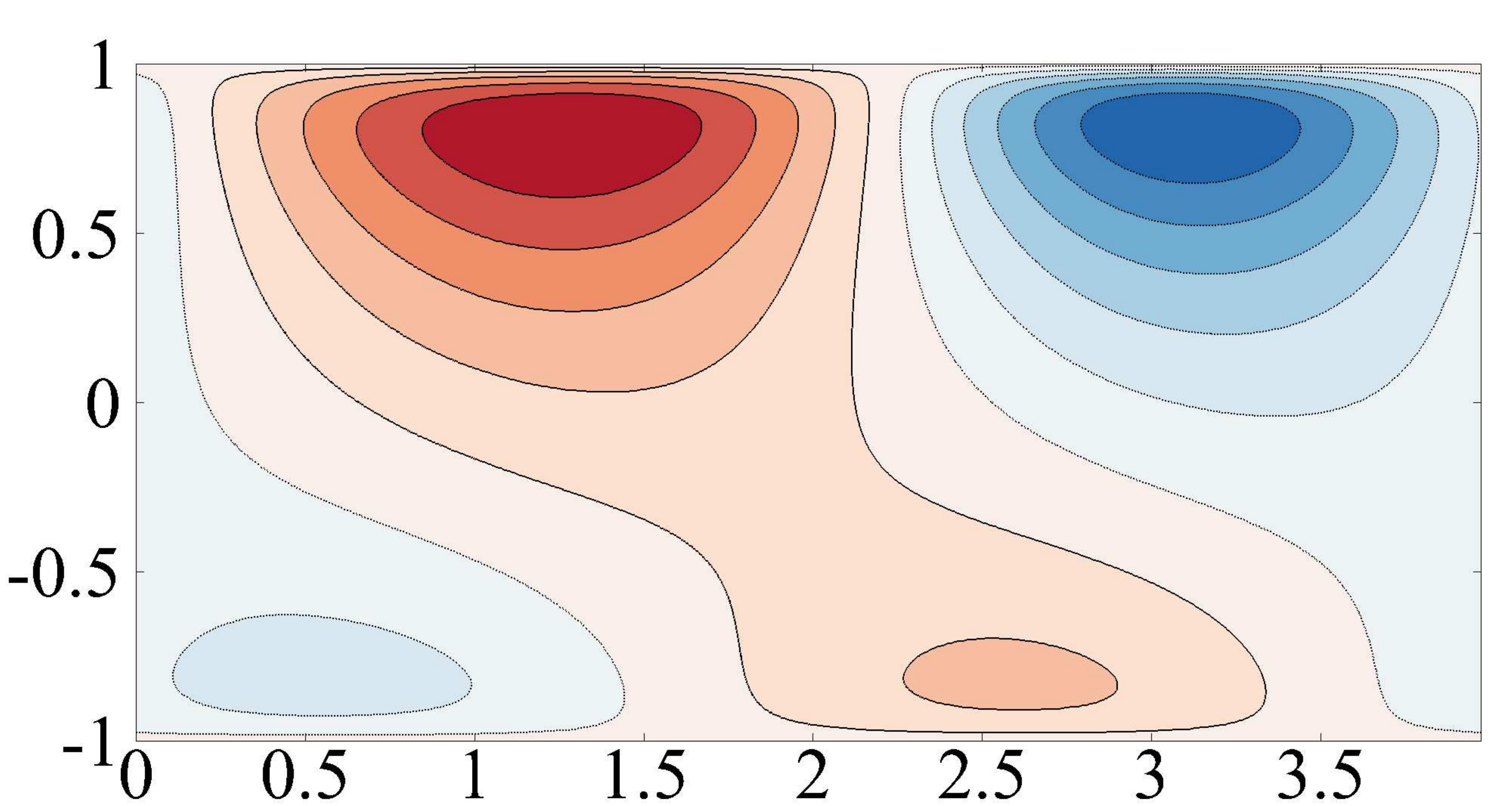}} &
\makecell{ \\  \vspace{7mm} \rotatebox{90}{\footnotesize{$y$}}} & \makecell{\includegraphics[width=0.458\textwidth]{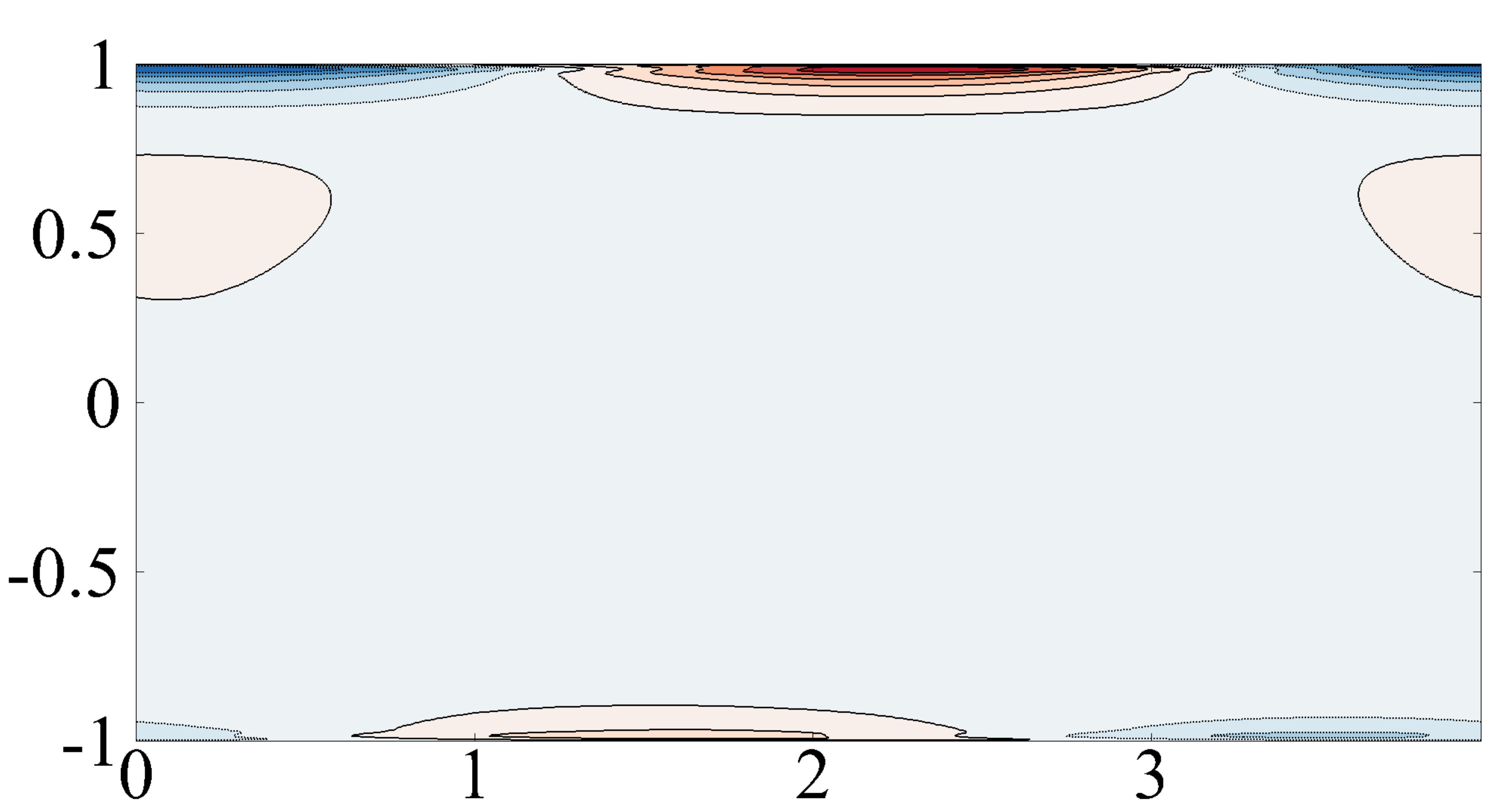}}  \\
 & \hspace{38mm} \footnotesize{$x$} & & \hspace{38mm} \footnotesize{$x$} \\
 \footnotesize{(c)} & \hspace{4mm} \footnotesize{$t = 1.70\times10^4$, $(-1.44 < \hat{v} < 1.63)\times10^{-1}$} &  \footnotesize{(d)} & \hspace{4mm} \footnotesize{$(-6.50 < \hat{u} < 3.87)\times10^{-1}$}\\
 \makecell{\\  \vspace{7mm} \rotatebox{90}{\footnotesize{$y$}}} & \makecell{\includegraphics[width=0.458\textwidth]{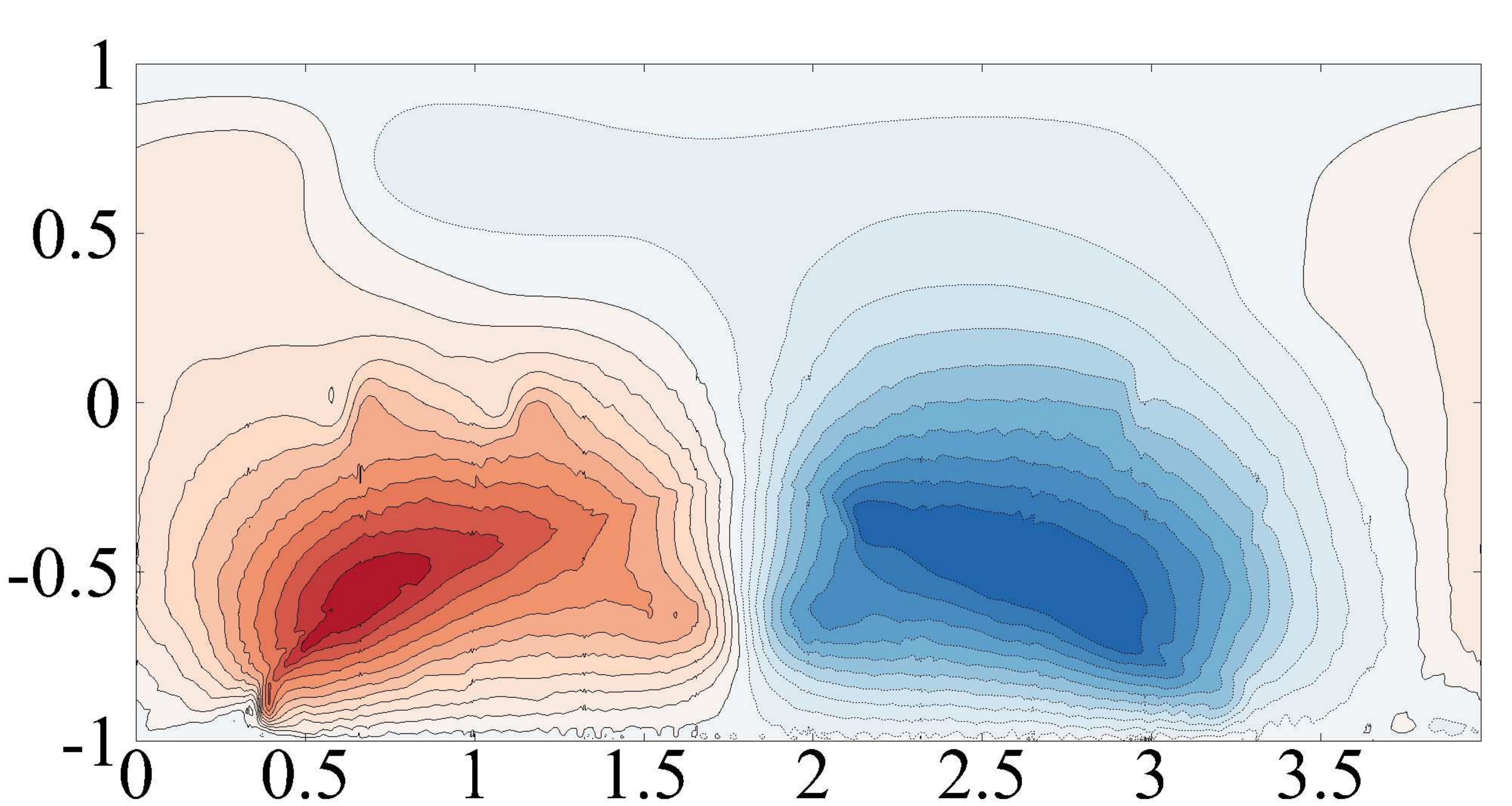}} &
\makecell{ \\  \vspace{7mm} \rotatebox{90}{\footnotesize{$y$}}} & \makecell{\includegraphics[width=0.458\textwidth]{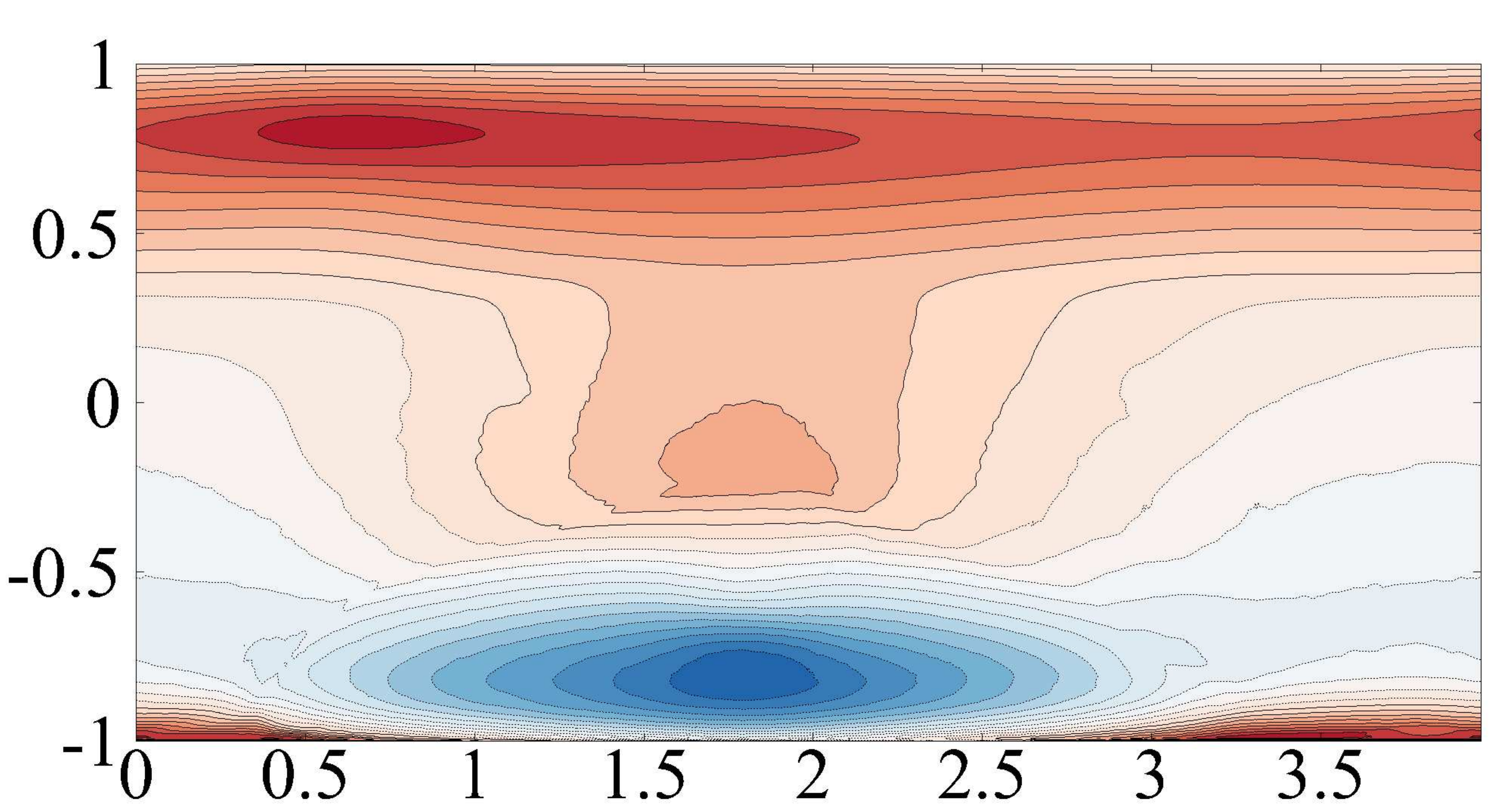}} \\
 & \hspace{38mm} \footnotesize{$x$} & & \hspace{38mm} \footnotesize{$x$} \\
\footnotesize{(e)} & \hspace{4mm} \footnotesize{$t = 2.15\times10^4$, $(-1.68 < \hat{v} < 1.58)\times10^{-3}$} &  \footnotesize{(f)} & \hspace{4mm} \footnotesize{$(-13.51 < \hat{u} < 6.33)\times10^{-2}$}\\
 \makecell{ \\  \vspace{7mm} \rotatebox{90}{\footnotesize{$y$}}} & \makecell{\includegraphics[width=0.458\textwidth]{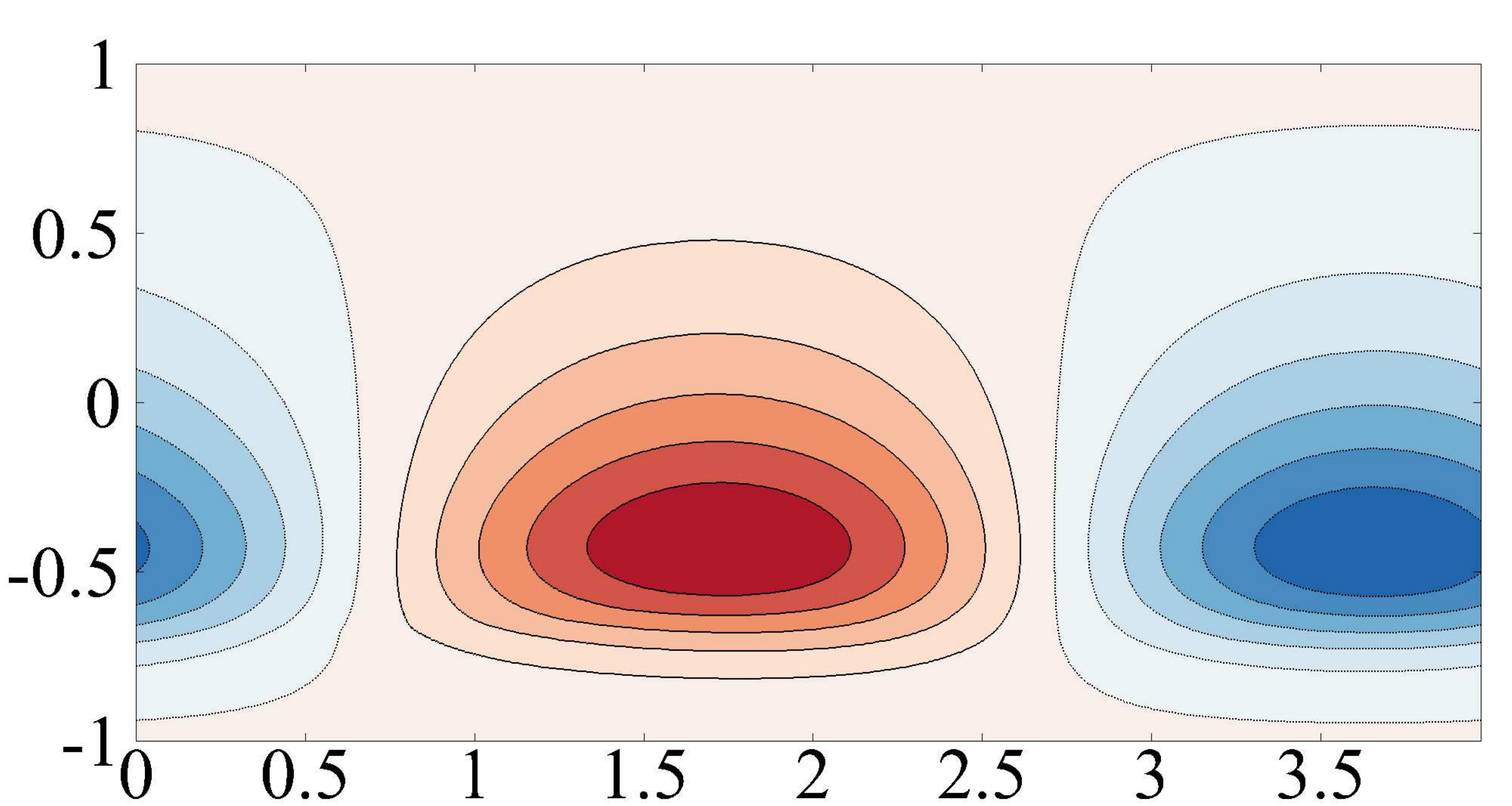}} &
\makecell{\\  \vspace{7mm} \rotatebox{90}{\footnotesize{$y$}}} & \makecell{\includegraphics[width=0.458\textwidth]{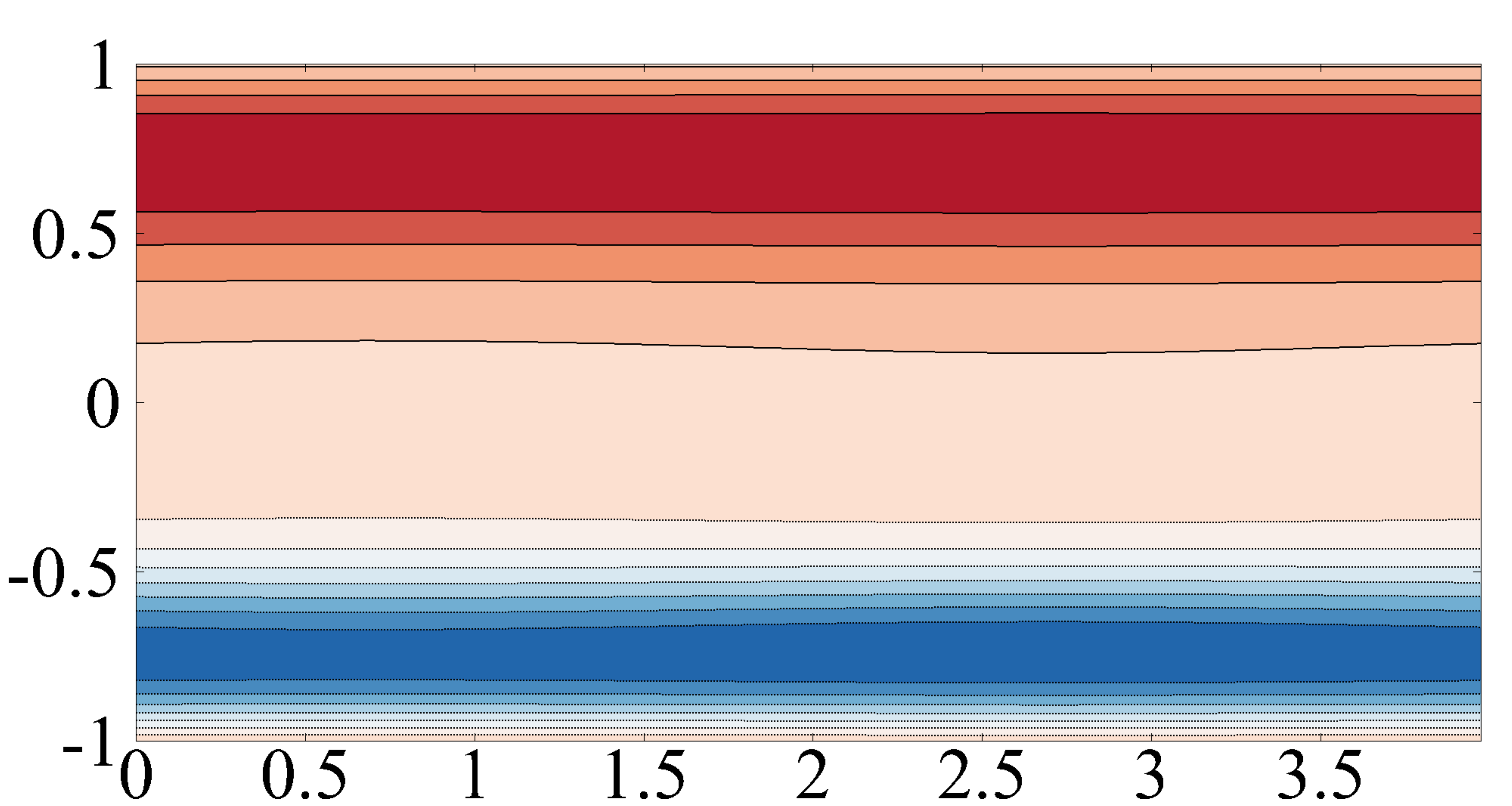}} \\
 & \hspace{38mm} \footnotesize{$x$} & & \hspace{38mm} \footnotesize{$x$} \\
\end{tabular}
\addtolength{\tabcolsep}{+2pt}
\addtolength{\extrarowheight}{+10pt}
\end{center}
    \caption{Temporal snapshots of MHD-Couette flow at $\Rey/\ReyCrit=1.1$, $H=100$, $E_0=10^{-2}$ (\fig\ \ref{fig:Couette_DNS}). Wall-normal velocity perturbation $\hat{v}$ (left); streamwise velocity perturbation $\hat{u}$ (right). Solid lines (red flooding) positive; dotted lines (blue flooding) negative. The linear \TS\ wave evolves into an arched \TS\ wave, leading to a turbulent state and the rapid growth of a streamwise independent structure. The \TS\ wave then flattens out and relaminarization occurs.}
    \label{fig:DNS_1e-02}
\end{figure}

Representative flow fields are depicted for MHD-Couette flow in \fig\ \ref{fig:DNS_1e-02}, at $H=100$, $\Rey/\ReyCrit=1.1$, $E_0=10^{-2}$ (energy growth depicted in \fig\ \ref{fig:Couette_DNS}). In the linear growth region, $2\times10^2 \lesssim  t\lesssim 1.24\times10^4$, a pattern of similar structure to the linear stability eigenvector  is observed in \fig\ \ref{fig:DNS_1e-02}(a) (recall \fig\ \ref{fig:lin_evecs_vary_H_sym}), although both the left and right running eigenvector are observable in the nonlinear computation. Most of the energy in $\hat{u}^2$ is located where gradients in $\hat{v}$ are largest, \emph{i.e}.\ very close to the walls. During the initial nonlinear growth period, $1.24\times10^4 \lesssim t\lesssim 1.26\times10^4$, the additional growth originates from the \TS\ wave arching, as visible in the $\hat{v}$ field in \fig\ \ref{fig:DNS_1e-02}(c), and the form of this dominant structure persists through the turbulent stage, $1.26\times10^4 \lesssim t\lesssim 1.65\times10^4$. Some underlying smaller scale features are also visible in \fig\ \ref{fig:DNS_1e-02}(c). This dominant modulated \TS\ wave can periodically break down and re-form (as energy is driven to larger scales) throughout the turbulent stages. Linear transient optimals were found to experience a secondary nonlinear growth through the same mechanism in isolated exponential boundary layers \citep{Camobreco2020role}, with a large-scale arched \TS\ wave structure similarly persistent. The appearance of $\hat{u}$ also starkly changes with the nonlinear growth and transition to turbulence, with two elongated streamwise structures rapidly forming at each wall, which tend to reduce the local shear. These structures store perturbation energy, as shown by the slow decay of the zeroth mode in \fig\ \ref{fig:fit53}(b), and by comparing \figs\ \ref{fig:Couette_DNS}(a) and (b). After relaminarization, $1.65\times10^4 \lesssim  t\lesssim 2.5\times10^4$, the \TS\ wave flattens out, \fig\ \ref{fig:DNS_1e-02}(e), is pushed away from the high shear region (by the streamwise independent structure), and cleanly decays. As the Reynolds number is supercritical, the linear mode is re-excited from noise at the numerical floor. 

\begin{figure}
\begin{center}
\addtolength{\extrarowheight}{-10pt}
\addtolength{\tabcolsep}{-2pt}
\begin{tabular}{ ll }
\footnotesize{(a)} & \hspace{5mm} \footnotesize{$H = 3$, $E_0 = 10^{-4}$, $\timeD = 1.11\times10^4$, $-0.733 < \hat{\omega}_z < 0.798$} \\  
\makecell{ \\  \vspace{8.8mm} \rotatebox{90}{\footnotesize{$y$}}} & \makecell{\includegraphics[width=0.875\textwidth]{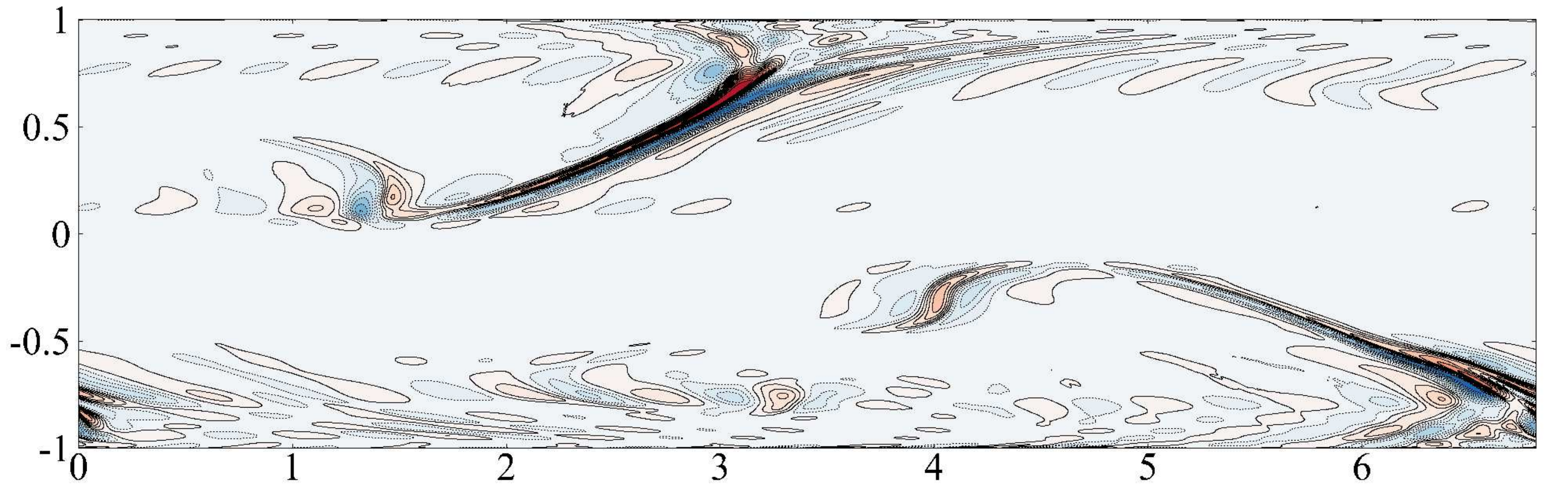}} \\
 & \hspace{72mm} \footnotesize{$x$} \\ 
\footnotesize{(b)} & \hspace{5mm} \footnotesize{$H=100$, $E_0 = 10^{-8}$, $\timeD = 1.28\times10^4$, $-7.788 < \mathrm{sgn}(\hat{v})\log_{10}(|\hat{v}|) < 7.876$}\\
 \end{tabular}
 \begin{tabular}{ ll }
\makecell{\\  \vspace{9.1mm} \rotatebox{90}{\footnotesize{$y$}}} & \makecell{\includegraphics[width=0.458\textwidth]{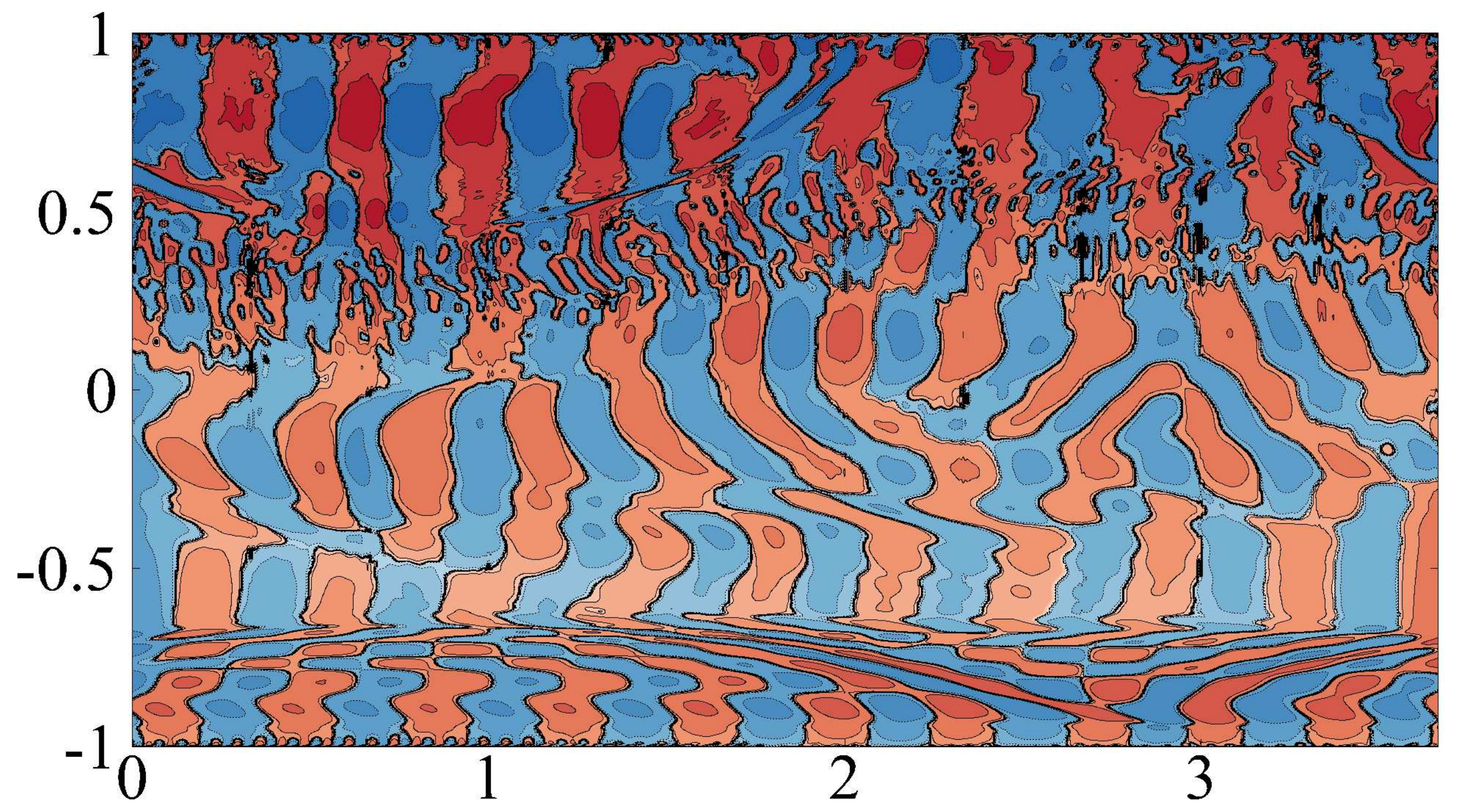}} \\
 & \hspace{38mm} \footnotesize{$x$}\\
\end{tabular}
\addtolength{\tabcolsep}{+2pt}
\addtolength{\extrarowheight}{+10pt}
\end{center}
    \caption{Streamwise high-pass-filtered snapshots of Shercliff flow shortly after the transition to turbulence at $\Rey/\ReyCrit=1.1$. Streamwise Fourier coefficients of modes $|\kappa| \leq 9$ have been removed. Solid lines (red flooding) positive; dotted lines (blue flooding) negative.}
    \label{fig:prove_tbt}
\end{figure}

However, the smaller turbulent scales in \fig\ \ref{fig:DNS_1e-02} are occluded by the dominance of the arched \TS\ wave. \Fig\ \ref{fig:prove_tbt} depicts two snapshots revealing key flow features present in the smaller scales. In these snapshots, a high-pass filter has been applied to remove streamwise Fourier modes $|\kappa| \leq 9$ from the flows. Strongly localized jets emanating from the side-walls entrain narrow shear layers, observable in the example at $H=3$, while at $H=100$ a myriad of smaller scale features are present.

\section{Conclusions}\label{sec:conclusions}
%
This work examined the influence of the base flow in the scenario of transition to 
turbulence in a \qtwod\ duct flow with a transverse magnetic field. The base flow is 
varied through the relative velocity of the two lateral walls. This is of particular importance in the context of recent developments in flow control, where turbulence is suppressed via the introduction of a friction effect to flatten the base flow 
\citep{Hof2010eliminating, Kuhnen2018destabilizing, Marensi2019stabilisation}. 
Ideas along the same lines can be conversely applied to the promotion, rather than the suppression, of turbulence. Promoting turbulence to enhance heat transfer is indeed necessary for one the motivations of this work: to assess the feasibility of dual-purpose liquid metal coolant ducts in magnetic confinement fusion reactors \citep{Smolentsev2008characterization}. Fluid structures have a strong tendency to two-dimensionalize within these ducts, which exhibit naturally flat base flows, due to the action of the Lorentz force.
The linear stability of \qtwod\ \CouPois\ base flows provided two key insights. First, the addition of any amount of antisymmetry to the base flow eventually leads to unconditional stability to infinitesimal perturbations at low enough friction parameters $H$. The reason is that the antisymmetric part of the 
base flow drives the \TS\ wave structures to destructively interfere, preventing growth. 
Conversely, an increasing friction parameter, beyond a critical value $H^\infty$, flattens the central region of the base flow and isolates the wave structures at each wall, limiting their interaction, allowing growth to occur at finite critical Reynolds numbers. 
$H^\infty$ increases with decreasing velocity of the bottom wall $\UsubR$, which controls the level of antisymmetry in the base flow (the top wall is at fixed velocity of unity). Only a symmetric, Shercliff base flow has finite $\ReyCrit$ for all non-zero $H$.
Second, the critical parameters collapse to those of an isolated exponential boundary layer at high $H$, which occurs with noticeably lower imposed friction for increasingly antisymmetric base flow profiles.
Antisymmetric profiles have a larger base flow velocity gradient at one wall than the other, leading the \TS\ wave instability to preferentially form at only the one wall where the mean shear is largest. In such cases, friction need only keep the instability sufficiently far from the other wall to avoid any interference. This requires comparatively less friction than isolating two waves from one another (the greatest constructive interference thereby occurs in the symmetric Shercliff flow). 

Conversely, the energetics of all Q2D \CouPois\ flows show little dependence on the degree of antisymmetry in the base flow. As such, the energetic Reynolds numbers are always finite. Furthermore, the transient growth of Q2D \CouPois\ flows is also not strongly dependent on the degree of antisymmetry in the base flow, with variations in growth between base flows only visible at $H \leq 10$. Destructive interference could explain the slight reduction in transient growth for more strongly antisymmetric base flows when $H$ is small enough to permit interference. At larger friction parameters, $H \gtrsim 30$, transient growth is almost identical for all base flows. The growth attained is equivalent to that of an isolated exponential boundary layer \citep{Camobreco2020role} and is increasingly damped with increasing  $H$. Given that $H$ would be of order $10^4$ in a realistic fusion environment, linear transient growth may not be very relevant in their context.

The weakly nonlinear analysis also compounds the difficulties in promoting turbulence in realistic fusion environments, given the scaling of the equilibrium amplitude with $H^{5/2}$ for all base flows. However, the weakly nonlinear analysis still indicates the possibility of subcritical transitions for any $H$.
Supercritical bifurcations are only found along the lower branch of the neutral curve, and only for $H \gg H^\infty$. At lower friction parameters, for base flows with any degree of antisymmetry, the entire computed neutral curve indicates a subcritical bifurcation. 

As the transient growth depicted little base flow dependence, and has been previously analysed in Ref.~\cite{Camobreco2020role}, direct numerical simulations target the exponential growth predicted by the linear stability analysis. There are two key findings. First, the relaminarization of turbulent states in symmetric Shercliff flows always occured through a monotonic decay, while MHD-Couette flows experienced re-excitation to a turbulent state, in some cases at amplitudes where nonlinearity was relevant. Second, the magnitude of the friction parameter seemed to largely dictate the ability to trigger turbulence. At low $H \leq 1$, the linear and nonlinear growth led only to a saturated state, without turbulence. At intermediate $3 \leq H \leq 10$, a transition to turbulence was observed, and at $H=10$ the turbulence state was maintained to the computed extent of simulations. Fourier analysis also indicated the presence of an intertial subrange, where the perturbation energy exhibited a wave number dependence of $\kappa^{-5/3}$.  At higher $H \geq 30$ (the bound above which transient growth is equivalent to that of an isolated exponential boundary layer), although transition was observed, the turbulent state quickly collapsed. In all cases, the nonlinear growth, and turbulence, was dominanted by a persistent large scale arched \TS\ wave. Streamwise independent structures also formed, which stored perturbation energy and which reduced the gradients in the boundary layers. Overall, the general features of the secondary nonlinear growth mirror the secondary nonlinear growth of the finite amplitude linear transient optimals simulated in Ref.~\cite{Camobreco2020role}, where nonlinear growth is due to the arching of the conventional \TS\ wave.

As a final word, the results of this paper indicate that it may be exceedingly difficult to obtain Q2D subcritical transitions with random, and even optimized, initial conditions. Future work may therefore be best focussed on directly reducing $\ReyCrit$, permitting supercritical transitions at lower Reynolds numbers. Inflection points, introduced to the base flow with increasing antisymmetry, were not beneficial in this work due to their location. However, investigating the capabilities of inflection points within the boundary layers remains as a promising avenue for destabilizing Q2D flows (an ongoing work), which can be achieved through the use of time-periodic, rather than steady, wall motion.




\begin{acknowledgements}
The authors are grateful to J\={a}nis Priede for discussions regarding the implementation of the weakly nonlinear stability analysis. C.J.C.\ receives an Australian Government Research Training Program (RTP) Scholarship. A.P.\ is supported by  the Royal Society (Wolfson Research Merit Award Scheme grant WM140032). This research was supported by the Australian Government via the Australian Research Council (Discovery Grants DP150102920 and DP180102647), the National Computational Infrastructure (NCI), Pawsey Supercomputing Centre (PSC), Monash University via the MonARCH cluster, and by the Royal Society under the International Exchange Scheme between the UK and Australia (grant E170034).
\end{acknowledgements}

\end{document}